\documentclass[journal,twocolumn]{IEEEtran}
\usepackage[T1]{fontenc}
\usepackage[latin9]{inputenc}
\usepackage{amsmath}
\usepackage{amsthm}
\usepackage{amssymb}
\usepackage{graphicx}
\usepackage[unicode=true,
 bookmarks=true,bookmarksnumbered=true,bookmarksopen=true,bookmarksopenlevel=1,
 breaklinks=false,pdfborder={0 0 0},pdfborderstyle={},backref=false,colorlinks=false]
 {hyperref}
\hypersetup{pdftitle={Your Title},
 pdfauthor={Your Name},
 pdfborderstyle={},pdfpagelayout=OneColumn,pdfnewwindow=true,pdfstartview=XYZ,plainpages=false}

\makeatletter

\providecommand{\tabularnewline}{\\}

  \theoremstyle{definition}
  \newtheorem{defn}{\protect\definitionname}
  \theoremstyle{remark}
  \newtheorem{rem}{\protect\remarkname}
  \theoremstyle{plain}
  \newtheorem{thm}{\protect\theoremname}
  \theoremstyle{plain}
  
  \theoremstyle{plain}
  \newtheorem{cor}{\protect\corollaryname}
  \theoremstyle{plain}
  \newtheorem{lem}{\protect\lemmaname}
  \theoremstyle{definition}
  \newtheorem{example}{\protect\examplename}

\usepackage{cite}
\usepackage{color}

\providecommand{\corollaryname}{Corollary}
\providecommand{\definitionname}{Definition}
\providecommand{\examplename}{Example}
\providecommand{\lemmaname}{Lemma}
\providecommand{\remarkname}{Remark}
\providecommand{\theoremname}{Theorem}

\@ifundefined{showcaptionsetup}{}{%
 \PassOptionsToPackage{caption=false}{subfig}}
\usepackage{subfig}
\makeatother

\providecommand{\corollaryname}{Corollary}
\providecommand{\definitionname}{Definition}
\providecommand{\examplename}{Example}
\providecommand{\lemmaname}{Lemma}
\providecommand{\remarkname}{Remark}
\providecommand{\theoremname}{Theorem}

\begin{document}

\title{A Unified Theory of Multiple-Access and Interference Channels via Approximate Capacity Regions for the MAC-IC-MAC}

\author{Yimin~Pang,~\IEEEmembership{Student Member,~IEEE,} and~Mahesh~Varanasi,~\IEEEmembership{Fellow,~IEEE}\thanks{This research was supported in part by NSF Grant 1423657 and by Qualcomm Gift 24868. It was presented
in part at the 51st Annual Allerton Conference on Communication, Control,
and Computing (Allerton), 2013.}\thanks{Yimin Pang is with the Department of Electrical, Computer and Energy
Engineering, University of Colorado Boulder, Boulder, US, e-mail:
\protect\protect\protect\href{mailto:yipa5803@colorado.edu}{yipa5803@colorado.edu}.}\thanks{Mahesh~K.~Varanasi is with the Department of Electrical, Computer
and Energy Engineering, University of Colorado Boulder, Boulder, US,
e-mail: \protect\protect\protect\href{mailto:varanasi@colorado.edu}{varanasi@colorado.edu}.}}
\maketitle
\begin{abstract}
Approximate capacity regions are established for a class of interfering
multiple access channels consisting of two multiple-access channels
(MACs), each with an arbitrary number of transmitters, with one transmitter in each MAC causing interference to the receiver of the other MAC,
a channel we refer to henceforth as the MAC-IC-MAC. For the discrete memoryless (DM) MAC-IC-MAC, two inner bounds are obtained that are generalizations of prior inner bounds for the two-user DM interference channel (IC) due to Chong {\em et al}.  For the semi-deterministic MAC-IC-MAC, it is shown that single-user coding at the non-interfering transmitters and superposition coding at the interfering
transmitter of each MAC achieves a rate region that is within a quantifiable
gap of the capacity region, thereby extending such a result for the two-user semi-deterministic IC by Telatar and Tse. For the Gaussian MAC-IC-MAC, an approximate capacity region that is within a constant gap of the capacity region is obtained, generalizing such a result for the two-user Gaussian IC by Etkin {\em et al}. Contrary to the aforementioned approximate capacity results for the two-user IC whose achievability requires the union of all admissible input distributions, our gap results on the semi-deterministic and the Gaussian MAC-IC-MAC are achievable by only a subset and one of all admissible coding distributions, respectively. The symmetric generalized
degrees of freedom (GDoF) of the symmetric Gaussian MAC-IC-MAC with
more than one user per cell, which is a function of the interference
strength (the ratio of INR to SNR at high SNR, both expressed in dB)
and the numbers of users in each cell, is V-shaped with flat shoulders. An analysis based on signal-level partitions shows that the non-interfering transmitters utilize the signal-level partitions at the receiver where they are intended that cannot be accessed by the interfering transmitters (due to the restriction of superposition coding), thereby
improving the sum symmetric GDoF to up to one degree of freedom per cell under a range of SINR exponent levels, which in turn becomes wider as the number of transmitters in each cell increases. Consequently, time-sharing between interfering and non-interfering transmitters is GDoF-suboptimal in general, as is time-sharing between the two embedded MAC-Z-MACs.

\end{abstract}


\section{Introduction}

\begin{figure}[h]
\begin{centering}
\subfloat[A cellular network\label{fig:mim-ex-1}]{\begin{centering}
\includegraphics[width=1.65in]{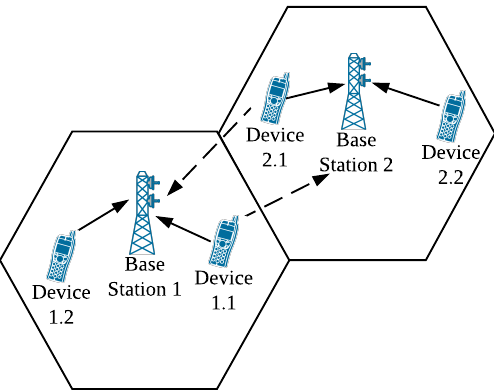} 
\par\end{centering}
}\subfloat[A femto cell network\label{fig:mim-ex-2}]{\begin{centering}
\includegraphics[width=1.65in]{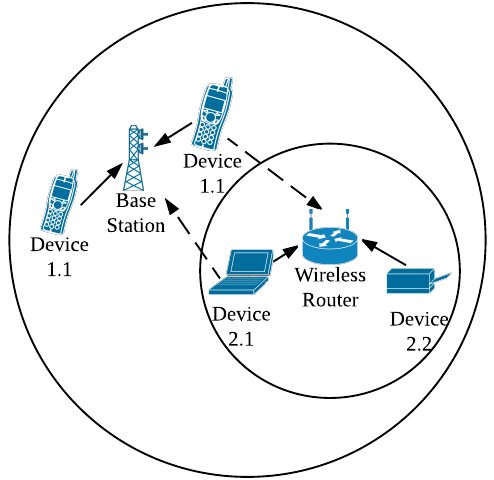} 
\par\end{centering}
}
\par\end{centering}
\caption{{\small{}Examples of the MAC-IC-MAC}\label{fig:mim-ex}}
\end{figure}

Due to the rapid increase of data demands in recent years, wireless
co-band communication has drawn significant interest in both theory
and practice. Bluetooth and Wi-Fi have both been established on 2.4
GHz band, and more recently 3GPP introduced LTE Licensed Assisted
Access (LAA) to offload LTE packets to unlicensed spectrum at 5 GHz,
which causes Wi-Fi and LAA to coexist. Such emerging technologies
motivate the study of co-band interference between cellular networks
in network information theory. In this paper, we obtain bounds on the capacity region and approximate
capacity regions for the discrete-memoryless, semi-deterministic and Gaussian classes of mutually interfering two-cell MAC networks
in which there is interference from one of the transmitters of the
MAC to the receiver of the other MAC. For brevity, we refer to this
two-cell network as the MAC-IC-MAC. As shown in Fig.\,\ref{fig:mim-ex},
the MAC-IC-MAC captures many practical communication scenarios. Fig.\,\ref{fig:mim-ex-1}
shows an uplink cellular network where devices 1.1 and 2.1 are located
on the cell edge so that interference paths exist between the two
cells as indicated by the dashed arrows, whereas devices 1.2 and
2.2 do not cause interference to the neighboring cell due to their
favorable access location in their own cell. Fig.\,\ref{fig:mim-ex-2}
represents a femtocell network, where similar partial interference
exists between the macrocell and femtocell. The study of the capacity
region of MAC-IC-MAC could therefore provide an approach to increasing
uplink throughput and cell edge spectrum efficiency for co-band networks.

From a theoretical point of view, this paper on the MAC-IC-MAC can be seen as the outcome  of an effort to unify and generalize the capacity results for the MAC \cite{liao1972multiple,ahlswede1973multi,wyner1974recent} on the one hand, and
the more recent capacity approximations for the two-user IC  \cite{carleial1978interference,han1981new,chong2006comparison,kobayashi2007further,telatar2007bounds,chong2008han,etkin2008gaussian} on the other. The main results of this paper
are described next.

\subsubsection*{1) Two achievable rate regions for the DM MAC-IC-MAC}

We unify and generalize the capacity
region of the DM MAC, seen as a tight inner bound, and the best known
inner bounds for the two-user DM IC due to \cite{han1981new,chong2006comparison,kobayashi2007further,chong2008han}, 
to obtain two inner bounds for the DM MAC-IC-MAC.

The coding scheme we employ is what would be deemed the most natural
for it: each of the two interfering transmitters employ rate-splitting
and superposition coding (the CMG scheme of \cite{chong2008han})
and the non-interfering transmitters in each MAC employ single-user
random coding. An analysis of such a coding scheme results in an inner
bound for the capacity region that is a polytope in a number of dimensions
that is two more than the total number of transmitters due to rate-splitting
at the two interfering transmitters. Such a polytope is defined by
an indeterminate number of sum-rate inequalities because there is no restriction
on the number of users in each cell, and the problem lies in
the elimination of the two auxiliary rate variables. Utilizing the
particular structure of this polytope, and using a form of
structured Fourier-Motzkin elimination, the two split rates are projected
out to obtain a explicit polyhedral description of the achievable rate region. The union of such polytopes over all admissible coding distributions is an inner bound as well but two sets of inequalities defining the polytopes are shown to be redundant in this case, thereby generalizing the CMG inner bound for the two-user DM IC in \cite{chong2008han}, named the compact HK region therein.

\subsubsection*{2) A quantifiable gap to the capacity region of the semi-deterministic MAC-IC-MAC}

We also unify and extend the capacity region of the DM MAC, viewed as a tight outer bound, and the Telatar-Tse outer bound for the semi-deterministic two-user DM IC of \cite{telatar2007bounds}\textemdash
which is within a quantifiable gap of the capacity region\textemdash to the semi-deterministic MAC-IC-MAC, to obtain an outer bound for the latter,
while also assuring a similar quantifiable gap to its capacity region. In extending the outer bound of \cite{telatar2007bounds}, certain set functions have to be identified in order to handle exponentially many partial sum rate restrictions and the appropriate genie information must be chosen to extend the genie-aided argument of \cite{telatar2007bounds} to the semi-deterministic MAC-IC-MAC in a natural way. 


\subsubsection*{3) Constant gap to capacity for the Gaussian MAC-IC-MAC}

The capacity region of the (scalar) Gaussian MAC is well-known \cite{wyner1974recent} and that of the Gaussian IC was characterized to within one bit in \cite{etkin2008gaussian}. In this paper, we obtain a one-bit gap to capacity region approximation for the Gaussian MAC-IC-MAC and a two-bit gap to capacity region attainable with a single coding scheme (i.e., with a single distribution). This clarifies that the previously obtained one-bit gap result on the Gaussian IC of \cite{etkin2008gaussian} involves the consideration of all possible coding distributions in its achievable scheme, while also demonstrating that a two-bit gap result for the two-user IC can be shown to be achievable by the single random coding scheme  proposed in \cite{etkin2008gaussian}. 

\subsubsection*{4) The generalized degrees of freedom (GDoF) region for the Gaussian
MAC-IC-MAC}

\begin{figure}[tbh]
\begin{centering}
\includegraphics{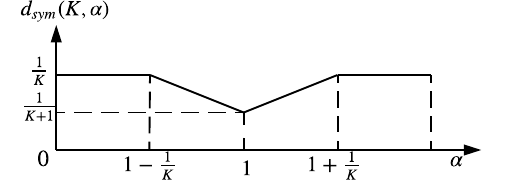} 
\par\end{centering}
\caption{Symmetric GDoF of Gaussian MAC-IC-MAC, when each MAC contains more
than one user. \label{fig:mim-sym-gdof}}
\end{figure}

The  constant gap capacity approximation for the Gaussian MAC-IC-MAC yields its GDoF region. In particular, the symmetric GDoF curve, shown in Fig.\,\ref{fig:mim-sym-gdof}, which is a function of $\alpha$, the interference strength (the ratio of INR to SNR at high SNR, both
expressed in dB) and the number of users in a cell, is V-shaped with
flat shoulders on both sides. It reveals that in a Gaussian
MAC-IC-MAC with $K \geq 2 $ users per cell with equal rates, all transmitters can send information at an approximately interference-free
rate in the high SNR regime when $\alpha=\frac{\log\mathsf{INR}}{\log\mathsf{SNR}}\in[0,1-\frac{1}{K}]\cup[1+\frac{1}{K},\infty)$. As a byproduct, we have that the sum symmetric DoF (i.e., GDoF at $\alpha =1 $) of the $K$-user symmetric Gaussian MAC-IC-MAC is $\frac{K}{K+1}$. 

When $K=1$, the MAC-IC-MAC is the two-user IC, and the symmetric GDoF is given by the well-known "W" curve [10]. For a comparison of the sum symmetric GDoF of the MAC-IC-MAC for $K=1, 2,3 $ and $4$, see Fig. \ref{fig:mim-sym-gdof-num}.

In Section \ref{subsec:mim-gdof}, a signal-level (or power partition) method is described that is inspired by the works of \cite{karmakar2012generalized}, using which a few examples are discussed to explain the attainability of the sum symmetric GDoF in the symmetric MAC-IC-MAC with two transmitters per cell, for various values
of $\alpha$. The insight gained here is that the non-interfering transmitters utilize the signal-level partitions at the receiver where they are intended that cannot be accessed by the interfering transmitters (due to the restriction of superposition coding), thereby improving the sum symmetric GDoF to up to one degree of freedom per cell under a range of SINR exponent levels, which in turn becomes wider as the number of transmitters in each cell increases. 


\subsection{Related Previous Works}

Previous related works are summarized as follows. Multiple access
channels are the best understood multi-terminal networks with the
capacity region determined by  Liao \cite{liao1972multiple}, Ahlswede \cite{ahlswede1973multi} and Wyner \cite{wyner1974recent}. Some of the key papers on two-user interference
channels are \cite{carleial1978interference,han1981new,chong2006comparison,kramer2006review,chong2008han,telatar2007bounds,etkin2008gaussian,karmakar2012generalized,karmakar2013capacity}.

For the interference channel, the Han-Kobayashi achievable scheme
(HK scheme) in \cite{han1981new}, as well as its alternative, the CMG scheme of \cite{chong2008han}, give the (same) best inner bound to the capacity region known to date. Telatar and Tse \cite{telatar2007bounds} found an outer
bound for the class of semi-deterministic interference channels and
quantified the gap to the CMG inner bound. For Gaussian scalar and
vector interference channels, Etkin et al \cite{etkin2008gaussian}
and Karmakar and Varanasi \cite{karmakar2013capacity} characterized
approximate capacity regions to within constant, channel coefficient-
and SNR-independent gaps, respectively. 

For the two-cell interfering multiple-access channel with an arbitrary
number of transmitters in each cell, the work by \cite{suh2008interference}
used interference alignment 
to achieve
interference-free degrees of freedom when the number of users in each
cell goes to infinity. Perron et al. \cite{perron2009interference}
defined a type of multiple-access interference channel with only four
nodes, where one of the receivers must decode the messages from both
transmitters. The capacity region within a quantifiable gap was obtained
for the semi-deterministic case. Chaaban and Sezgin studied a
fully connected two-cell channel in which a two-user
MAC interferes with a point-to-point link \cite{chaaban2011capacity}.
The capacity region is found for very strong and some cases of strong
interference and upper and lower bounds on the sum-rate in the weak
interference regime (with the lower bound achievable by treating interference
as noise) are also obtained. Subsequently, in \cite{gherekhloo2014sub},
they showed that when the interference is weak, treating interference
as noise in their model is sub-optimal. The cognitive radio version
of that model was studied by the same authors in \cite{chaaban2011capacitya}.
Buhler and Wunder \cite{buhler2012multiple} derived upper bounds
on the sum rate and an achievable scheme for the linear deterministic
version of the model in \cite{chaaban2011capacity}. Fritschek and
Wunder obtain a result on the reciprocity between the two-cell deterministic
interfering MAC (IMAC) and the two-cell deterministic interfering
broadcast channel (IBC) in \cite{fritschek2014upper}, and obtain
an achievable region under a weak interference condition for both
those channels. In \cite{fritschek2015constant}, the deterministic
IMAC was revisited using the lower triangular deterministic model
introduced by \cite{niesen2013interference}, and a constant gap sum
capacity as well as the sum GDoF were obtained. For the Gaussian IMAC,
Fritschek and Wunder \cite{fritschek2014enabling} close the gap between
the achievable sum rate regions for Gaussian IMAC and the deterministic
IMAC. Their coding scheme employs signal scale alignment and lattice
coding. Zhu et al. \cite{zhu2014capacity} studied the interference
Z MAC, a special case of the channel studied in \cite{chaaban2011capacity}, with only one-sided interference links from the two MAC transmitters to the point-to-point link receiver. The authors therein obtained the capacity region or the sum capacity under certain channel conditions using superposition
encoding and joint decoding. In the conference version of this paper
\cite{pang2013bounds}, an approximate capacity region was presented
for a special case of the MAC-IC-MAC with a two-user MAC in the first
cell and a point-to-point link in the second cell\footnote{Since we are studying a more general channel model here than the conference
version of this paper in \cite{pang2013bounds}, the missing proofs
in \cite{pang2013bounds} can be completed by setting $K_{1}=2$ and
$K_{2}=1$ in the results and proofs in Section III and Appendices \ref{sec:mim-errors}.
Typographical errors in \cite{pang2013bounds} are also corrected
therein.}.

\subsection{Notations}

The notations used throughout the paper are summarized as follows.
The $j$-th user in the $i$-th cell is indexed as $i.j$, where $i\in\{1,2\}$,
$j\in\{1,\cdots,K_{i}\}$ and $K_{i}$ is the number of the user in
cell-$i$. Hence, the $j$-th transmitter in the $i$-th cell is denoted
as Tx$i.j$, whose message, transmit symbol, and rate are denoted
as $M_{i.j}$, $X_{i.j}$ and $R_{i.j}$, respectively.

Let $\varTheta_{i}$ be the set of indices of all users in the $i$-th
cell, i.e. $\varTheta_{i}=\{i.1,\cdots,i.K_{i}\}$. For the sake of
convenience, the $K_{i}$-tuples of messages, input symbols and rates
of users in cell $i$ be denoted as $M_{\varTheta_{i}}$, $X_{\varTheta_{i}}$
and $R_{\varTheta_{i}}$. For example, the input symbols of cell-1
$\{X_{1.1},\cdots,X_{1.K_{1}}\}$ are denoted simply as $X_{\varTheta_{1}}$.
Similarly, $M_{\varTheta_{1}}$ denotes the $K_{1}$-tuple of messages
$\{M_{1.1},\cdots,M_{1.K_{1}}\}$, and $R_{\varTheta_{1}}$ denotes
the $K_{1}$-tuple of their rates $\{R_{1.1},\cdots,R_{1.K_{1}}\}$.

Throughout, we let $\varOmega_{i}$ denote any non-empty subset of
$\varTheta_{i}$, i.e., $\varOmega_{i}\in2^{\varTheta_{i}}\backslash\emptyset$,
where $2^{\varTheta_{i}}$ is the power set of $\varTheta_{i}$. Moreover,
we let $\varUpsilon_{i}$ denote any non-empty subset of $\varTheta_{i}$
that necessarily contains the element $i.1$. The sets $\bar{\varUpsilon}_{i}$
and $\bar{\varOmega}_{i}$ are defined as the complements of $\varUpsilon_{i}$
and $\varOmega_{i}$ relative to $\varTheta_{i}$. Furthermore, the
collection of input symbols of users indexed by elements of $\varUpsilon_{i}$
or $\varOmega_{i}$ are written as $X_{\varUpsilon_{i}}$ and $X_{\varOmega_{i}}$.

We use capital letters to denote random variables or sequences, such
as $X_{i.j}$ and $X_{i.j}^{n}$, where $X_{i.j}^{n}=(X_{i.j1},\cdots,X_{i.jn})$,
so that the $t$-th random variable of the random sequence $X_{i.j}^{n}$
is denoted by $X_{i.jt}^{n}$. The underlying alphabets are denoted
by $\mathcal{X}_{i.j}$ and $\mathcal{X}_{i.j}^{n}$, and specific
values of $X_{i.j}$ and $X_{i.j}^{n}$ by $x_{i.j}$ and $x_{i.j}^{n}$.
Unless specified explicitly, we will use the usual short hand notation
for (conditional) probability distributions where the lower case arguments
also denote the random variables whose (conditional) distribution
is being considered. For example, $p(y_{i}|x_{i.j})$ denotes $p_{{\rm Y}_{i}|\mathrm{X}_{i,j}}(y_{i}|x_{i,j})$.

If random variables $X$, $Y$ and $Z$ form a Markov chain, we denote
it as $X-\circ-Y-\circ-Z$.

In the Gaussian MAC-IC-MAC to be defined in the next section, a signal
path from the transmitter Tx$i.j$ to the receiver Rx$i$ is represented
as $i.j\rightarrow i$, so that $h_{i.j\rightarrow i}$ denotes the
path attenuation from Tx$i.j$ to Rx$i$. Similarly, the signal-to-noise
ratio (SNR) and interference-to-noise ratio (INR) from Tx$i.j$ and
Tx$i'.j$ to Rx$i$ are written as $\mathsf{SNR}_{i.j\rightarrow i}$
and $\mathsf{INR}_{i'.j\rightarrow i}$, respectively, where $i,i^{'}\in\{1,2\}$
and $i\neq i^{'}$.

The achievable schemes of this paper involve message splitting at
the two transmitters that cause interference at their unintended receiver.
A common sub-message sent by Tx$i.1$ and decoded at both receivers
is denoted as $m_{i.1c}$. The private sub-message of Tx$i.1$ to
be decoded only at the intended receiver Rx$i$ is denoted as $m_{i.1p}$.
The rate of $m_{i.1c}$ and $m_{i.1p}$ are written as $R_{i.1c}$
and $R_{i.1p}$, respectively.

We use $\mathbb{C}$ to denote the set of complex numbers, $X\sim\mathcal{CN}(0,\sigma^{2})$
to denote zero-mean, circularly symmetric complex Gaussian random
variable with variance $\sigma^{2}$, and $|\cdot|$ to denote the
magnitude of a complex number.


The rest of the paper is organized as follows. Section \ref{sec:mim-channel-models}
introduces the three classes of channel models of the MAC-IC-MAC and
formulates the problem. Section \ref{sec:mim-main-results} presents
the main results of this paper with outlines of proofs. Section \ref{sec:mim-conclusions}
concludes the paper. Some detailed proofs are relegated to the appendices.

\section{Channel Models and Problem Formulation\label{sec:mim-channel-models}}

In this section, we introduce three classes of MAC-IC-MACs, namely,
the general discrete memoryless MAC-IC-MAC, the semi-deterministic
MAC-IC-MAC and the Gaussian MAC-IC-MAC. 

\subsection{Discrete Memoryless MAC-IC-MAC (DM MAC-IC-MAC)\label{subsec:mim-dm-model}}

\begin{figure}[tbh]
\centering{}\includegraphics{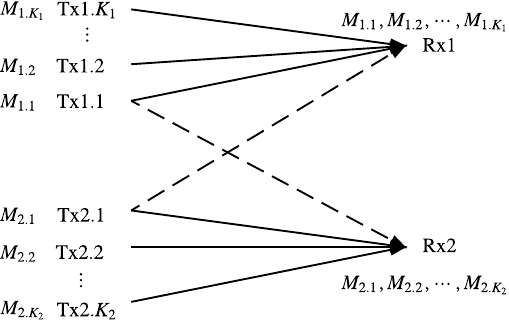}\caption{The Discrete Memoryless MAC-IC-MAC\label{fig:mim-dm-model}}
\end{figure}

In a DM MAC-IC-MAC, as shown in Fig.\,\ref{fig:mim-dm-model}, there
are two uplink communication cells: (Tx$1.1$, $\cdots$ ,Tx$1.K_{1}$$\longrightarrow$Rx$1$)
and (Tx$2.1$, $\cdots$,Tx$2.K_{2}$$\longrightarrow$Rx$2$). Two
interference links exist between these two cells between the first
users of each cell as shown in Fig.\,\ref{fig:mim-dm-model}.
The definition of DM MAC-IC-MAC is given next.

\begin{defn}
\label{def: mim-dm-model}A ($K_{1},K_{2}$) discrete memoryless MAC-IC-MAC
is a $(K_{1}+K_{2})$-transmitter and 2-receiver network $(\mathcal{X}_{\varTheta_{1}}\times\mathcal{X}_{\varTheta_{2}},p(y_{1},y_{2}|x_{\varTheta_{1}},x_{\varTheta_{2}}),\text{\ensuremath{\mathcal{Y}}}_{1}\times\mathcal{Y}_{2})$
with transition probability satisfying 
\begin{align}
 & p(y_{1}^{n},y_{2}^{n}|x_{\varTheta_{1}}^{n},x_{\varTheta_{2}}^{n})\nonumber \\
 & \qquad=\prod_{t=1}^{n}\left(p(y_{1t}|x_{\varTheta_{1}t},x_{2.1t})p(y_{2t}|x_{\varTheta_{2}t},x_{1.1t})\right)\label{eq:mim-dm-model}
\end{align}
The input and output symbols $X_{i.j}$ and $Y_{i}$ are taken from
discrete alphabets $\mathcal{X}_{i.j}$ and $\mathcal{Y}_{i}$ respectively,
where $j\in\{1,\cdots,K_{i}\}$. Message $M_{i.j}$ is generated from
set $\mathcal{M}_{i.j}$ uniformly at random, and encoded at transmitter
Tx$i.j$. Receiver Rx$i$ decodes $M_{\varTheta_{i}}$ as $\hat{M}_{\varTheta_{i}}$. 
\end{defn}
Given the channel as defined in Definition \ref{def: mim-dm-model},
a $(n,R_{\varTheta_{1}},R_{\varTheta_{2}},P_{e}^{(n)})$ coding scheme
for a DM MAC-IC-MAC consists of 
\begin{itemize}
\item $M_{i.j}$, the message of transmitter Tx$i.j$, assumed to be uniformly
distributed over $\mathcal{M}_{i.j}\in\{1,\cdots,2^{nR_{i.j}}\}$,
for each $i.j\in\varTheta_{i}$ and $i\in\{1,2\}$; 
\item Encoding functions $f_{i.j}(\cdot)$ such that 
\begin{align*}
f_{i.j}(\cdot): & \;\mathcal{M}_{i.j}\longmapsto\mathcal{X}_{i.j}^{n},\;m_{i.j}\longmapsto x_{i.j}^{n}(m_{i.j}).
\end{align*}
\item Decoding functions $g_{i}(\cdot)$ such that 
\[
g_{i}(\cdot):\;\mathcal{Y}_{i}^{n}\longmapsto\prod_{j=1}^{K_{i}}\mathcal{M}_{i.j},\;y_{i}^{n}\longmapsto\hat{m}_{\varTheta_{i}}(y_{i}^{n}).
\]
\end{itemize}
The probability of error $P_{e}^{(n)}$ is defined to be 
\[
P_{e}^{(n)}=P\left\{ M_{\varTheta_{1}}\neq\hat{M}_{\varTheta_{1}}\thinspace\mathrm{or}\thinspace M_{\varTheta_{2}}\neq\hat{M}_{\varTheta_{2}}\right\} .
\]

A $K_{1}+K_{2}$ rate-tuple $(R_{\varTheta_{1}},R_{\varTheta_{2}})$
is said to be achievable if there exists a sequence of $(n,R_{\varTheta_{1}},R_{\varTheta_{2}},P_{e}^{(n)})$
coding schemes for which $P_{e}^{(n)}\rightarrow0$ as $n\rightarrow\infty$.

The capacity region, denoted as $\mathcal{C}$, is the closure of
all achievable rate-tuples.


\subsection{Semi-Deterministic MAC-IC-MAC\label{subsec:mim-sd-model}}

\begin{figure}[tbh]
\begin{centering}
\includegraphics{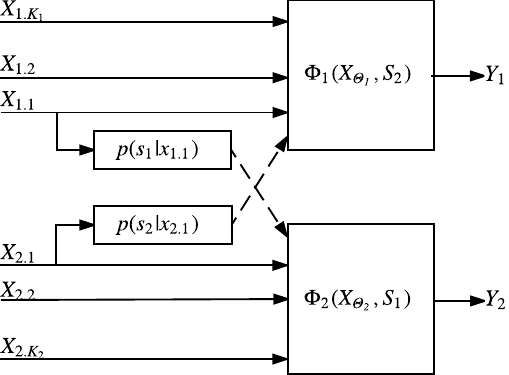} 
\par\end{centering}
\caption{Semi-deterministic MAC-IC-MAC\label{fig:mim-sd-model}}
\end{figure}

In a semi-deterministic MAC-IC-MAC, the received interference has
a special structure: the output $Y_{i}$ is determined by $X_{\varTheta_{i}}$
and the output $S_{i^{'}}$ resulting from passing $X_{i^{'}.1}$ through
a DM point-to-point channel $p(s_{i^{'}}|x_{i^{'}.1})$, as shown
in Fig.\,\ref{fig:mim-sd-model}. A formal definition is given next.

\begin{defn}
\label{def:mim-sd-model}Let $S_{i}$ be a random variable over the
alphabet $\mathcal{S}_{i}$ for $i\in\{1,2\}$. A MAC-IC-MAC is semi-deterministic
if the outputs $Y_{i}$ satisfy 
\begin{align}
Y_{i} & =\phi_{i}(X_{\varTheta_{i}},S_{i^{'}})\label{eq:mim-sd-model}
\end{align}
where $i^{'}\in\{1,2\}$, $i^{'}\neq i$, $\phi_{i}$ is a deterministic
function and for any fixed inputs $x_{\varTheta_{i}}$, the mapping
\begin{align}
\phi_{i}(x_{\varTheta_{i}},\cdot) & : \mathcal{S}_{i^{'}}\rightarrow\mathcal{Y}_{i},  s_{i^{'}}\rightarrow\phi_{i}(x_{\varTheta_{i},}s_{i^{'}})\label{eq:mim-sd-mapping}
\end{align}
is invertible. 
\end{defn}
This structure is seen, for instance, in the Gaussian MAC-IC-MAC. 

Let $\mathcal{C}^{{\rm sd}}$ denote the capacity region of the semi-deterministic MAC-IC-MAC. 


\subsection{Gaussian MAC-IC-MAC\label{subsec:mim-gs-model}}

Consider an additive Gaussian MAC-IC-MAC, whose input-output relation
can be written as 
\begin{align}
Y_{1} & =\sum_{j=1}^{K_{1}}h_{1.j\rightarrow1}X_{1.j}+h_{2.1\rightarrow1}X_{2.1}+Z_{1}\label{eq:mim-gs-mdoel-1}\\
Y_{2} & =\sum_{j=1}^{K_{2}}h_{2.j\rightarrow2}X_{2.j}+h_{1.1\rightarrow2}X_{1.1}+Z_{2}\label{eq:mim-gs-model-2}
\end{align}
where $X_{i.j}$ and $Y_{i}$ are complex input and output symbols,
$h_{i.j\rightarrow i},h_{i.j\rightarrow i^{'}}\in\mathbb{C}$ are
path attenuations from Tx$i.j$ to Rx$i$ and Rx$i^{'}$, respectively,
and $Z_{i}\in\mathcal{CN}(0,1)$ is the additive Gaussian noise. The
transmitted codeword $x_{i.j}^{n}\in\mathcal{X}_{i.j}^{n}$ at Tx$i.j$
should meet the average per-codeword power constraints:

\begin{equation}
\frac{1}{n}\sum_{t=1}^{n}|x_{i.j,t}|^{2}\leq P_{i.j}.\label{eq:mim-power-constraint}
\end{equation}
The SNRs and INRs at receiver Rx$i$ are defined to be 
\begin{align}
\mathsf{SNR}_{i.j\rightarrow i} & =P_{i.j}|h_{i.j\rightarrow i}|^{2}\label{eq:mim-snr}\\
\mathsf{INR}_{i'.1\rightarrow i} & =P_{i'.1}|h_{i'.1\rightarrow i}|^{2}\label{eq:mim-inr}
\end{align}
where $P_{i.j}$ is the maximum average transmission power at Tx$i.j$.
We denote the capacity region of Gaussian MAC-IC-MAC by $\mathcal{C}^{{\rm G}}$.

Other than not being a discrete alphabet channel and having input (power) constraints, the Gaussian MAC-IC-MAC
is a special case of the semi-deterministic MAC-IC-MAC, as can be seen by choosing $S_{i}$
in \eqref{eq:mim-sd-model} to be $S_{i}=h_{i.1\rightarrow i^{'}}X_{i.1}+Z_{i^{'}}$
and the functions $\phi_{i}$ to be weighted sum with weights determined
by the channel coefficients.

\section{Results\label{sec:mim-main-results}}

In this section, we present the main results of this paper, namely,
inner and outer bounds on the capacity region of the DM, semi-deterministic and Gaussian MAC-IC-MACs.

First, we define two generic regions in Section \ref{subsec:mim-generic}
which we will use repeatedly in the following sections to describe inner
and outer bounds on the capacity region of the three MAC-IC-MAC models.
Section \ref{subsec:mim-inner} states two inner bounds in Theorems  \ref{thm:mim-dm-inner-fix-dist} and \ref{thm:mim-dm-inner-compact} on the capacity
region $\mathcal{C}$ of the general DM MAC-IC-MAC, supported by a random coding argument.
The first bound is a single polytope bound 
obtained by fixing one ``coding\textquotedbl{}
distribution of auxiliary/input random variables 
and the second is a union of polytopes (compact region) bound 
with the union taken over certain admissible coding distributions.

In Section \ref{subsec:mim-sd-bounds}, we extend the genie-aided arguments of \cite{telatar2007bounds} for the two-user semi-deterministic IC to the semi-deterministic MAC-IC-MAC to obtain an
outer bound for $\mathcal{C}^{{\rm sd}}$ in Theorem \ref{thm:mim-sd-outer} of Section \ref{subsec:mim-sd-bounds}. Moreover, we also quantify the gap between the single-polytope and union-of-polytopes inner bounds of Theorems  \ref{thm:mim-dm-inner-fix-dist} and \ref{thm:mim-dm-inner-compact} (when specialized to the semi-deterministic channel) and the outer bound in Theorems \ref{lem:mim-sd-gap-2} and \ref{thm:mim-sd-gap}.
Our quantifiable gap analysis reveals insights beyond those available even in the previously studied two-user semi-deterministic
IC, and also clarifying past studies in the process.

In Section \ref{subsec:mim-gs-gap}, the Gaussian MAC-IC-MAC is investigated. In particular, in Theorem \ref{thm:mim-gs-inner}, we obtain an inner bound based on a universal coding scheme where the interfering transmitters perform power splitting that is independent of the rate-pair to be achieved (but dependent on channel coefficients) and the non-interfering transmitters encode with single-user Gaussian codebooks. 
Moreover, a single-polytope, and hence explicit, outer bound for $\mathcal{C}^{{\rm G}}$ is obtained in Theorem \ref{thm:.mim-gs-outer}. These inner and outer bounds are shown to be within a two-bit gap of each other, and hence of $\mathcal{C}^{{\rm G}}$, in Theorem \ref{thm:mim-gs-gap}. If the power split can be chosen based on the rate-tuple to be achieved in addition to the channel coefficients  then a one-bit gap to capacity can be achieved as shown in Lemma \ref{lem: mim-gs-gap-2}. 

Finally, in Section \ref{subsec:mim-gdof}, the GDoF region of the
Gaussian MAC-IC-MAC is obtained in Theorem \ref{thm:mim-gdof} and specialized to the symmetric GDoF region in Corollary \ref{cor:mim-sym-gdof}. A careful study of it shows how the non-interfering transmitters improve the sum symmetric GDoF as a function of the relative interference strength parameter $\alpha$, defined as the ratio of the INR to SNR (with both expressed in the dB scale), at high SNR.

\begin{rem}
The study of the Gaussian MAC-IC-MAC at high SNR is also relevant
to the fully-connected $(K_{1}+K_{2})$-transmitter, 2-receiver, two-cell
interference networks when the interference from all but one transmitter
in each cell are sufficiently weak so that they are received at the
unintended receiver at the noise level (and are therefore treated
as noise). 
\end{rem}


\subsection{Two Generic Regions\label{subsec:mim-generic}}

For brevity, we provide the definitions of two generic regions in this
subsection in terms of which all key results of this paper are succinctly expressed. 
\begin{defn}
\label{def:mim-generic-set-functions}For any sets $\varOmega_{i}\subseteq\varTheta_{i}=\{i.1,\cdots,i.K_{i}\}$,
$\varUpsilon_{i}^{'}\subseteq\varTheta_{i}\backslash\{i.1\}$ and
$\varUpsilon_{i}=\varUpsilon_{i}^{'}\cup\{i.1\}$, where $i\in\{1,2\}$,
let $\mathcal{A}$ and $\mathcal{E}$ be non-negative, real-valued
functions of set $\varUpsilon_{i}$ so that, 
\begin{align}
\mathcal{A}: & \quad \{ \varUpsilon_{i} \in 2^{\varTheta_i}: i.1\in \varUpsilon_{i}\} \rightarrow \mathbb{R}^+,\quad\varUpsilon_{i}\rightarrow\mathcal{A}_{\varUpsilon_{i}}\label{eq:mim-set-fcn-a}\\
\mathcal{E}: & \quad \{ \varUpsilon_{i} \in 2^{\varTheta_i}: i.1\in \varUpsilon_{i}\} \rightarrow \mathbb{R}^+,\quad\varUpsilon_{i}\rightarrow\mathcal{E}_{\varUpsilon_{i}}\label{eq:mim-set-fcn-e}
\end{align}
and $\mathcal{B}_{\varOmega_{i}}$ and $\mathcal{G}_{\varOmega_{i}}$
be non-negative real-valued functions of set $\varOmega_{i}$ so that,
\begin{align}
\mathcal{B}: & \quad2^{\varTheta_{i}}\backslash\emptyset\rightarrow\mathbb{R}^{+},\quad\varOmega_{i}\rightarrow\mathcal{B}_{\varOmega_{i}}\label{eq:mim-set-fcn-b}\\
\mathcal{G}: & \quad2^{\varTheta_{i}}\backslash\emptyset\rightarrow\mathbb{R}^{+},\quad\varOmega_{i}\rightarrow\mathcal{G}_{\varOmega_{i}}.\label{eq:mim-set-fcn-g}
\end{align}
\end{defn}
There is a slight abuse of notation in the definitions of set functions $\mathcal{A}$, $\mathcal{B}$,
$\mathcal{E}$ and $\mathcal{G}$ in \eqref{eq:mim-set-fcn-a}-\eqref{eq:mim-set-fcn-g}, which we clarify. The domains $\{ \varUpsilon_{i} \in 2^{\varTheta_i}: i.1\in \varUpsilon_{i}\}$
of functions $\mathcal{A}$ and $\mathcal{E}$ or $2^{\varTheta_{i}}\backslash\emptyset$
of $\mathcal{B}$ and $\mathcal{G}$, depend on $i\in\{1,2\}$, which
means we are actually using the same function name for two different
functions.
However, which input domain a set
function is referring to can be easily identified by the subscript
of its argument (such as in $\varUpsilon_{1}$ or $\varOmega_{2}$,
respectively). 
\begin{defn}
\label{def:mim-generic-region} Let the Cartesian product of the domains
of $(\varUpsilon_{1},\varOmega_{1},\varUpsilon_{2},\varOmega_{2})$
\begin{align}
\Xi & \triangleq\left\{\varUpsilon_{1} \in 2^{\varTheta_1}: 1.1\in \varUpsilon_{1}\right\} \times\left\{ 2^{\varTheta_{1}}\backslash\emptyset\right\} \nonumber \\
 & \qquad\times\left\{\varUpsilon_{2} \in 2^{\varTheta_2}: 2.1\in \varUpsilon_{2}\right\} \times\left\{ 2^{\varTheta_{2}}\backslash\emptyset\right\} \label{eq:mim-set-domains}
\end{align}
Define the region $\mathcal{R}(\mathcal{A},\mathcal{B},\mathcal{E},\mathcal{G})$
to be 
\begin{align}
\mathcal{R}(\mathcal{A},\mathcal{B},\mathcal{E},\mathcal{G})=\bigg\{(R_{\varTheta_{1}},R_{\varTheta_{2}}) & \in\mathbb{R}_{+}^{K_{1}+K_{2}}:\nonumber \\
 & \forall(\varUpsilon_{1},\varOmega_{1},\varUpsilon_{2},\varOmega_{2})\in\Xi\nonumber \\
\sum_{1.j\in\varOmega_{1}}R_{1.j} & \leq\mathcal{B}_{\varOmega_{1}}\label{eq:mim-generic-1}\\
\sum_{1.j\in\varUpsilon_{1}}R_{1.j} & \leq\mathcal{A}_{\varUpsilon_{1}}+\mathcal{E}_{\varUpsilon_{2}}\label{eq:mim-generic-2}\\
\sum_{2.j\in\varOmega_{2}}R_{2.j} & \leq\mathcal{B}_{\varOmega_{2}}\label{eq:mim-generic-3}\\
\sum_{2.j\in\varUpsilon_{2}}R_{2.j} & \leq\mathcal{E}_{\varUpsilon_{1}}+\mathcal{A}_{\varUpsilon_{2}}\label{eq:mim-generic-4}\\
\sum_{1.j\in\varUpsilon_{1}}R_{1.j}+\sum_{2.j\in\varOmega_{2}}R_{2.j} & \leq\mathcal{A}_{\varUpsilon_{1}}+\mathcal{G}_{\varOmega_{2}}\label{eq:mim-generic-5}\\
\sum_{1.j\in\varOmega_{1}}R_{1.j}+\sum_{2.j\in\varUpsilon_{2}}R_{2.j} & \leq\mathcal{G}_{\varOmega_{1}}+\mathcal{A}_{\varUpsilon_{2}}\label{eq:mim-generic-6}\\
\sum_{1.j\in\varUpsilon_{1}}R_{1.j}+\sum_{2.j\in\varUpsilon_{2}}R_{2.j} & \leq\mathcal{E}_{\varUpsilon_{1}}+\mathcal{E}_{\varUpsilon_{2}}\label{eq:mim-generic-7}\\
\sum_{1.j\in\varUpsilon_{1}}R_{1.j}+\sum_{1.j\in\varOmega_{1}}R_{1.j}\qquad\nonumber \\
+\sum_{2.j\in\varUpsilon_{2}}R_{2.j} & \leq\mathcal{A}_{\varUpsilon_{1}}+\mathcal{G}_{\varOmega_{1}}+\mathcal{E}_{\varUpsilon_{2}}\label{eq:mim-generic-8}\\
\sum_{1.j\in\varUpsilon_{1}}R_{1.j}+\sum_{2.j\in\varUpsilon_{2}}R_{2.j}\qquad\nonumber \\
+\sum_{2.j\in\varOmega_{2}}R_{2.j} & \leq\mathcal{E}_{\varUpsilon_{1}}+\mathcal{A}_{\varUpsilon_{2}}+\mathcal{G}_{\varOmega_{2}}\bigg\}\label{eq:mim-generic-9}
\end{align}
\end{defn}
Thus, $ \mathcal{R}(\mathcal{A},\mathcal{B},\mathcal{E},\mathcal{G}) $ is a polytope defined via nine classes of partial sum-rate inequalities.

\begin{defn}
\label{def:mim-generic-compact-region}Let us also define $\mathcal{R}_{\mathrm{c}}(\mathcal{A},\mathcal{B},\mathcal{E},\mathcal{G})\subseteq\mathbb{R}_{+}^{K_{1}+K_{2}}$
for given set functions $\mathcal{A}$, $\mathcal{B}$, $\mathcal{E}$
and $\mathcal{G}$ to be the rate region 
\begin{align}
\mathcal{R}_{\mathrm{c}}(\mathcal{A},\mathcal{B},\mathcal{E},\mathcal{G})=\bigg\{(R_{\varTheta_{1}},R_{\varTheta_{2}}) & \in\mathbb{R}_{+}^{K_{1}+K_{2}}:\nonumber \\
 & \forall(\varUpsilon_{1},\varOmega_{1},\varUpsilon_{2},\varOmega_{2})\in\Xi\nonumber \\
\sum_{1.j\in\varOmega_{1}}R_{1.j} & \leq\mathcal{B}_{\varOmega_{1}}\label{eq:mim-generic-c-1}\\
\sum_{2.j\in\varOmega_{2}}R_{2.j} & \leq\mathcal{B}_{\varOmega_{2}}\label{eq:mim-generic-c-2}\\
\sum_{1.j\in\varUpsilon_{1}}R_{1.j}+\sum_{2.j\in\varOmega_{2}}R_{2.j} & \leq\mathcal{A}_{\varUpsilon_{1}}+\mathcal{G}_{\varOmega_{2}}\label{eq:mim-generic-c-3}\\
\sum_{1.j\in\varOmega_{1}}R_{1.j}+\sum_{2.j\in\varUpsilon_{2}}R_{2.j} & \leq\mathcal{G}_{\varOmega_{1}}+\mathcal{A}_{\varUpsilon_{2}}\label{eq:mim-generic-c-4}\\
\sum_{1.j\in\varUpsilon_{1}}R_{1.j}+\sum_{2.j\in\varUpsilon_{2}}R_{2.j} & \leq\mathcal{E}_{\varUpsilon_{1}}+\mathcal{E}_{\varUpsilon_{2}}\label{eq:mim-generic-c-5}\\
\sum_{1.j\in\varUpsilon_{1}}R_{1.j}+\sum_{1.j\in\varOmega_{1}}R_{1.j}\qquad\nonumber \\
+\sum_{2.j\in\varUpsilon_{2}}R_{2.j} & \leq\mathcal{A}_{\varUpsilon_{1}}+\mathcal{G}_{\varOmega_{1}}+\mathcal{E}_{\varUpsilon_{2}}\label{eq:mim-generic-c-6}\\
\sum_{1.j\in\varUpsilon_{1}}R_{1.j}+\sum_{2.j\in\varUpsilon_{2}}R_{2.j}\qquad\nonumber \\
+\sum_{2.j\in\varOmega_{2}}R_{2.j} & \leq\mathcal{E}_{\varUpsilon_{1}}+\mathcal{A}_{\varUpsilon_{2}}+\mathcal{G}_{\varOmega_{2}}\bigg\}.\label{eq:mim-generic-c-7}
\end{align}
\end{defn}
The subscript 'c' in $\mathcal{R}_{\mathrm{c}}(\mathcal{A},\mathcal{B},\mathcal{E},\mathcal{G})$
indicates that this region is in \textit{compact} form since it can be obtained
by removing the two classes of inequalities \eqref{eq:mim-generic-2}
and \eqref{eq:mim-generic-4} from $\mathcal{R}(\mathcal{A},\mathcal{B},\mathcal{E},\mathcal{G})$. Thus, $ \mathcal{R}_{\mathrm{c}}(\mathcal{A},\mathcal{B},\mathcal{E},\mathcal{G}) $ is a polytope defined via just seven classes of partial sum-rate inequalities.

Note that in Definitions \ref{def:mim-generic-set-functions} and \ref{def:mim-generic-compact-region}, for
those sets of inequalities whose bounds don't depend on all 
of the subsets $\varUpsilon_{1}$, $\varOmega_{1}$, $\varUpsilon_{2}$
and $\varOmega_{2}$, duplicated copies will be produced when either
$\mathcal{R}$ or $\mathcal{R}_{\mathrm{c}}$ is written exhaustively according to those definitions. However, we adopt those definitions for the analytical convenience of using the single domain $\Xi$, instead of individually specifying the domains for each set of inequalities in the two definitions. Duplicated copies are of course to be ejected in any explicit specification of the regions.
\begin{rem}
As will be observed in the later section, each set function $\mathcal{A}$,
$\mathcal{B}$, $\mathcal{E}$ or $\mathcal{G}$ represents a mutual
information term for a certain set $\varOmega_{i}$ or $\varUpsilon_{i}$.
Their exact form will be assigned according to the channel model
\textendash{} DM, semi-deterministic or Gaussian \textendash{} and
whether the bound under consideration is an inner bound or an
outer bound. When a particular region needs to be specified, we will
specialize each of the four set functions, and a rate region can be
described by replacing the generic set functions with the specified
ones. 
\end{rem}

\subsection{Inner Bounds on the Capacity Region of DM MAC-IC-MAC\label{subsec:mim-inner}}

As mentioned earlier, the HK scheme, and its alternative, the CMG
scheme, leads to the best inner bound to date for the two-user interference
channel. In this section, we use the CMG scheme at the interfering
transmitters of the MAC-IC-MAC. More specifically, we employ the 
superposition coding of the ``the cloud-satellite'' type at Tx$i.1$ and use a single-user random codebook for transmitter Tx$i.j$, $j\neq1$. Applying the joint-typicality
decoding argument we derive two inner bounds for the DM MAC-IC-MAC,
as stated in Theorems \ref{thm:mim-dm-inner-fix-dist} and \ref{thm:mim-dm-inner-compact}. 
\begin{defn}
\label{def:mim-dm-inner-distribution}Let $\mathcal{P}_{\mathrm{in}}$
be the set of distributions $P_{\mathrm{in}}$ of joint random variables
$(Q,U_{1},X_{\varTheta_{1}},U_{2},X_{\varTheta_{2}})$ that can be
factored as 
\begin{align}
 & p(q,u_{1},x_{\varTheta_{1}},u_{2},x_{\varTheta_{2}})\nonumber \\
 & \qquad=p(q)p(u_{1},x_{1.1}|q)\prod_{1.j\in\varTheta_{1}\backslash\{1.1\}}p(x_{1.j}|q)\nonumber \\
 & \qquad\qquad\cdot p(u_{2},x_{2.1}|q)\prod_{2.j\in\varTheta_{2}\backslash\{2.1\}}p(x_{2.j}|q). \label{eq:mim-p-in}
\end{align}
\end{defn}
\begin{defn}
\label{def:mim-dm-inner-set-functions}Given any jointly distributed
random variables $(Q,U_{1},X_{\varTheta_{1}},U_{2},X_{\varTheta_{2}})$
with distribution $P_{\mathrm{in}}\in\mathcal{P}_{\mathrm{in}}$,
define the non-negative real-valued set functions $\mathsf{A}$, $\mathsf{B}$,
$\mathsf{E}$ and $\mathsf{G}$ as 
\begin{align}
\mathcal{\mathsf{A}}_{\varUpsilon_{i}} & \triangleq I(X_{\varUpsilon_{i}};Y_{i}|X_{\bar{\varUpsilon}_{i}},U_{i},U_{i^{'}},Q)\label{eq:mim-dm-inner-a}\\
\mathcal{\mathsf{B}}_{\varOmega_{i}} & \triangleq I(X_{\varOmega_{i}};Y_{i}|X_{\bar{\varOmega}_{i}},U_{i^{'}},Q)\label{eq:mim-dm-inner-b}\\
\mathcal{\mathsf{E}}_{\varUpsilon_{i}} & \triangleq I(X_{\varUpsilon_{i}},U_{i^{'}};Y_{i}|X_{\bar{\varUpsilon}_{i}},U_{i},Q)\label{eq:mim-dm-inner-e}\\
\mathcal{\mathsf{G}}_{\varOmega_{i}} & \triangleq I(X_{\varOmega_{i}},U_{i^{'}};Y_{i}|X_{\bar{\varOmega}_{i}},Q) \label{eq:mim-dm-inner-g}
\end{align}
and, using Definition \ref{def:mim-generic-region}, define the associated
region 
\begin{align}
\mathcal{R}_{\mathrm{in}}(Q,U_{1},X_{\varTheta_{1}},U_{2},X_{\varTheta_{2}}) & \triangleq\mathcal{R}_{\mathrm{in}}(P_{\mathrm{in}}) \triangleq\mathcal{R}(\mathsf{A},\mathsf{B},\mathsf{E},\mathsf{G}).\label{eq:mim-inner-fix-dist}
\end{align}
\end{defn}
\begin{thm}
\label{thm:mim-dm-inner-fix-dist}For DM MAC-IC-MAC, 
\[
\mathcal{R}_{\mathrm{in}}(P_{\mathrm{in}})\subseteq\mathcal{C}.
\]
\end{thm}
\begin{IEEEproof}
We outline the proof here, and relegate the details to Appendix \ref{sec:mim-dm-inner-fix-dist-proof}.
The proof is given in two parts.

First, with the interfering transmitters employing the CMG rate-splitting
and superposition coding scheme and the non-interfering transmitters using independent
random coding (as in a MAC), and each receiver decoding all its desired
messages and the common sub-message of the interfering transmitter
using simultaneous non-unique decoding, standard random coding analysis
leads to an intermediate rate region containing two auxiliary random
variables which represent the common sub-message
(also referred as ``cloud\textquotedbl{} codewords) of interfering
transmitters, and two auxiliary rates corresponding to these codebooks.
In particular, 1) as in a MAC, the non-interfering transmitter
Tx$i.j$, $j\neq1$ sends information $m_{i.j}$ by some codeword
$x_{i.j}^{n}(m_{i.j})$ using a single-user random codebook; 2) as
in the CMG achievable scheme \cite{chong2008han}, the interfering
transmitter Tx$i.1$ splits its message $m_{i.1}$ into common and
private sub-messages $m_{i.1c}$ and $m_{i.1p}$, respectively. The
common information is first encoded into the cloud center codeword
$u_{i}^{n}(m_{i.1c})$, and then, based on the private sub-message
$m_{i.1p}$, the entire message $m_{i.1}$ is encoded into the codeword
$x_{i.1}^{n}(u_{i}^{n}(m_{i.1c}),m_{i.1p})$ for transmission; 3)
Rx$i$ uniquely decodes the intended messages $m_{i.j}$, $j\neq1$
and the public and private information $m_{i.1c}$ and $m_{i.1p}$,
all from its intended transmitters, and non-uniquely the common information
$m_{i^{'}.1c}$ from the non-intended transmitter Tx$i^{'}.1$. Though
the number of inequalities in each class is indeterminate because
the number of users in each cell are arbitrary, it turns out that
only four appropriately chosen classes of error events are needed to obtain conditions on rates to assure reliable communication for all users.

Second, the Fourier-Motzkin method is used to eliminate the two auxiliary
rates analytically to project the rate region onto $\mathbb{R}_{+}^{K_{1}+K_{2}}$
of message rates, despite the numbers of users in both cells being
arbitrary. In this step, the combinatorial structure of the inequality
system created in the first step is utilized. The detailed proof
is given in Appendix \ref{sec:mim-dm-inner-fix-dist-proof}. 
\end{IEEEproof}
\begin{rem}
Without a recognition
of the presence of the structure in the reliability conditions, one would have to resort to a hand-crafted
Fourier-Motzkin elimination procedure which would severely restrict
the values of $K_{1}$ and $K_{2}$ for which it could be carried
out with reasonable effort as noted in the next two remarks.
\end{rem}
\begin{rem}
When $K_{1}=K_{2}=1$, the only admissible 4-tuple $(\varUpsilon_{1},\varOmega_{1},\varUpsilon_{2},\varOmega_{2})$
is $(\{1.1\},\{1.1\},\{2.1\},\{2.1\})$. Hence, Theorem \ref{thm:mim-dm-inner-fix-dist}
recovers (as it should) the inner bound of the two-user DM IC for
fixed input distribution as obtained via 
Fourier-Motzkin elimination by Kobayashi and Han in \cite[Theorem D]{kobayashi2007further} and stated by Chong {\em et al} in \cite[Lemma 4]{chong2008han}.
\end{rem}
\begin{rem}
In the $K_{1}=2$ and $K_{2}=1$ case, Fourier-Motzkin elimination was again used to obtain the rate region in the conference version of this paper in \cite{pang2013bounds}. A ``compact form'' of that rate region is given in Corollary \ref{cor:mim-PMAIC} (to follow).
\end{rem}
Next, Theorem 1 implies that the union over admissible distributions
\begin{equation}
\mathcal{R}_{\mathrm{in}}\triangleq\bigcup_{P_{\mathrm{in}}\in\mathcal{P}_{\mathrm{in}}}\mathcal{R}_{{\rm in}}(P_{\mathrm{in}})\label{eq:mim-inner-union-fix-dist}
\end{equation}
is also achievable. That is, $\mathcal{R}_{\mathrm{in}}\subseteq\mathcal{C}$.
However, when considering such a union, it turns out that the sets
of inequalities \eqref{eq:mim-generic-2} and \eqref{eq:mim-generic-4}
become redundant as stated in Theorem \ref{thm:mim-dm-inner-compact} to follow.
\begin{defn}
\label{def:mim-inner-compact}Let the non-negative real-valued set
functions $\mathsf{A}$, $\mathsf{B}$, $\mathsf{E}$ and $\mathsf{G}$
be as defined in \eqref{eq:mim-dm-inner-a}-\eqref{eq:mim-dm-inner-g},
and define the region 
\begin{equation}
\mathcal{R}_{{\rm c}}(P_{\mathrm{in}})\triangleq\mathcal{R}_{\mathrm{c}}(\mathsf{A},\mathsf{B},\mathsf{E},\mathsf{G})\label{eq:mim-inner-compact}
\end{equation}
for any fixed $P_{\mathrm{in}}\in\mathcal{P}_{\mathrm{in}}$. We also
define the region $\mathcal{R}_{{\rm c}}$ as the union of $\mathcal{R}_{\mathrm{c}}(P_{\mathrm{in}})$
over $\mathcal{P}_{\mathrm{in}}$, i.e., 
\begin{align}
\mathcal{R}_{{\rm c}} & \triangleq\bigcup_{P_{\mathrm{in}}\in\mathcal{P}_{\mathrm{in}}}\mathcal{R}_{{\rm c}}(P_{\mathrm{in}}).\label{eq:mim-inner-union-compact}
\end{align}
\end{defn}
\begin{thm}
\label{thm:mim-dm-inner-compact}For a $(K_{1},K_{2})$ DM MAC-IC-MAC,
\[
\mathcal{R}_{\mathrm{c}}=\mathcal{R}_{\mathrm{in}}\subseteq\mathcal{C}.
\]
\end{thm}
\begin{IEEEproof}
Clearly, $\mathcal{R}_{\mathrm{in}}\subseteq\mathcal{C}$ by Theorem
\ref{thm:mim-dm-inner-fix-dist}. Moreover, since $\mathcal{R}_{\mathrm{c}}(P_{\mathrm{in}})$ involves two sets of inequalities fewer than does $\mathcal{R}_{\mathrm{in}}(P_{\mathrm{in}})$, we have $\mathcal{R}_{\mathrm{in}}(P_{\mathrm{in}})\subseteq\mathcal{R}_{\mathrm{c}}(P_{\mathrm{in}})$
for each $P_{\mathrm{in}}\in\mathcal{P}_{\mathrm{in}}$. Hence, $\mathcal{R}_{\mathrm{in}}\subseteq\mathcal{R}_{\mathrm{c}}$.
It remains to show that $\mathcal{R}_{\mathrm{in}}\supseteq\mathcal{R}_{\mathrm{c}}$.
We show in Appendix \ref{sec:mim-dm-inner-proof} that indeed $\mathcal{R}_{\mathrm{c}}\backslash\mathcal{R}_{\mathrm{in}}$
is empty. 
\end{IEEEproof}
\begin{rem}
When $K_{1}=1$, $K_{2}=1$, the only admissible 4-tuple $(\varUpsilon_{1},\varOmega_{1},\varUpsilon_{2},\varOmega_{2})$
is $(\{1.1\},\{1.1\},\{2.1\},\{2.1\})$. It can be verified that Theorem \ref{thm:mim-dm-inner-compact}
recovers (as it should) the main result of \cite{chong2008han} which is the CMG representation of the celebrated HK region of \cite{han1981new}. In particular, Theorem \ref{thm:mim-dm-inner-compact} becomes the inner bound of \cite[Theorem 2]{chong2008han},
referred to therein as the ``compact'' HK region, which is union over admissible distributions of polyhedra described by 7 inequalities.
\end{rem}
\begin{cor}
\label{cor:mim-PMAIC} When $K_{1}=2$, $K_{2}=1$, Theorem \ref{thm:mim-dm-inner-compact} gives a generalization of the CMG representation for the two-user IC to the $(2,1)$ MAC-IC-MAC, which is a union over admissible distribution of polyhedra described by 19 inequalities in 7 classes, namely\footnote{We group inequalities of the same class in curly brackets with the
corresponding equation number from the set \eqref{eq:mim-generic-c-1}-\eqref{eq:mim-generic-c-7}
as subscript.},
\begin{align}
\mathcal{R}_{\mathrm{c}}= \bigcup_{P_{\mathrm{in}}\in\mathcal{P}_{\mathrm{in}}} \bigg\{(R_{1.1},R_{1.2},R_{2.1}) & \in\mathbb{R}_{+}^{3}:\nonumber \\
\big\{ R_{\{1.2\}} & \leq\mathsf{B}_{\{1.2\}}\nonumber \\
R_{\{1.1\}} & \leq\mathsf{B}_{\{1.1\}}\nonumber \\
R_{\{1.2\}}+R_{\{1.1\}} & \leq\mathsf{B}_{\{1.1,1.2\}}\big\}{}_{\mathrm{\eqref{eq:mim-generic-c-1}}}\nonumber \\
\big\{ R_{\{2.1\}} & \leq\mathsf{B}_{\{2.1\}}\big\}_{\eqref{eq:mim-generic-c-2}}\nonumber \\
\big\{ R_{\{1.1\}}+R_{\{2.1\}} & \leq\mathsf{A}_{\{1.1\}}+\mathsf{G}_{\{2.1\}}\nonumber \\
R_{\{1.1\}}+R_{\{1.2\}}+R_{\{2.1\}} & \leq\mathsf{A}_{\{1.1,1.2\}}+\mathsf{G}_{\{2.1\}}\big\}_{\mathrm{\eqref{eq:mim-generic-c-3}}}\nonumber \\
\big\{ R_{\{1.1\}}+R_{\{2.1\}} & \leq\mathsf{G}_{\{1.1\}}+\mathsf{A}_{\{2.1\}}\nonumber \\
R_{\{1.2\}}+R_{\{2.1\}} & \leq\mathsf{G}_{\{1.2\}}+\mathsf{A}_{\{2.1\}}\nonumber \\
R_{\{1.1\}}+R_{\{1.2\}}+R_{\{2.1\}} & \leq\mathsf{G}_{\{1.1,1.2\}}+\mathsf{A}_{\{2.1\}}\big\}_{\mathrm{\eqref{eq:mim-generic-c-4}}}\nonumber \\
\big\{ R_{\{1.1\}}+R_{\{2.1\}} & \leq\mathsf{E}_{\{1.1\}}+\mathsf{E}_{\{2.1\}}\nonumber \\
R_{\{1.1\}}+R_{\{1.2\}}+R_{\{2.1\}} & \leq\mathsf{E}_{\{1.1,1.2\}}+\mathsf{E}_{\{2.1\}}\big\}_{\mathrm{\eqref{eq:mim-generic-c-5}}}\nonumber \\
\big\{2R_{\{1.1\}}+R_{\{2.1\}} & \leq\mathsf{A}_{\{1.1\}}+\mathsf{G}_{\{1.1\}}+\mathsf{E}_{\{2.1\}}\nonumber \\
R_{\{1.1\}}+R_{\{1.2\}}+R_{\{2.1\}} & \leq\mathsf{A}_{\{1.1\}}+\mathsf{G}_{\{1.2\}}+\mathsf{E}_{\{2.1\}}\nonumber \\
2R_{\{1.1\}}+R_{\{1.2\}}+R_{\{2.1\}} & \leq\mathsf{A}_{\{1.1\}}+\mathsf{G}_{\{1.1,1.2\}}+\mathsf{E}_{\{2.1\}}\nonumber \\
2R_{\{1.1\}}+R_{\{1.2\}}+R_{\{2.1\}} & \leq\mathsf{A}_{\{1.1,1.2\}}+\mathsf{G}_{\{1.1\}}+\mathsf{E}_{\{2.1\}}\nonumber \\
R_{\{1.1\}}+2R_{\{1.2\}}+R_{\{2.1\}} & \leq\mathsf{A}_{\{1.1,1.2\}}+\mathsf{G}_{\{1.2\}}+\mathsf{E}_{\{2.1\}}\nonumber \\
2R_{\{1.1\}}+2R_{\{1.2\}}+R_{\{2.1\}} & \leq\mathsf{A}_{\{1.1,1.2\}}+\mathsf{G}_{\{1.1,1.2\}}\nonumber \\
 & \qquad+\mathsf{E}_{\{2.1\}}\big\}_{\mathrm{\eqref{eq:mim-generic-c-6}}}\nonumber \\
\big\{ R_{\{1.1\}}+2R_{\{2.1\}} & \leq\mathsf{A}_{\{2.1\}}+\mathsf{G}_{\{2.1\}}+\mathsf{E}_{\{1.1\}}\nonumber \\
R_{\{1.1\}}+R_{\{1.2\}}+2R_{\{2.1\}} & \leq\mathsf{A}_{\{2.1\}}+\mathsf{G}_{\{2.1\}}\nonumber \\
 & \qquad+\mathsf{E}_{\{1.1,1.2\}}\big\}_{\mathrm{\eqref{eq:mim-generic-c-7}}}\bigg\}\label{eq:mim-PMAIC}
\end{align}
The inner bound of \eqref{eq:mim-PMAIC} for the $(2,1)$ MAC-IC-MAC coincides with the result given in the conference version of
this paper \cite[Theorem 1]{pang2013bounds}.
\end{cor}
\begin{rem}
Since we are studying a more general channel model here than in \cite{pang2013bounds},
the missing proofs in \cite{pang2013bounds} can be completed by setting
$K_{1}=2$ and $K_{2}=1$ in the results and proofs in Section \ref{sec:mim-main-results}
and the appendices. Some clarifications are provided regarding \cite{pang2013bounds}
in Appendix \ref{sec:mim-errors}.
\end{rem}
\begin{rem}
\label{remark:blah}
Theorems \ref{thm:mim-dm-inner-fix-dist} and \ref{thm:mim-dm-inner-compact}
assert that $\mathcal{R}_{\mathrm{in}}(P_{\mathrm{in}})$ (hence $\mathcal{R}_{\mathrm{in}}$)
and $\mathcal{R}_{\mathrm{c}}$ are achievable by fixed (and union
of) input distribution(s) and by the union over admissible input distributions,
respectively. However, the per coding distribution region in $\mathcal{R}_{{\rm c}}$, namely $\mathcal{R}_{{\rm c}}(P_{\mathrm{in}})$, is
not known to be achievable by its associated single input distribution
$P_{\mathrm{in}}$ (cf. Example \ref{exa:mim-counter-ex} for a case
when $\mathcal{R}_{{\rm c}}(P_{\mathrm{in}})\supsetneq\mathcal{R}_{{\rm in}}(P_{\mathrm{in}})$
). 
\end{rem}
\begin{rem}
In the special case where the two MACs are non-interacting (i.e., there is no interference from either MAC to the other), it is easily verified that $ \mathcal{R}_{{\rm in}} $ reduces to the Cartesian product of the capacity regions of the two MACs, as it should.
\end{rem}

\subsection{Bounds for the Semi-Deterministic MAC-IC-MAC\label{subsec:mim-sd-bounds}}

The inner bounds
stated in the previous subsection are of course applicable to the semi-deterministic MAC-IC-MAC. However,
our goal is, as it was in the work of Telatar and Tse for the two-user
semi-deterministic IC \cite{telatar2007bounds}, to obtain inner and outer bounds for it so that the gap between the two (and hence gap to the capacity region) is {\em quantifiable}.

Since this section is long, we give a summary of it first.

In Section \ref{sec-sd-inner}, to obtain an inner bound, we consider a region contained in the inner bound of Theorem
\ref{thm:mim-dm-inner-compact} obtained by restricting the union of regions
therein to a class $\mathcal{P}_{\mathrm{in}}^{\mathrm{sd}}\subseteq\mathcal{P}_{\mathrm{in}}$ (to be defined in \eqref{eq:mim-sd-union-compact})
that depends on the semi-deterministic property. 
Denote the resulting inner bound as $\mathcal{R}_{\mathrm{c}}^{\mathrm{sd}}$.

In Section \ref{sec-sd-outer}, we use the semi-deterministic property given in \eqref{eq:mim-sd-model} to determine an outer bound $\mathcal{R}_{\mathrm{o}}^{\mathrm{sd}}$.
The key idea is to allow a genie to give Rx$i$ side information $T_{i^{'}.1}$
which has the same distribution as $S_{i^{'}}$ given $X_{\varTheta_{i^{'}}}$
to help Rx$i$'s decoding and obtain an outer bound for this genie-aided
semi-deterministic MAC-IC-MAC. Since providing side information to
the receivers does not decrease capacity, an outer bound to the original semi-deterministic MAC-IC-MAC can therefore be characterized. This extends the outer bound of \cite{telatar2007bounds}
for the two-user semi-deterministic IC to the semi-deterministic MAC-IC-MAC.

We will then show in Section \ref{subsubsec:mim-sd-gap} that the gap between the inner bound $\mathcal{R}_{\mathrm{c}}(P_{\mathrm{in}}^{\mathrm{sd}})$
and the outer bound $\mathcal{R}_{\mathrm{o}}^{\mathrm{sd}}(P_{\mathrm{in}}^{\mathrm{sd}})$
is quantifiable for each $(Q,X_{\varTheta_{1}},X_{\varTheta_{2}})\in\mathcal{P}_{\mathrm{in}}^{\mathrm{sd}}$,
which leads to a quantifiable gap between the union inner and outer
bounds $\mathcal{R}_{\mathrm{c}}^{\mathrm{sd}}$ and $\mathcal{R}_{\mathrm{o}}^{\mathrm{sd}}$,
thereby extending the main result of \cite{telatar2007bounds} to
the semi-deterministic MAC-IC-MAC.

Furthermore, in Section \ref{sec:big-gap}, we quantify the gap between $\mathcal{R}_{\mathrm{in}}(P_{\mathrm{in}}^{\mathrm{sd}})$
and $\mathcal{R}_{\mathrm{c}}(P_{\mathrm{in}}^{\mathrm{sd}})$ to investigate if the rate region $\mathcal{R}_{\mathrm{in}}^{\mathrm{sd}}$ (defined in \eqref{eq:mim-sd-inner}) contributed by the restricted set of coding distributions $\mathcal{P}_{\mathrm{in}}^{\mathrm{sd}}$ could also be
guaranteed to be within a quantifiable gap to the outer bound. 
This question is answered in the affirmative by Lemma
\ref{lem:mim-sd-gap-1} in Section \ref{sec:big-gap}. The motivation for this is given next.

As mentioned in Remark \ref{remark:blah}, it is unclear that $\mathcal{R}_{\mathrm{c}}(P_{\mathrm{in}}^{\mathrm{sd}})$
is achievable by its associated distribution $P_{\mathrm{in}}^{\mathrm{sd}}$.
Similarly, there is no guarantee that any rate-tuple $\mathcal{R}_{\mathrm{c}}^{\mathrm{sd}}$
will be achievable by some $P_{\mathrm{in}}^{\mathrm{sd}}\in\mathcal{P}_{\mathrm{in}}^{\mathrm{sd}}$,
i.e., it is not clear if the inequalities \eqref{eq:mim-generic-2}
and \eqref{eq:mim-generic-4} will still be redundant when only the
union over $\mathcal{P}_{\mathrm{in}}^{\mathrm{sd}}$ is considered.
However, we know from Theorem \ref{thm:mim-dm-inner-fix-dist} that
$\mathcal{R}_{\mathrm{in}}(P_{\mathrm{in}}^{\mathrm{sd}})$ is achievable
by its coding distribution $P_{\mathrm{in}}^{\mathrm{sd}}$. Moreover, in Lemma
\ref{lem:mim-sd-gap-1}, we will show that for any fixed $P_{\mathrm{in}}^{\mathrm{sd}}\in\mathcal{P}_{\mathrm{in}}^{\mathrm{sd}}$,
the gap between $\mathcal{R}_{\mathrm{in}}(P_{\mathrm{in}}^{\mathrm{sd}})$
and $\mathcal{R}_{\mathrm{c}}(P_{\mathrm{in}}^{\mathrm{sd}})$ is
quantifiable for a given $p(q,x_{\varTheta_{1}},x_{\varTheta_{2}})$.
Hence the gap between $\mathcal{R}_{\mathrm{in}}(P_{\mathrm{in}}^{\mathrm{sd}})$
and $\mathcal{R}_{\mathrm{o}}(P_{\mathrm{in}}^{\mathrm{sd}})$ (and hence between $\mathcal{R}_{\mathrm{in}}^{\mathrm{sd}}$ and $\mathcal{R}_{\mathrm{o}}^{\mathrm{sd}}$) is
quantifiable as well, although it is larger. Thus, in contrast to $\mathcal{R}_{\mathrm{c}}^{\mathrm{sd}}$ which requires for its achievability the consideration of {\em all} distributions in $ \mathcal{P}_{\mathrm{in}} $, the achievability of $\mathcal{R}_{\mathrm{in}}^{\mathrm{sd}}$ requires the consideration of only distributions in $ \mathcal{P}_{\mathrm{in}}^{\mathrm{sd}} $.

\subsubsection{Inner bound}
\label{sec-sd-inner}

As mentioned previously, we specify the joint distribution of the
auxiliary random variables $U_{1}$ and $U_{2}$ given $Q$, $X_{\varTheta_{1}}$
and $X_{\varTheta_{2}}$ in the inner bounds of Theorems \ref{thm:mim-dm-inner-fix-dist}
and \ref{thm:mim-dm-inner-compact} as follows: The cloud symbols
$(U_{1},U_{2})$ take values in $\mathcal{S}_{1}\times\mathcal{S}_{2}$
according to the conditional distribution

\begin{align}
p(u_{1},u_{2}|x_{\varTheta_{1}},x_{\varTheta_{2}},q) & =p_{S_{1}|X_{1.1}}(u_{1}|x_{1.1},q)\times\nonumber \\
 & \qquad p_{S_{2}|X_{2.1}}(u_{2}|x_{2.1},q)\label{eq:mim-sd-inner-dist}
\end{align}
but independently of $(S_{1},S_{2})$ conditioned on $Q$, $X_{\varTheta_{1}}$
and $X_{\varTheta_{2}}$. The choice $p(q,x_{\varTheta_{1}},x_{\varTheta_{2}})$
is taken to be arbitrary. We denote $\mathcal{P}_{\mathrm{in}}^{\mathrm{sd}}$
as the set of distributions $P_{\mathrm{in}}^{\mathrm{sd}}$ that
satisfies \eqref{eq:mim-sd-inner-dist} so that 
\begin{align}
\mathcal{P}_{\mathrm{in}}^{\mathrm{sd}}= & \big\{ P_{\mathrm{in}}\in\mathcal{P}_{\mathrm{in}}:p(q,u_{1},x_{\varTheta_{1}},u_{2},x_{\varTheta_{2}})\nonumber \\
 & =p(q)p(x_{1.1}|q)p_{S_{1}|X_{1.1}}(u_{1}|x_{1.1},q)\nonumber \\
 & \qquad\prod_{1.j\in\varTheta_{1}\backslash\{1.1\}}p(x_{1.j}|q)p(x_{2.1}|q)p_{S_{2}|X_{2.1}}(u_{2}|x_{2.1},q)\nonumber \\
 & \qquad\cdot\prod_{2.j\in\varTheta_{2}\backslash\{2.1\}}p(x_{2.j}|q)\big\}.\label{eq:mim-p-in-sd}
\end{align}
Based on $\mathcal{P}_{\mathrm{in}}^{\mathrm{sd}}$,
we define the set functions $\underline{\mathsf{A}}$, $\underline{\mathsf{B}}$,
$\underline{\mathsf{E}}$ and $\underline{\mathsf{G}}$ for region
$\mathcal{R}_{\mathrm{c}}(P_{\mathrm{in}}^{\mathrm{sd}})$ that have
the same function maps as $\mathsf{A}$, $\mathsf{B}$, $\mathsf{E}$
and $\mathsf{G}$ given by \eqref{eq:mim-dm-inner-a}-\eqref{eq:mim-dm-inner-g}
of Definition \ref{def:mim-dm-inner-set-functions}.
\begin{defn}
\label{def:mim-sd-inner-set-functions}Given any jointly distributed
random variables $(Q,U_{1},X_{\varTheta_{1}},U_{2},X_{\varTheta_{2}})$
with distribution $P_{\mathrm{in}}^{\mathrm{sd}}\in\mathcal{P}_{\mathrm{in}}^{\mathrm{sd}}$,
define the non-negative real-valued set functions $\underline{\mathsf{A}}$,
$\underline{\mathsf{B}}$, $\underline{\mathsf{E}}$ and $\underline{\mathsf{G}}$
as 
\begin{align}
\underline{\mathcal{\mathsf{A}}}_{\varUpsilon_{i}} & \triangleq H(Y_{i}|X_{\bar{\varUpsilon}_{i}},U_{i},U_{i^{'}},Q)-H(S_{i^{'}}|U_{i^{'}},Q)\label{eq:mim-sd-inner-a}\\
\underline{\mathcal{\mathsf{B}}}_{\varOmega_{i}} & \triangleq H(Y_{i}|X_{\bar{\varOmega}_{i}},U_{i^{'}},Q)-H(S_{i^{'}}|U_{i^{'}},Q)\label{eq:mim-sd-inner-b}\\
\underline{\mathcal{\mathsf{E}}}_{\varUpsilon_{i}} & \triangleq H(Y_{i}|X_{\bar{\varUpsilon}_{i}},U_{i},Q)-H(S_{i^{'}}|U_{i^{'}},Q)\label{eq:mim-sd-inner-e}\\
\underline{\mathcal{\mathsf{G}}}_{\varOmega_{i}} & \triangleq H(Y_{i}|X_{\bar{\varOmega}_{i}},Q)-H(S_{i^{'}}|U_{i^{'}},Q).\label{eq:mim-sd-inner-g}
\end{align}
\end{defn}
The set functions $\underline{\mathsf{A}}$, $\underline{\mathsf{B}}$,
$\underline{\mathsf{E}}$ and $\underline{\mathsf{G}}$ are derived
by specializing the set functions
$\mathsf{A}$, $\mathsf{B}$, $\mathsf{E}$ and $\mathsf{G}$ given
in Definition \ref{def:mim-dm-inner-set-functions} by substituting the distribution in \eqref{eq:mim-sd-inner-dist} in place of that in \eqref{eq:mim-inner-fix-dist}.

Since $\mathcal{P}_{\mathrm{in}}^{\mathrm{sd}}\subseteq\mathcal{P}_{\mathrm{in}}$,
Theorem \ref{thm:mim-dm-inner-compact} implies that
\begin{align}
\mathcal{R}_{\mathrm{c}}^{\mathrm{sd}} & \triangleq\bigcup_{P_{\mathrm{in}}^{\mathrm{sd}}\in\mathcal{P}_{\mathrm{in}}^{\mathrm{sd}}}\mathcal{R}_{\mathrm{c}}(P_{\mathrm{in}}^{\mathrm{sd}})\label{eq:mim-sd-union-compact}\\
 & \triangleq\bigcup_{P_{\mathrm{in}}^{\mathrm{sd}}\in\mathcal{P}_{\mathrm{in}}^{\mathrm{sd}}}\mathcal{R}_{\mathrm{c}}(\underline{\mathsf{A}},\mathsf{\underline{B}},\underline{\mathsf{E}},\underline{\mathsf{G}})\nonumber 
\end{align}
is achievable.

In the next section, we obtain an outer bound to the capacity region
that has the same structure as that of the inner bound $\mathcal{R}_{\mathrm{c}}^{\mathrm{sd}}$
so that the gap between the two can be quantified.



\subsubsection{Outer bound}
\label{sec-sd-outer}

To describe the outer bound, consider the following definition. 
\begin{defn}
Let $\mathcal{P}_{\mathrm{o}}^{\mathrm{sd}}$ be the set of distributions
$P_{\mathrm{o}}^{\mathrm{sd}}$ of the ensemble $(Q,T_{1},X_{\varTheta_{1}},T_{2},X_{\varTheta_{2}})$
which can be factored as 
\begin{align}
 & p(q,t_{1},x_{\varTheta_{1}},t_{2},x_{\varTheta_{2}})\nonumber \\
 & \qquad=p(q)p_{S_{1}|X_{1}}(t_{1}|x_{1.1})\prod_{1.j\in\varTheta_{1}}p(x_{1.j}|q)\nonumber \\
 & \qquad\qquad\times p_{S_{2}|X_{2}}(t_{2}|x_{2.1})\prod_{2.j\in\varTheta_{2}}p(x_{2.j}|q).\label{eq:mim-p-out}
\end{align}
\end{defn}
\begin{defn}
\label{def:mim-sd-outer}Given any set of random variables $(Q,X_{\varTheta_{1}},T_{1},X_{\varTheta_{2}},T_{2})$
with joint distribution $P_{\mathrm{o}}^{\mathrm{sd}}$, let the set
functions $\overline{\mathsf{A}}$, $\overline{\mathsf{B}}$, $\overline{\mathsf{E}}$
and $\overline{\mathsf{G}}$ be 
\begin{align}
\overline{\mathcal{\mathsf{A}}}_{\Upsilon_{i}} & \triangleq H(Y_{i}|X_{\bar{\varUpsilon}_{i}},T_{i},X_{i^{'}.1},Q)-H(S_{i^{'}}|X_{i^{'}.1},Q)\label{eq:mim-sd-outer-A}\\
\overline{\mathsf{B}}_{\varOmega_{i}} & \triangleq H(Y_{i}|X_{\bar{\varOmega}_{i}},X_{i^{'}.1},Q)-H(S_{i^{'}}|X_{i^{'}.1},Q)\label{eq:mim-sd-outer-B}\\
\overline{\mathcal{\mathsf{E}}}_{\Upsilon_{i}} & \triangleq H(Y_{i}|X_{\bar{\varUpsilon}_{i}},T_{i},Q)-H(S_{i^{'}}|X_{i^{'}.1},Q)\label{eq:mim-sd-outer-E}\\
\overline{\mathsf{G}}_{\varOmega_{i}} & \triangleq H(Y_{i}|X_{\bar{\varOmega}_{i}},Q)-H(S_{i^{'}}|X_{i^{'}.1},Q)\label{eq:mim-sd-outer-G}
\end{align}
where the auxiliary random variables $(T_{1},T_{2}$) take values
in $\mathcal{S}_{1}\times\mathcal{S}_{2}$ according to 
\begin{align}
p(t_{1},t_{2}|x_{\varTheta_{1}},x_{\varTheta_{2}},q) & =p_{S_{1}|X_{1}}(t_{1}|x_{1.1})\cdot p_{S_{2}|X_{2}}(t_{2}|x_{2.1})\label{eq:mim-sd-genie}
\end{align}
and are conditionally independent of $S_{i}$ given $X_{\varTheta_{i}}$.
\end{defn}
\begin{thm}
\label{thm:mim-sd-outer} For the semi-deterministic MAC-IC-MAC, defining
the region $\mathcal{R}_{{\rm o}}^{{\rm sd}}(P_{\mathrm{o}}^{\mathrm{sd}})\triangleq\mathcal{R}_{\mathrm{c}}(\overline{\mathsf{A}},\overline{\mathsf{B}},\overline{\mathsf{E}},\overline{\mathsf{G}}),$
we have 
\begin{align*}
\mathcal{C}^{{\rm sd}} & \subseteq\mathcal{R}_{{\rm o}}^{{\rm sd}}\\
 & \triangleq\bigcup_{P_{\mathrm{o}}^{\mathrm{sd}}\in\mathcal{P}_{\mathrm{o}}^{\mathrm{sd}}}\mathcal{R}_{{\rm o}}^{{\rm sd}}(P_{\mathrm{o}}^{\mathrm{sd}}).
\end{align*}
\end{thm}
\begin{IEEEproof}
For the two-user semi-deterministic IC, a proof is given in \cite[Theorem 1]{telatar2007bounds} for a similar outer bound
with a genie-aided argument. When extending their proof technique, the
intra-cell sum rates $R_{\varUpsilon_{i}}$ or $R_{\varOmega_{i}}$
for a given user subset $\varUpsilon_{i}$ or $\varOmega_{i}$, instead
of each individual rate, are upper bounded and for which purpose the
transmitted signals of unconsidered transmitters in the same MAC,
i.e., $X_{\mathrm{\bar{\varUpsilon}_{i}}}$ or $X_{\mathrm{\bar{\varOmega}_{i}}}$,
are fed to the receiver Rx$i$ as part of side information. The fact
that each transmitter in a MAC encodes its own message independently
is utilized, and consequently, we obtain four classes of intra-cell
sum-rate upper bounds with the same structure as shown in the proof
of \cite[Theorem 1]{telatar2007bounds}. Linearly combining these
upper bounds across two MACs cancels the negative entropy terms which
are not in the form of $H(\mathrm{output}|\mathrm{input})$ in several
different ways, which leads to the seven classes of inequalities as
given in the theorem. Please refer to Appendix \ref{sec:mim-outer-proof}
for details.
\end{IEEEproof}

\subsubsection{Quantifiable gap\label{subsubsec:mim-sd-gap}}

Next, we show that the inner bound $\mathrm{\mathcal{R}_{\mathrm{c}}^{\mathrm{sd}}}$
is within a quantifiable gap of the outer bound $\mathcal{R}_{\mathrm{o}}^{\mathrm{sd}}$.
\begin{thm}
\label{lem:mim-sd-gap-2} Consider the semi-deterministic MAC-IC-MAC.
For any rate tuple $(R_{\varTheta_{1}},R_{\varTheta_{2}})\in\mathcal{R}_{{\rm o}}^{{\rm sd}}(P_{\mathrm{o}}^{\mathrm{sd}})$,
let $\tilde{R}_{\varTheta_{i}}$ be the rate tuple 
\begin{align*}
\tilde{R}_{\varTheta_{i}}=\bigg(\left(R_{i.1}-I(X_{i^{'}.1};S_{i^{'}}|U_{i^{'}},Q)\right)^{+},\\
\left(R_{i.2}-I(X_{i^{'}.1};S_{i^{'}}|U_{i^{'}},Q)\right)^{+},\\
\cdots,\left(R_{i.K_{i}}-I(X_{i^{'}.1};S_{i^{'}}|U_{i^{'}},Q)\right)^{+}\bigg).
\end{align*}
Then there exists $P_{\mathrm{in}}^{\mathrm{sd}}\in\mathcal{P}_{\mathrm{in}}^{\mathrm{sd}}$,
such that
\[
(\tilde{R}_{\varTheta_{1}},\tilde{R}_{\varTheta_{2}})\in\mathcal{R}_{{\rm c}}^{\mathrm{sd}}(P_{\mathrm{in}}^{\mathrm{sd}}).
\]
\end{thm}
\begin{IEEEproof}
The inequality systems of outer and compact form inner bounds have
the same algebraic structure. Also there is one-to-one correspondence
between the involved intra-cell sum rate term on the left hand side
and the set function on the right hand side of each inequality in
both bounds. Hence, we need only show that the difference between every corresponding pair
of set functions from the two bounds, i.e., $\overline{\mathsf{A}}_{\varUpsilon_{i}}-\underline{\mathsf{A}}_{\varUpsilon_{i}}$,
$\overline{\mathsf{B}}_{\varOmega_{i}}-\underline{\mathsf{B}}_{\varOmega_{i}}$,
etc, is within $I(X_{i^{'}.1};S_{i^{'}}|U_{i^{'}},Q)$. For example, the difference between $\overline{\mathsf{A}}_{\varUpsilon_{1}}$
and $\underline{\mathsf{A}}_{\varUpsilon_{1}}$ can be quantified
as follows:
\begin{align*}
\overline{\mathcal{\mathsf{A}}}_{\Upsilon_{1}} & \leq H(Y_{1}|X_{\bar{\varUpsilon}_{1}},T_{1},X_{2.1},Q)-H(S_{2}|X_{2.1},Q)\\
 & \overset{(a)}{=}H(Y_{1}|X_{\bar{\varUpsilon}_{1}},U_{1},X_{2.1},Q)-H(S_{2}|X_{2.1},Q)\\
 & \stackrel{(b)}{\leq}H(Y_{1}|X_{\bar{\varUpsilon}_{1}},U_{1},U_{2},Q)-H(S_{2}|X_{2.1},Q)\\
 & \stackrel{(c)}{\leq}H(Y_{1}|X_{\bar{\varUpsilon}_{1}},U_{1},U_{2},Q)-H(S_{2}|U_{2},Q)\\
 & \qquad+I(X_{2.1};S_{2}|U_{2},Q)\\
 & =\underline{\mathsf{A}}_{\varUpsilon_{1}}+I(X_{2.1};S_{2}|U_{2},Q).
\end{align*}
The steps (a), (b) and (c) hold true because (a) random variables
$U_{i}$, $T_{i}$ and $S_{i}$ are i.i.d. conditioned on $X_{i.1}$
as stated by \eqref{eq:mim-p-in-sd} and \eqref{eq:mim-p-out}. Hence,
as long as we choose $P_{\mathrm{in}}^{\mathrm{sd}}$ so that $P_{\mathrm{in}}^{\mathrm{sd}}$
and $P_{\mathrm{o}}^{\mathrm{sd}}$ have identical marginal distribution
$p(q,x_{\varTheta_{1}},x_{\varTheta_{2}})$, we can replace $U_{1}$
with $T_{1}$ in the conditional entropy term $H(Y_{1}|X_{\bar{\varUpsilon}_{1}},U_{1},X_{2.1},Q)$.
(b) $Q-\circ-U_{2}-\circ-X_{2.1}$ (form a Markov chain) (c)
given $X_{\varTheta_{1}}$, there is a one-to-one correspondence between
$Y_{1}$ and $S_{2}$. Other terms can be verified similarly, which
proves the theorem.
\end{IEEEproof}
\begin{rem}
In the MAC-DIC-MAC (DIC stands for deterministic IC) where the channel
side information $S_{i}$ is a deterministic function of the respective
$X_{i.1}$, we have $T_{i}=S_{i}$ and hence the gap becomes zero.
So, the inner bound $\mathcal{R}_{\mathrm{in}}^{\mathrm{sd}}$ is
the capacity region for the MAC-DIC-MAC. When $K_{1}=K_{2}=1$, this
result recovers the result for the DIC of El Gamal and Costa in \cite{gamal1982capacity}.
\end{rem}

\subsubsection{Achievability by coding distributions in $ \mathcal{P}_{\mathrm{in}}^{\mathrm{sd}}$}
\label{sec:big-gap}
In the previous section we quantified the gap to capacity of the inner
bound $\mathcal{R}_{\mathrm{c}}^{\mathrm{sd}}$ of \eqref{eq:mim-sd-union-compact}.
Note however that it is unclear if each region in the union that defines
it, namely $\mathcal{R}_{{\rm c}}(P_{\mathrm{in}}^{\mathrm{sd}})$,
is achievable with its single associated coding distribution $P_{\mathrm{in}}^{\mathrm{sd}}$.
Hence, it is unclear if any
rate-tuple in $\mathcal{R}_{\mathrm{c}}^{\mathrm{sd}}$ is achievable
by some $P_{\mathrm{in}}^{\mathrm{sd}}\in\mathcal{P}_{\mathrm{in}}^{\mathrm{sd}}$.
In particular, recall that in the proof of Theorem \ref{thm:mim-dm-inner-compact}
(cf. Appendix \ref{sec:mim-dm-inner-proof}), we showed that $\mathcal{R}_{\mathrm{c}}\backslash\mathcal{R}_{\mathrm{in}}$ 
is empty by constructing an auxiliary inner bound inside $\mathcal{R}_{\mathrm{in}}$
with some $U_{i}=\emptyset$. By choosing $p(u_{i}|x_{i.1})$ as \eqref{eq:mim-sd-inner-dist},
and with $U_{i}$ and $S_{i}$ are i.i.d. conditioned on $X_{i.1}$,
there is no guarantee that $U_{i}$ could be chosen to be $\emptyset$. Hence, it is unclear if $\mathcal{R}_{\mathrm{c}}^{\mathrm{sd}}\backslash\mathcal{R}_{\mathrm{in}}^{\mathrm{sd}}$ is empty.

The fact that any rate-tuple in $\mathcal{R}_{\mathrm{c}}^{\mathrm{sd}}$
may not be achievable by some $P_{\mathrm{in}}^{\mathrm{sd}}\in\mathcal{P}_{\mathrm{in}}^{\mathrm{sd}}$
might be seen as an undesirable feature. In this section, we analyze
the gap to capacity of an inner bound that, even though subsumed by
$\mathcal{R}_{\mathrm{c}}^{\mathrm{sd}}$, does not have that undesirable
feature.

Since $P_{\mathrm{in}}^{\mathrm{sd}}\in\mathcal{P}_{\mathrm{in}}$,
Theorem \ref{thm:mim-dm-inner-fix-dist} implies that $\mathcal{R}_{{\rm in}}(P_{\mathrm{in}}^{\mathrm{sd}})$
is an inner bound to $\mathcal{C}$ and, moreover, it is achievable
by its associated single coding distribution $P_{\mathrm{in}}^{\mathrm{sd}}$.
Also, it is clear that 
\begin{equation}
\mathcal{R}_{\mathrm{in}}^{\mathrm{sd}}\triangleq\bigcup_{P_{\mathrm{in}}^{\mathrm{sd}}\in\mathcal{P}_{\mathrm{in}}^{\mathrm{sd}}}\mathcal{R}_{{\rm in}}(P_{\mathrm{in}}^{\mathrm{sd}})\subseteq\mathcal{C}\label{eq:mim-sd-inner}
\end{equation}
and that any rate-tuple in $\mathcal{R}_{\mathrm{in}}^{\mathrm{sd}}$
is achievable by some distribution in $\mathcal{P}_{\mathrm{in}}^{\mathrm{sd}}$.
Clearly, $\mathcal{R}_{\mathrm{in}}^{\mathrm{sd}}\subseteq\mathcal{R}_{\mathrm{c}}^{\mathrm{sd}}$.
We will quantify in this section the gap to capacity of the smaller
inner bound $\mathcal{R}_{\mathrm{in}}^{\mathrm{sd}}$ to capacity.
The gap is naturally expected to be larger than that found between
$\mathcal{R}_{\mathrm{c}}^{\mathrm{sd}}$ and capacity in the previous
section.




As stated previously, the strategy is to first quantify the gap between $\mathcal{R}_{\mathrm{in}}(P_{\mathrm{in}}^{\mathrm{sd}})$
and $\text{\ensuremath{\mathcal{R}}}_{\mathrm{c}}(P_{\mathrm{in}}^{\mathrm{sd}})$
and then use the result of Theorem \ref{lem:mim-sd-gap-2} on the
gap between $\mathcal{R}_{\mathrm{c}}(P_{\mathrm{in}}^{\mathrm{sd}})$
and $\text{\ensuremath{\mathcal{R}}}_{\mathrm{o}}(P_{\mathrm{o}}^{\mathrm{sd}})$
to quantify the gap between $\mathcal{R}_{\mathrm{in}}(P_{\mathrm{in}}^{\mathrm{sd}})$
and $\text{\ensuremath{\mathcal{R}}}_{\mathrm{o}}(P_{\mathrm{o}}^{\mathrm{sd}})$.

\begin{lem}
\label{lem:mim-sd-gap-1}Given $P_{\mathrm{in}}^{\mathrm{sd}}\in\mathcal{P}_{\mathrm{in}}^{\mathrm{sd}}$,
for any rate tuple $(R_{\varTheta_{1}},R_{\varTheta_{2}})\in\mathcal{R}_{\mathrm{c}}(P_{\mathrm{in}}^{\mathrm{sd}})$,
let $\tilde{R}_{\varTheta_{i}}$ be the rate tuple 
\begin{align*}
\bigg(\left(R_{i.1}-I(X_{i.1};S_{i}|U_{i})\right)^{+},\left(R_{i.2}-I(X_{i.1};S_{i}|U_{i})\right)^{+},\\
\cdots,\left(R_{i.K_{i}}-I(X_{i.1};S_{i}|U_{i})\right)^{+}\bigg).
\end{align*}
Then we have 
\[
(\tilde{R}_{\varTheta_{1}},\tilde{R}_{\varTheta_{2}})\in\mathcal{R}_{{\rm in}}(P_{\mathrm{in}}^{\mathrm{sd}}).
\]
\end{lem}
\begin{IEEEproof}
The bounds $\text{\ensuremath{\mathcal{R}}}_{\mathrm{in}}(P_{\mathrm{in}}^{\mathrm{sd}})$
and $\text{\ensuremath{\mathcal{R}}}_{c}(P_{\mathrm{in}}^{\mathrm{sd}})$
differ from each other in that $\text{\ensuremath{\mathcal{R}}}_{\mathrm{c}}(P_{\mathrm{in}}^{\mathrm{sd}})$
has intra-cell sum rate $\sum_{i.j\in\varOmega_{i}}R_{i.j}$ only
bounded by $\mathsf{B}_{\varOmega_{i}}$, whereas $\text{\ensuremath{\mathcal{R}}}_{\mathrm{in}}(P_{\mathrm{in}}^{\mathrm{sd}})$
has $\sum_{i.j\in\varOmega_{i}}R_{i.j}$ additionally bounded by $\mathsf{A}_{\varUpsilon_{i}}+\mathsf{E}_{\varUpsilon_{i^{'}}}$.
The semi-deterministic structure of channel \eqref{eq:mim-sd-model}
and the coding distribution \eqref{eq:mim-sd-inner-dist} are fully
considered in the proof. Please refer to Appendix \ref{sec:mim-sd-gap-1-proof}
for details.
\end{IEEEproof}
\begin{rem}
If $K_1=K_2=1$, Lemma \ref{lem:mim-sd-gap-1} quantifies the gap between
the non-compact and compact form inner bounds for two-user semi-deterministic
IC. To the best of our knowledge, this gap was not observed heretofore in the literature on the two-user IC.
\end{rem}
Now we are ready to state the main result of this section.
\begin{thm}
\label{thm:mim-sd-gap}For any rate tuple $(R_{\varTheta_{1}},R_{\varTheta_{2}})\in\mathcal{R}_{{\rm o}}^{{\rm sd}}$,
let $I_{0}=I(X_{1.1};S_{1}|U_{1},Q)+I(X_{2.1};S_{2}|U_{2},Q)$ and
$\tilde{R}_{\varTheta_{i}}$ be the rate tuple $\left(\left(R_{i.1}-I_{0}\right)^{+},\left(R_{i.2}-I_{0}\right)^{+},\cdots,\left(R_{i.K_{i}}-I_{0}\right)^{+}\right)$.
Then we have 
\[
(\tilde{R}_{\varTheta_{1}},\tilde{R}_{\varTheta_{2}})\in\mathcal{R}_{{\rm in}}^{\mathrm{sd}}.
\]
\end{thm}
\begin{IEEEproof}
Combining Lemma \ref{lem:mim-sd-gap-1} and Theorem \ref{lem:mim-sd-gap-2},
it can be inferred that for any $\mathcal{R}_{{\rm o}}^{{\rm sd}}(P_{\mathrm{o}}^{\mathrm{sd}})$,
$\exists P_{\mathrm{in}}^{\mathrm{sd}}\in\mathcal{P}_{\mathrm{in}}^{\mathrm{sd}}$
so that the gap between $\mathcal{R}_{\mathrm{in}}^{\mathrm{sd}}(P_{\mathrm{in}}^{\mathrm{sd}})$
and $\mathcal{R}_{\mathrm{o}}^{\mathrm{sd}}(P_{\mathrm{o}}^{\mathrm{sd}})$
will not exceed the sum of the gaps from $\mathcal{R}_{\mathrm{in}}(P_{\mathrm{in}}^{\mathrm{sd}})$
to $\mathcal{R}_{\mathrm{c}}^{\mathrm{sd}}(P_{\mathrm{in}}^{\mathrm{sd}})$
and from $\mathcal{R}_{\mathrm{c}}^{\mathrm{sd}}(P_{\mathrm{in}}^{\mathrm{sd}})$
to $\mathcal{R}_{\mathrm{o}}^{\mathrm{sd}}(P_{\mathrm{o}}^{\mathrm{sd}})$.
This sum gap is hence at most $I_{0}=I(X_{1.1};S_{1}|U_{1},Q)+I(X_{2.1};S_{2}|U_{1},Q)$.
Taking the union over $\mathcal{P}_{\mathrm{in}}^{\mathrm{sd}}$ and
$\mathcal{P}_{\mathrm{o}}^{\mathrm{sd}}$ for $\mathcal{R}_{\mathrm{in}}^{\mathrm{sd}}(P_{\mathrm{in}}^{\mathrm{sd}})$
and $\mathcal{R}_{\mathrm{o}}^{\mathrm{sd}}(P_{\mathrm{o}}^{\mathrm{sd}})$,
we can conclude that the gap between the union bounds $\mathcal{R}_{{\rm in}}^{{\rm sd}}$
and $\mathcal{R}_{{\rm o}}^{{\rm sd}}$ is also quantifiable and within
$I_{0}$.
\end{IEEEproof}

\subsection{The Approximate Capacity Region of the Gaussian MAC-IC-MAC to Within
Two Bits\label{subsec:mim-gs-gap}}

The quantifiable gap for the semi-deterministic MAC-IC-MAC requires the
inner bound in Theorem \ref{thm:mim-dm-inner-compact}, which is a
union bound over $\mathcal{P}_{\mathrm{in}}^{\mathrm{sd}}$. When
the channel gets further specialized to the Gaussian case, can we specify a universal (single distribution)
coding scheme whose achievable rate region is within a constant gap
of the capacity region for the Gaussian MAC-IC-MAC? In this subsection, we will answer this question in the affirmative. Indeed,
we will give inner and outer bounds for the Gaussian
MAC-IC-MAC as explicit single polyhedral regions, and show these are within two bits of each other, independently of all channel parameters.

\subsubsection{Inner bound}

We provide an inner bound for a simple coding scheme first. We will show in Section \ref{sec:twobitgapg} that such a strategy has an achievable rate region that is within a two-bit gap to the capacity region.
 
\begin{defn}
\label{def:mim-gs-inner}Suppose $\mathsf{C}(P)=\log(1+P)$ for some
$P\geq0$. Define the coefficient 
\begin{equation}
\mu_{i}\triangleq\min\left\{ 1,\frac{1}{\mathsf{INR}_{i.1\rightarrow i^{'}}}\right\} ,\label{eq:mim-gs-inner-mu}
\end{equation}
set functions $A$, $B$, $E$ and $G$ as 
\begin{align}
A_{\varUpsilon_{i}} & \triangleq\mathsf{C}\left(\frac{\mu_{i}\mathsf{SNR}_{i.1\rightarrow i}+\sum_{i.j\in\varUpsilon_{i}\backslash\{i.1\}}\mathsf{SNR}_{i.j\rightarrow i}}{1+\mu_{i^{'}}\mathsf{INR}_{i^{'}.1\rightarrow i}}\right)\label{eq:mim-gs-inner-a}\\
B_{\varOmega_{i}} & \triangleq\mathsf{C}\left(\frac{\sum_{i.j\in\varOmega_{i}}\mathsf{SNR}_{i.j\rightarrow i}}{1+\mu_{i^{'}}\mathsf{INR}_{i^{'}.1\rightarrow i}}\right)\label{eq:mim-gs-inner-b}\\
E_{\varUpsilon_{i}} & \triangleq\mathsf{C}\left(\frac{\mu_{i}\mathsf{SNR}_{i.1\rightarrow i}+\sum_{i.j\in\varUpsilon_{i}\backslash\{i.1\}}\mathsf{SNR}_{i.j\rightarrow i}}{1+\mu_{i^{'}}\mathsf{INR}_{i^{'}.1\rightarrow i}}\right.\nonumber \\
 & \qquad\qquad\qquad\qquad\qquad\left.+\frac{(1-\mu_{i^{'}})\mathsf{INR}_{i^{'}.1\rightarrow i}}{1+\mu_{i^{'}}\mathsf{INR}_{i^{'}.1\rightarrow i}}\right)\label{eq:mim-gs-inner-e}\\
G_{\varOmega_{i}} & \triangleq\mathsf{C}\left(\frac{\sum_{i.j\in\varOmega_{i}}\mathsf{SNR}_{i.j\rightarrow i}+(1-\mu_{i^{'}})\mathsf{INR}_{i^{'}.1\rightarrow i}}{1+\mu_{i^{'}}\mathsf{INR}_{i^{'}.1\rightarrow i}}\right)\label{eq:mim-gs-inner-g}
\end{align}
and the region
\[
\mathcal{R}_{\mathrm{in}}^{\mathrm{G}}\triangleq\mathcal{R}(A,B,E,G)
\]
The superscript ``G'' denotes Gaussian. 
\end{defn}
\begin{thm}
\label{thm:mim-gs-inner}For the Gaussian MAC-IC-MAC, $\mathbf{\mathcal{R}}_{{\rm in}}^{{\rm G}}\subseteq\mathcal{C}^{{\rm G}}$. 
\end{thm}
\begin{IEEEproof}
Consider a single coding scheme in which all transmitters use all available power, Tx$i.j$, $j\neq 1$, uses a Gaussian codebook, and the interfering transmitter Tx$i.1$ splits its message into common and private sub-messages and employs additive superposition coding with Gaussian codebooks with powers $P_{i.1c}$ and $P_{i.1p}$, respectively. In particular, the Etkin-Tse-Wang (ETW) power-split \cite{etkin2008gaussian} 
\begin{align}
P_{i.1p} & =\min(P_{i.1},\frac{1}{|h_{i.1\rightarrow i^{'}}|^{2}})\label{eq:mim-private-power}\\
P_{i.1c} & =P_{i,1}-P_{i.1p}.\label{eq:mim-common-power}
\end{align}
is used so that the unintended private message from Tx$i$ arrives
at Rx$i^{'}$ at no more than noise power. 

This scheme can be viewed as the CMG scheme with a coding distribution from $\mathcal{P}_{\mathrm{in}}$ where both the distribution of $U_{i}$ and $X_{i}$ are fixed to be Gaussian, and
\[
p(x_{i}|u_{i})=p(u_{i}+v_{i}|u_{i}).
\]
Thus, Theorem \ref{thm:mim-dm-inner-fix-dist} applies. All we need is the explicit evaluation of the mutual information terms in its bounds for our particular coding scheme, which gives  $A_{\varUpsilon_{i}}$,
$B_{\varOmega_{i}}$, $E_{\varUpsilon_{i}}$ and $G_{\varOmega_{i}}$ of \eqref{eq:mim-gs-inner-g}, hence the result.

\end{IEEEproof}
\begin{rem}
Note that the Gaussian MAC-IC-MAC is semi-deterministic. However, the coding
distribution used to obtain Theorem \ref{thm:mim-gs-inner} is not in $\mathcal{P}_{\mathrm{in}}^{\mathrm{sd}}$. Hence, we cannot directly apply the results in Section \ref{subsec:mim-sd-bounds} due to the average power constraint \eqref{eq:mim-power-constraint} on the transmitted codeword $X_{i.1}$. In other words, choosing $U_{i}$ according to $p(u_{i}|x_{i})=p(s_{i}|x_{i})$ will not always be feasible since the side information $S_{i}$ may have arbitrarily large power.
\end{rem}
\begin{rem}
Define the region
\[
\mathbf{\mathcal{R}}_{\mathrm{c}}^{{\rm G}}\triangleq\mathcal{R}_{\mathrm{c}}(A,B,E,G).
\]
Obviously, $\mathcal{R}_{\mathrm{in}}^{\mathrm{G}}\subseteq\mathcal{R}_{\mathrm{c}}^{\mathrm{G}}$,
but the achievability of $\mathcal{R}_{\mathrm{c}}^{\mathrm{G}}$
by ETW coding scheme is not clear. In fact, the power split scheme
\eqref{eq:mim-private-power} and \eqref{eq:mim-common-power} we
used for $\mathcal{R}_{\mathrm{in}}^{\mathrm{G}}$ is not enough to
achieve $\mathcal{R}_{\mathrm{c}}^{\mathrm{G}}$ for all channel settings,
as shown in the next example.
\end{rem}
\begin{example}
\label{exa:mim-counter-ex} We choose $K_{1}=K_{2}=1$ (two-user IC),
$\mathsf{SNR}_{1.1\rightarrow1}=50$, $\mathsf{SNR}_{2.1\rightarrow2}=20$,
$\mathsf{INR}_{1.1\rightarrow2}=10$, $\mathsf{INR}_{2.1\rightarrow1}=15$.
The computation of the set functions in Definition \ref{def:mim-gs-inner}
gives the achievable region $\mathcal{R}^{\mathrm{G}}_{\mathrm{in}}$ to be
\begin{align}
R_{1.1} & \leq B_{\{1.1\}}=4.7004\label{eq:mim-gs-ex-1}\\
R_{1.1} & \leq A_{\{1.1\}}+E_{\{2.1\}}=4.4318\label{eq:mim-gs-ex-2}\\
R_{2.1} & \leq B_{\{2.1\}}=3.4594\label{eq:mim-gs-ex-3}\\
R_{2.1} & \leq A_{\{2.1\}}+E_{\{1.1\}}=4.1293\label{eq:mim-gs-ex-4}\\
R_{1.1}+R_{2.1} & \leq A_{\{1.1\}}+G_{\{2.1\}}=5.7616\label{eq:mim-gs-ex-5}\\
R_{1.1}+R_{2.1} & \leq G_{\{1.1\}}+A_{\{2.1\}}=5.7814\label{eq:mim-gs-ex-6}\\
R_{1.1}+R_{2.1} & \leq E_{\{1.1\}}+E_{\{2.1\}}=6.0168\label{eq:mim-gs-ex-7}\\
2R_{1.1}+R_{2.1} & \leq A_{\{1.1\}}+G_{\{1.1\}}+E_{\{2.1\}}=9.4762\label{eq:mim-gs-ex-8}\\
R_{1.1}+2R_{2.1} & \leq E_{\{1.1\}}+A_{\{2.1\}}+G_{\{2.1\}}=8.0835\label{eq:mim-gs-ex-9}
\end{align}
Hence, due to inequality \eqref{eq:mim-gs-ex-2}, $R_{1.1}$ can be at most 4.4318 bits per channel use,  while in $\mathcal{R}_{\mathrm{c}}^{\mathrm{G}}$
which is defined by all inequalities of $\mathcal{R}^{\mathrm{G}}_{\mathrm{in}}$ except \eqref{eq:mim-gs-ex-2} and \eqref{eq:mim-gs-ex-4}, $R_{1.1}$ can take values
up to $4.7004$ bits per channel use. Nevertheless, we will
show in the next lemma that the gap between $\mathcal{R}_{\mathrm{in}}^{\mathrm{G}}$
and $\mathcal{R}_{\mathrm{c}}^{\mathrm{G}}$ is no more than one bit.
\end{example}
\begin{lem}
\label{lem:mim-gs-gap-1}There is no more than a one-bit gap between
$\mathcal{R}_{\mathrm{in}}^{\mathrm{G}}$ and $\mathcal{R}_{\mathrm{c}}^{\mathrm{G}}$. 
\end{lem}
\begin{IEEEproof}
The extra bounds in $\mathcal{R}_{\mathrm{in}}^{\mathrm{G}}$ compared to those in $\mathcal{R}_{\mathrm{c}}^{\mathrm{G}}$
are only on the intra-cell sum rates $\sum_{1.j\in\varUpsilon_{1}}R_{1.j}$ and $\sum_{2.j\in\varUpsilon_{2}}R_{2.j}$. Without loss
of generality, it is sufficient to show that for any $\varUpsilon_{1},\varUpsilon_{2}\in\left\{ 2^{\varTheta_{1}\backslash\{1.1\}}\cup\{1.1\}\right\} \times\left\{ 2^{\varTheta_{2}\backslash\{2.1\}}\cup\{2.1\}\right\} $
, $\exists\varOmega_{1}\in2^{\varTheta_{1}}\backslash\emptyset$ such
that the gap between $B_{\varOmega_{1}}$ and $A_{\varUpsilon_{1}}+E_{\varUpsilon_{2}}$
is within one bit. In particular,
\begin{align*}
 & \quad\thinspace A_{\varUpsilon_{1}}+E_{\varUpsilon_{2}}\\
 & \geq\mathsf{C}\left(\frac{\mu_{1}\mathsf{SNR}_{1.1\rightarrow1}+\sum_{1.j\in\varUpsilon_{1}\backslash\{1.1\}}\mathsf{SNR}_{1.j\rightarrow1}}{1+\mu_{2}\mathsf{INR}_{2.1\rightarrow1}}\right)\\
 & \qquad+\mathsf{C}\left(\frac{(1-\mu_{1})\mathsf{INR}_{1.1\rightarrow2}}{1+\mu_{1}\mathsf{INR}_{1.1\rightarrow2}}\right)\\
 & =\mathsf{C}\left(\frac{\mu_{1}\mathsf{SNR}_{1.1\rightarrow1}+\sum_{1.j\in\varUpsilon_{1}\backslash\{1.1\}}\mathsf{SNR}_{1.j\rightarrow1}}{1+\mu_{2}\mathsf{INR}_{2.1\rightarrow1}}\right)\\
 & \qquad+\log\left(\frac{1+\mathsf{INR}_{1.1\rightarrow2}}{1+\mu_{1}\mathsf{INR}_{1.1\rightarrow2}}\right)\\
 & \geq\mathsf{C}\left(\frac{\mathsf{SNR}_{1.1\rightarrow1}\frac{\mu_{1}\mathsf{INR}_{1.1\rightarrow2}}{1+\mu_{1}\mathsf{INR}_{1.1\rightarrow2}}+\sum_{1.j\in\varUpsilon_{1}\backslash\{1.1\}}\mathsf{SNR}_{1.j\rightarrow1}}{1+\mu_{2}\mathsf{INR}_{2.1\rightarrow1}}\right)
\end{align*}
Note the power splitting scheme ensures $\mu_{1}\mathsf{INR}_{1.1\rightarrow2}\leq1$,
hence the above term should satisfy 
\begin{align*}
 & \quad\thinspace A_{\varUpsilon_{1}}+E_{\varUpsilon_{2}}\\
 & \geq\mathsf{C}\left(\frac{\mathsf{SNR}_{1.1\rightarrow1}/2+\sum_{1.j\in\varUpsilon_{1}\backslash\{1.1\}}\mathsf{SNR}_{1.j\rightarrow1}}{1+\mu_{2}\mathsf{INR}_{2.1\rightarrow1}}\right)\\
 & >\mathsf{C}\left(\frac{\mathsf{SNR}_{1.1\rightarrow1}+\sum_{1.j\in\varUpsilon_{1}\backslash\{1.1\}}\mathsf{SNR}_{1.j\rightarrow1}}{1+\mu_{2}\mathsf{INR}_{2.1\rightarrow1}}\right)-1\\
 & =B_{\varUpsilon_{1}}-1=B_{\varOmega_{1}}-1
\end{align*}
Recall $\varUpsilon_{1}$ is defined to be a user index subset which
contains user $1.1$ while $\varOmega_{1}$ an arbitrary user subset,
so for a given $\varUpsilon_{1}$ we can always find a corresponding
$\varOmega_{1}$ such that $\varOmega_{1}=\varUpsilon_{1}$. 
\end{IEEEproof}

\subsubsection{Outer bound}
\label{sec:outboundg}
Our goal is to obtain a single region outer bound. First, using
the idea from the semi-deterministic MAC-IC-MAC that the genie information
$T_{i}$ is chosen to be i.i.d with $S_{i}$ conditioned on $X_{i}$,
we let $T_{i}=h_{i.1\rightarrow i^{'}}X_{i}+Z_{i^{'}}$, where
$Z_{i^{'}}\perp Z_{i}$. Then we show that Gaussian input distribution with full power for all inputs gives a single region that is also an outer bound (i.e., it equals the union of regions outer bound of Theorem \ref{thm:mim-sd-outer}, when applied to the Gaussian channel).

We define the outer bound first.
\begin{defn}
\label{def:mim-gs-outer}
Define $\mathcal{R}_{{\rm o}}^{{\rm G}}\triangleq\mathcal{R}_{\mathrm{c}}(\overline{A},\overline{B},\overline{E},\overline{G})$ where
the set functions $\overline{A}$, $\overline{B}$,
$\overline{E}$ and $\overline{G}$ are given as
\begin{align}
\overline{A}_{\varUpsilon_{i}} & \triangleq\mathsf{C}\left(\sum_{i.j\in\varUpsilon_{i}\backslash\{i.1\}}\mathsf{SNR}_{i.j\rightarrow i}+\frac{\mathsf{SNR}_{i.1\rightarrow i}}{1+\mathsf{INR}_{i.1\rightarrow i^{'}}}\right)\label{eq:mim-gs-outer-a}\\
\overline{B}_{\varOmega_{i}} & \triangleq\mathsf{C}\left(\sum_{i.j\in\varOmega_{i}}\mathsf{SNR}_{i.j\rightarrow i}\right)\label{eq:mim-gs-outer-b}\\
\overline{E}_{\varUpsilon_{i}} & \triangleq\mathsf{C}\bigg(\sum_{i.j\in\varUpsilon_{i}\backslash\{i.1\}}\mathsf{SNR}_{i.j\rightarrow i}+\frac{\mathsf{SNR}_{i.1\rightarrow i}}{1+\mathsf{INR}_{i.1\rightarrow i^{'}}}\nonumber \\
 & \qquad+\mathsf{INR}_{i^{'}.1\rightarrow i}\bigg)\label{eq:mim-gs-outer-e}\\
\overline{G}_{\varOmega_{i}} & \triangleq\mathsf{C}\left(\sum_{i.j\in\varOmega_{i}}\mathsf{SNR}_{i.j\rightarrow i}+\mathsf{INR}_{i^{'}.1\rightarrow i}\right).\label{eq:mim-gs-outer-g}
\end{align}
 
\end{defn}
\begin{thm}
\label{thm:.mim-gs-outer}For the Gaussian MAC-IC-MAC, $\mathcal{C}^{{\rm G}}\subseteq\mathcal{R}_{{\rm o}}^{{\rm G}}$. 
\end{thm}
\begin{IEEEproof}
Please see Appendix \ref{sec:mim-gs-outer-proof}. 
\end{IEEEproof}

\subsubsection{Two-bit gap}
\label{sec:twobitgapg}
Lastly, we show that the gap between $\mathcal{R}_{{\rm in}}^{{\rm G}}$
and $\mathcal{R}_{{\rm o}}^{{\rm G}}$ is within two bits, which implies
the inner bound of Theorem \ref{thm:mim-gs-inner} (and hence its associated universal coding scheme) is within a two-bit gap of the capacity region. To this end, we first show in Lemma \ref{lem: mim-gs-gap-2} that the gap between $\mathcal{R}_{\mathrm{c}}^{\mathrm{G}}$
and $\mathcal{R}_{\mathrm{o}}^{\mathrm{G}}$ is not more than one bit.
\begin{lem}
\label{lem: mim-gs-gap-2}There is no more than a one-bit gap between
$\mathcal{R}_{{\rm c}}^{{\rm G}}$ and $\mathcal{R}_{{\rm o}}^{{\rm G}}$. 
\end{lem}
\begin{IEEEproof}
It is enough to show the gap between each corresponding pair of sets of weighted sum rate bounds of $\mathcal{R}_{\mathrm{c}}^{\mathrm{G}}$
and $\mathcal{R}_{\mathrm{o}}^{\mathrm{G}}$ are within one bit of each other.
We bound, for instance, the gap between $E_{\varUpsilon_{i}}$ and $\overline{E}_{\varUpsilon_{i}}$ as
\begin{align*}
E_{\varUpsilon_{i}} & =\log\bigg(2+\mu_{i}\mathsf{SNR}_{i.1\rightarrow i}+\sum_{i.j\in\varUpsilon_{i}\backslash\{i.1\}}\mathsf{SNR}_{i.j\rightarrow i}\\
 & \qquad+(1-\mu_{i^{'}})\mathsf{INR}_{i^{'}.1\rightarrow i}\bigg)-\log\left(1+\mu_{i^{'}}\mathsf{INR}_{i^{'}.1\rightarrow i}\right)\\
 & \geq\log\bigg(2+\mu_{i}\mathsf{SNR}_{i.1\rightarrow i}+\sum_{i.j\in\varUpsilon_{i}\backslash\{i.1\}}\mathsf{SNR}_{i.j\rightarrow i}\\
 & \qquad+(1-\mu_{i^{'}})\mathsf{INR}_{i^{'}.1\rightarrow i}\bigg)-1\\
 & =\log\bigg(2+\min\left\{ 1,\frac{1}{\mathsf{INR}_{i.1\rightarrow i^{'}}}\right\} \mathsf{SNR}_{i.1\rightarrow i}+\mathsf{INR}_{i^{'}.1\rightarrow i}\\
 & \qquad+\sum_{i.j\in\varUpsilon_{i}\backslash\{i.1\}}\mathsf{SNR}_{i.j\rightarrow i}-\min\left\{ \mathsf{INR}_{i^{'}.1\rightarrow i},1\right\} \bigg)-1\\
 & \geq\mathsf{C}\bigg(\sum_{i.j\in\varUpsilon_{i}\backslash\{i.1\}}\mathsf{SNR}_{i.j\rightarrow i}+\frac{\mathsf{SNR}_{i.1\rightarrow i}}{1+\mathsf{INR}_{i.1\rightarrow i^{'}}}\\
 & \qquad+\mathsf{INR}_{i^{'}.1\rightarrow i}\bigg)-1\\
 & =\overline{E}_{\varUpsilon_{i}}-1.
\end{align*}
Other gaps can be verified similarly, which proves the lemma. 
\end{IEEEproof}
This leads us to the main result of this section.
\begin{thm}
\label{thm:mim-gs-gap} The universal coding scheme for the Gaussian MAC-IC-MAC of Theorem \ref{thm:mim-gs-inner} achieves a rate region that is within a two-bit gap of its capacity region.
\end{thm}
\begin{IEEEproof}
To prove the result, it is sufficient to show that there is no more than a two-bit gap between $\mathcal{R}_{{\rm in}}^{{\rm G}}$ and $\mathcal{R}_{{\rm o}}^{{\rm G}}$. 
The gap between $\mathcal{R}_{{\rm in}}^{{\rm G}}$ and $\mathcal{R}_{{\rm o}}^{{\rm G}}$ should not exceed the sum of gaps from $\mathcal{R}_{\mathrm{in}}^{\mathrm{G}}$
to $\mathcal{R}_{\mathrm{c}}^{\mathrm{G}}$ and from $\mathcal{R}_{{\rm c}}^{{\rm G}}$
to $\mathcal{R}_{{\rm o}}^{{\rm G}}$, both of which are equal to 1 bit by Lemmas \ref{lem:mim-gs-gap-1} and \ref{lem: mim-gs-gap-2}, respectively. Hence the result.
\end{IEEEproof}

\subsection{The GDoF Region \label{subsec:mim-gdof}}

The GDoF region
characterizes the simultaneously accessible signal-level
dimensions (per channel use) by the users of a network in the limit
of high SNR, while the ratios of the SNRs and INRs relative to a reference
SNR, each expressed in the dB scale, are held constant, with each
constant taken, in the most general case, to be arbitrary. 

Given the Gaussian MAC-IC-MAC model as defined by \eqref{eq:mim-gs-mdoel-1}
and \eqref{eq:mim-gs-model-2}, let $\rho$ be a nominal value for
SNR or INR, define $\bar{\alpha}=(\alpha_{1.1\rightarrow1},\cdots,\alpha_{1.K_{1}\rightarrow1},\alpha_{2.1\rightarrow2},\cdots,\alpha_{2.K_{2}\rightarrow2},\alpha_{1.1\rightarrow2},\alpha_{2.1\rightarrow1})$
with
{\small{}{} 
\begin{align*}
\alpha_{i.j\rightarrow i}=\frac{\log(\mathsf{SNR}_{i.j\rightarrow i})}{\log\rho} \; \; {\rm and} & \; \; \alpha_{i.1\rightarrow i^{'}}=\frac{\log(\mathsf{INR}_{i.1\rightarrow i^{'}})}{\log\rho}
\end{align*}
}{\small \par}

\begin{defn}
\label{def:mim-gdof-def}The GDoF region of a\textit{ $(K_{1},K_{2})$ }Gaussian MAC-IC-MAC
$\mathcal{D}(K_{1},K_{2},\bar{\alpha})\in R_{+}^{K_{1}+K_{2}}$ 
is defined as 
\begin{align}
\left\{ (d_{\varTheta_{1}},d_{\varTheta_{2}}):\right. & d_{i.j}=\lim_{\rho\rightarrow\infty}\frac{R_{i.j}}{\log\rho},i\in\{1,2\}, \; j\in\{1,\cdots,K_{i}\}, 
\nonumber \\
 & \left.\text{and\thinspace}(R_{\varTheta_{1}},R_{\varTheta_{2}})\in\mathcal{C}^{{\rm G}}(K_{1},K_{2},\bar{\alpha})\right\} \label{eq:mim-gdof-def}
\end{align}
and where $\mathcal{C}^{{\rm G}}(K_{1},K_{2},\bar{\alpha})$ denotes its capacity region.
\end{defn}
Since we have the capacity region to within a two-bit gap for the
Gaussian MAC-IC-MAC, computing the GDoF region can be easily done by substituting
$\mathsf{SNR}_{i.j\rightarrow i}=\rho^{\alpha_{i.j\rightarrow i}}$
and $\mathsf{INR}_{i^{'}.j\rightarrow i}=\rho^{\alpha_{i^{'}.j\rightarrow i}}$
in \eqref{eq:mim-gs-outer-a}-\eqref{eq:mim-gs-outer-g}, and computing
the limits of each set function under $\rho\rightarrow\infty$. 
\begin{defn}
\label{def:mim-gdof}For any two real numbers $x$ and $y$, let $(x-y)^{+}\triangleq\max\{0,x-y\}$.
Define the set functions $a$, $b$, $e$ and $g$ as 
\begin{align}
a_{\varUpsilon_{i}} & =\max\left\{ \max_{i.j\in\Upsilon_{i}\backslash\{i.1\}}\alpha_{i.j\rightarrow i},(\alpha_{i.1\rightarrow i}-\alpha_{i.1\rightarrow i^{'}})^{+}\right\} \label{eq:mim-gdof-a}\\
b_{\varOmega_{i}} & =\max_{i.j\in\varOmega_{i}}\alpha_{i.j\rightarrow i}\label{eq:mim-gdof-b}\\
e_{\varUpsilon_{i}} & =\max\bigg\{\max_{i.j\in\Upsilon_{i}\backslash\{i.1\}}\alpha_{i.j\rightarrow i},(\alpha_{i.1\rightarrow i}-\alpha_{i.1\rightarrow i^{'}})^{+}, \nonumber \\
 & \qquad \qquad \qquad\alpha_{i^{'}.1\rightarrow i}\bigg\}\label{eq:mim-gdof-e}\\
g_{\varOmega_{i}} & =\max\left\{ \max_{i.j\in\varOmega_{i}}\alpha_{i.j\rightarrow i},\alpha_{i^{'}.1\rightarrow i}\right\} .\label{eq:mim-gdof-g}
\end{align}
\end{defn}
\begin{thm}
\label{thm:mim-gdof}The GDoF region $\mathcal{D}(K_{1},K_{2},\bar{\alpha})$ of the $(K_{1},K_{2})$ Gaussian MAC-IC-MAC is equal to
\begin{align}
\mathcal{D}_{\mathrm{c}}(a,b,e,g)=\bigg\{(d_{\varTheta_{1}},d_{\varTheta_{2}}) & \in\mathbb{R}_{+}^{K_{1}+K_{2}}:\nonumber \\
 & \forall(\varUpsilon_{1},\varOmega_{1},\varUpsilon_{2},\varOmega_{2})\in\Xi\nonumber \\
\sum_{1.j\in\varOmega_{1}}d_{1.j} & \leq b_{\varOmega_{1}}\label{eq:mim-generic-cgdof-1}\\
\sum_{2.j\in\varOmega_{2}}d_{2.j} & \leq b_{\varOmega_{2}}\label{eq:mim-generic-cgdof-2}\\
\sum_{1.j\in\varUpsilon_{1}}d_{1.j}+\sum_{2.j\in\varOmega_{2}}d_{2.j} & \leq a_{\varUpsilon_{1}}+g_{\varOmega_{2}}\label{eq:mim-generic-cgdof-3}\\
\sum_{1.j\in\varOmega_{1}}d_{1.j}+\sum_{2.j\in\varUpsilon_{2}}d_{2.j} & \leq\ g_{\varOmega_{1}}+ a_{\varUpsilon_{2}}\label{eq:mim-generic-cgdof-4}\\
\sum_{1.j\in\varUpsilon_{1}}d_{1.j}+\sum_{2.j\in\varUpsilon_{2}}d_{2.j} & \leq e_{\varUpsilon_{1}}+ e_{\varUpsilon_{2}}\label{eq:mim-generic-cgdof-5}\\
\sum_{1.j\in\varUpsilon_{1}}d_{1.j}+\sum_{1.j\in\varOmega_{1}}d_{1.j}\qquad\nonumber \\
+\sum_{2.j\in\varUpsilon_{2}}d_{2.j} & \leq a_{\varUpsilon_{1}}+g_{\varOmega_{1}}+e_{\varUpsilon_{2}}\label{eq:mim-generic-cgdof-6}\\
\sum_{1.j\in\varUpsilon_{1}}d_{1.j}+\sum_{2.j\in\varUpsilon_{2}}d_{2.j}\qquad\nonumber \\
+\sum_{2.j\in\varOmega_{2}}d_{2.j} & \leq\ e_{\varUpsilon_{1}}+a_{\varUpsilon_{2}}+g_{\varOmega_{2}}\bigg\}.\label{eq:mim-generic-cgdof-7}
\end{align}
\end{thm}

\begin{rem}
The GDoF of the Gaussian MAC-IC-MAC applies also to a fully connected
$(K_{1}+K_{2})$-transmitter-2-receiver interference network (referred to as interfering multiple-access channel or IMAC, in which there are two MACs but all transmitters in one MAC cause interference to the receiver of the other MAC), when the interference from all but one transmitter in each cell is sufficiently weak to be at the noise level, or mathematically, when the corresponding INR exponents of these links are equal to zero. 
\end{rem}

\begin{rem}
\label{imac-discuss-gdof}
The GDoF of the Gaussian MAC-IC-MAC can be used to compute an outer bound for the IMAC. This outer bound would be an intersection of the GDoF regions of the $K_1 K_2$ embedded MAC-IC-MACs within the IMAC obtained by removing all but two interference links (since removing interference links doesn't decrease capacity). It would be interesting to describe the set of SNR and INR exponents $\bar{\alpha}$ for which simple achievable schemes achieve that outer bound. For instance, one could use the achievable scheme of this work over the IMAC (i.e., treat the signal from all but one interfering transmitter as noise at each receiver) for each of the $K_1 K_2$ embedded MAC-IC-MACs within the IMAC. The union of the corresponding GDoF regions gives an inner bound to the GDoF region of the fully connected IMAC. For what set of $\bar{\alpha}$ do the above outer and inner bounds coincide, and hence yield the GDoF region of the IMAC?
\end{rem}

In order to get some intuitive understanding of the general GDoF region of Theorem \ref{thm:mim-gdof}, we consider the symmetric case.  A Gaussian MAC-IC-MAC is said to be symmetric if $K_{1}=K_{2}=K$ and $\mathsf{SNR}_{i.j\rightarrow i}=\mathsf{SNR}$
and $\mathsf{INR}_{i.1\rightarrow i^{'}}=\mathsf{INR}$, i.e. all
SNRs are identical and all INRs are identical. Note this does not
mean the channel attenuations $h_{i.j\rightarrow i}$s or $h_{i^{'}.1\rightarrow i}$s
are identical, since transmitters may choose different transmission
power. SNR and INR terms can thus be normalized as 
\[
\mathsf{INR}=\mathsf{SNR}^{\alpha}=\rho^{\alpha}
\]
which implies $\alpha_{i.j\rightarrow i}=1$ and $\alpha_{i.1\rightarrow i^{'}}=\alpha$. In this symmetric case, we write the general GDoF region $\mathcal{D}(K_{1},K_{2},\bar{\alpha})$ simply as $ \mathcal{D}(K, \alpha) $ since $K_1=K_2=K$ and since the dependence on $ \bar{\alpha} $ is only through $\alpha $.

Furthermore, for the symmetric Gaussian MAC-IC-MAC, a further simplified yet instructive metric,
the \textit{symmetric GDoF}, denoted as $ d_{{\rm sym}}(K,\alpha) $ is defined as follows. 
\begin{defn}
\label{def:mim-sym-gdof}For a symmetric $K$-user (per cell) Gaussian
MAC-IC-MAC with GDoF region $\mathcal{D} (K,\alpha)$,
the \textit{symmetric generalized degrees-of-freedom} $d_{{\rm sym}}(K,\alpha)$
is defined as the solution to the following equation 
\[
d_{{\rm sym}}(K,\alpha)\triangleq\max_{\substack{d=d_{1.1}=\cdots=d_{1.K_{1}}=d_{2.1}=\cdots=d_{2.K_{2}}\\
(d_{\varTheta_{1}}, d_{\varTheta_{2}})\in\mathcal{D} (K,\alpha)
}
}d.
\]
The symmetric GDoF of Gaussian MAC-IC-MAC can be easily obtained by specializing the GDoF region given in Theorem \ref{thm:mim-gdof}.
When $K=1$ (the MAC-IC-MAC is the 2-user IC), Theorem \ref{thm:mim-gdof} recovers the well-known "W"-shaped symmetric GDoF curve established by Etkin {\em et al} in \cite{etkin2008gaussian}, as it must; for $K\geq2$, the symmetric GDoF
is stated next. 
\end{defn}
\begin{cor}
\label{cor:mim-sym-gdof} The symmetric GDoF
$d_{{\rm sym}}(K,\alpha)$ for $K\geq2$ of the symmetric $K$-user Gaussian MAC-IC-MAC is given as
 
\[
d_{{\rm sym}}(K,\alpha)=\begin{cases}
1 & 0\leq\alpha<1-\frac{1}{K}\\
-\frac{1}{K+1}\alpha+\frac{2}{K+1} & 1-\frac{1}{K}\leq\alpha<1\\
\frac{1}{K+1}\alpha & 1\leq\alpha<1+\frac{1}{K}\\
1 & \alpha\geq1+\frac{1}{K}
\end{cases}.
\]
\end{cor}
Remarkably, the symmetric GDoF curve for the $K\geq 2$ case for any $K$ is a simpler, raised "V"-shaped curve with flat shoulders on both sides of the curve (as shown in Fig.\,\ref{fig:mim-sym-gdof}), illustrating the benefit of adding
non-interfering transmitters to the cell. Indeed, maximum symmetric GDoF (which is $\frac{1}{K}$) can be achieved for $\alpha\in[0,1- \frac{1}{K}]\cup[1+ \frac{1}{K},\infty)$ compared to the "W" shape obtained in \cite{etkin2008gaussian} for $K=1$.

\begin{figure}[tbh]
\begin{centering}
\includegraphics[width=3.4in]{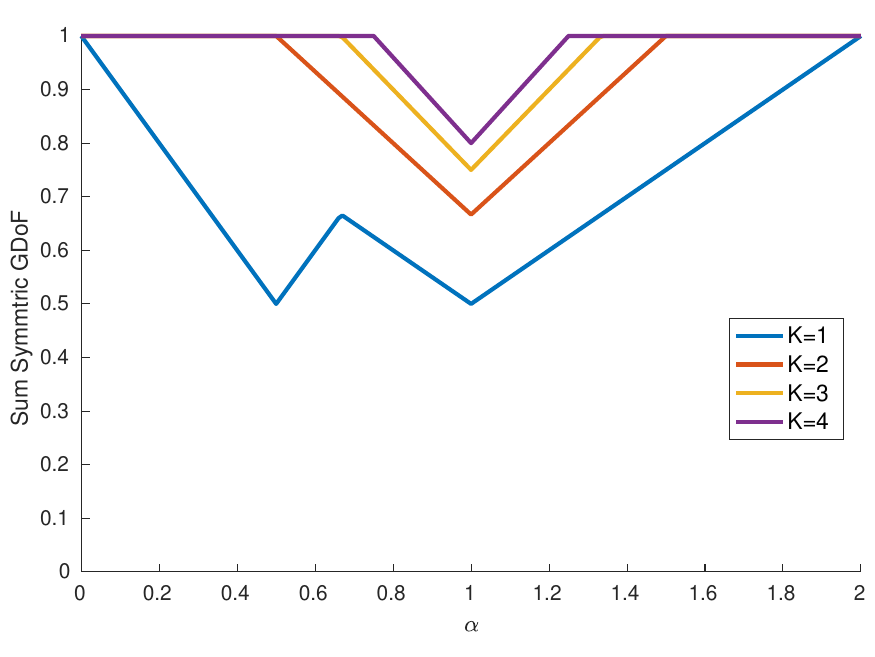} 
\par\end{centering}
\caption{$Kd_{{\rm sym}}(K,\alpha)$, the sum symmetric GDoF per cell versus $\alpha$ for
$K=1,2,3,4$\label{fig:mim-sym-gdof-num}}
\end{figure}

To see the overall improvement with increasing $K$ we plot the numerical
results of the {\em sum-symmetric} GDoF, i.e. $Kd_{{\rm sym}}(K,\alpha)$ for $K= 1, 2, 3, 4$ in Fig.\,\ref{fig:mim-sym-gdof-num}. When $K=1,$ the sum-symmetric GDoF is just the symmetric GDoF and we see the familiar "W" curve of the symmetric
GDoF curve for two-user IC of \cite{etkin2008gaussian}. As $K$ increases, the sum-symmetric GDoF increases for all $\alpha$, while the width of the
flat shoulder also increases, leading to maximum GDoF of 1 for an increasing range of $\alpha$. Thus, Fig.\,\ref{fig:mim-sym-gdof-num} gives the clearest indication of the benefit of adding interference-free transmitters to the entire cell GDoF, which is a measure of cell spectrum efficiency when all users are provided the same GDoF.  

\begin{rem}
Note that the sum symmetric DoF (i.e., GDoF at $\alpha =1 $) is $\frac{K}{K+1}$, approaching 1 as $K$ becomes large. This is also the worst case sum symmetric GDoF over $\alpha$. This is reminiscent of the sum-symmetric DoF of the two-cell fully-connected IMAC obtained in \cite{suh2008interference} but that result applies to a multi-carrier setting.
\end{rem}


Returning to the symmetric MAC-IC-MAC, we see that adding even just one non-interfering transmitter to the two-user IC improves GDoF significantly. This can be easily observed at $\alpha=\frac{1}{2}$
in Fig.\,\ref{fig:mim-sym-gdof-num}. Because of interference, a
two-user IC only allows each user to achieve $\frac{1}{2}$ DoF, while
a $(2,2)$ MAC-IC-MAC could achieve full GDoF while providing $\frac{1}{2}$
DoF to each of the four transmitters, thus doubling sum  GDoF.

\subsection{Signal-Level Partitioning}
\label{subsec:signal-part}

We explain the GDoF benefit of having interference-free transmitters in the MAC-IC-MAC by formalizing the development of signal level partitioning in the context of the MIMO interference channel by Karmakar and Varanasi in \cite{karmakar2012generalized}. That work also includes signal space partitioning which is not needed here.
We start by describing signal level partitioning in a Gaussian point-to-point link first followed by the two-user Gaussian MAC and then the MAC-IC-MAC. In the latter case, we restrict the discussion to symmetric GDoF, with the extension to asymmetric settings being straightforward, in principle.

\begin{figure}[tbh]
\begin{centering}
\includegraphics[width=3.4in]{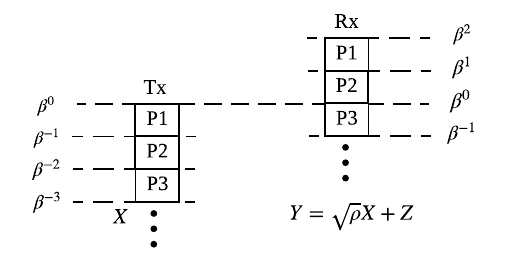}
\par\end{centering}
\caption{Signal level partitioning at the transmitter and receiver of a Gaussian link $Y = \sqrt{\rho} X + Z $ with $E[|X|^2] \leq 1 $ and zero-mean Gaussian noise $Z$ with unit variance. There are $q \alpha = q $ partitions at the transmitter but only $p \alpha = p = 2$ of them are seen above the noise level at the receiver.
\label{fig:mim-dtn-p2p}}
\end{figure}

Consider a single interference-free Gaussian link $Y=hX+Z$ in a network with unit transmit power constraint and unit noise variance with coefficient $|h|^2=\rho^\alpha$, with $\rho $ denoting a nominal or reference $ \mathsf{SNR}$ which can be regarded as say the $ \mathsf{SNR}$ of a reference link when we generalize the discussion to a network. Let $\alpha$ be a rational number.
Define $\beta$ so that $\beta^p = \rho$ for some positive rational number $p>1$ such that $n= p \alpha \in \mathbb{N}$.  The transmit signal $X$ can be decomposed into $m$ signal-level partitions (in the dB scale) such that $m=q \alpha \in \mathbb{N}$, with $q \geq p $ being another rational number. Assign the power $\beta^{-i} - \beta^{-i-1}$ to the $i^{th}$ signal level
for $i \in [0:m-1]$. This is shown in the left hand side of Fig. \ref{fig:mim-dtn-p2p} where the top of each partition is aligned with its power level, denoted by the dashed lines. For instance, Partition 1 (P1 in the figure) has power $1-\beta^{-1}$, partition 2 has power $\beta^{-1}-\beta^{-2}$, etc. Since for large $\mathsf{SNR}$, we have $\beta^{-i} - \beta^{-i-1} \approx \beta^{-i}$ we denote the top of $i^{th}$ partition simply as $\beta^{-i}$ in Fig. \ref{fig:mim-dtn-p2p}. Independent information is encoded in each signal partition and the encoded signals are combined using additive superposition. At the receiver the power in the $m$ signal levels are amplified by the factor $\rho^\alpha = \beta^{n}$. Hence, the top $n $ partitions are above the noise level (i.e., they are "heard" by the receiver). 
Successive decoding is used at the receiver from the top level to the last one that is above the noise level, i.e., information in each partition is decoded by treating all the signals in the lower partitions as noise. It is easy to verify that this way one obtains $\frac{1}{p}$ DoF for each of the $n $ signal levels, since for any $i \in [0,n -1]$
\begin{align*}
\lim_{\rho\rightarrow\infty}\frac{\log\big(1+\frac{\beta^{-i}-\beta^{-i-1}}{\beta^{-i-1}-\beta^{-m} }\big)}{{\log\rho}} = \frac{1}{p}.
\end{align*}
Hence the overall DoF of $\alpha $ is achieved. 

In Fig. \ref{fig:mim-dtn-p2p}, we consider $\alpha = 1$ in the point-to-point Gaussian link so that $|h|^2= \rho $ with the received $ \mathsf{SNR}$ being $\rho $ itself. Let $p=2$ and any $q \geq p$. The receiver's perspective is shown on the right hand side of the figure where it is seen that all signal levels are lifted by $n =2 $ levels, with the rest being below the noise level. The information in the two levels can be successively decoded to achieve 1/2 DoF per level since $\beta = \sqrt{\rho}$. 

Note that in the simple point-to-point link there is no need to send information over two signals levels or consider values of $\alpha $ other than 1. However, when that link is considered as one out of several links in a multi-user channel that has some reference link with nominal SNR that is $\rho $ and with the other links having disparate strengths of the form $\rho^\alpha $ (with $\alpha $ in general depending on the link being considered) the above model is much more useful. 

\begin{figure}[tbh]
\begin{centering}
\includegraphics[width=3.4in]{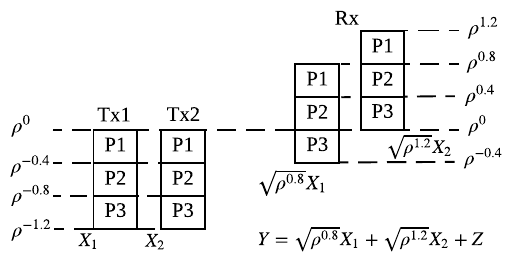}
\par\end{centering}
\caption{Signal level partitioning at the transmitters and receiver in a Gaussian MAC\label{fig:mim-dtn-mac}}
\end{figure}

To analyze a multi-terminal network with disparate $\mathsf{SNR}$ and $\mathsf{INR}$ exponents (relative to nominal $\rho $), we extend the aforementioned exponent partitioning with a sufficiently fine resolution (i.e, with $p$ chosen sufficiently large) so that the signal partitions from different transmitters that arrive at a receiver will be {\em aligned} at that receiver, for all receivers. Consider for instance the simple example of the Gaussian MAC $Y=h_{1}X_{1}+h_{2}X_{2}+Z$ where $|h_1|^2=\rho^{{0.8}}$ and $|h_2|^2=\rho^{{1.2}}$. Here, we choose $p= \frac{5}{2}$ so that $\beta = \rho^{0.4}$, i.e., the exponent resolution of the signal level is set at 0.4. The resulting signals levels at the two transmitters and at the receiver are illustrated in  Fig.\,\ref{fig:mim-dtn-mac} on its left and right sides, respectively. Note that all the partitions of the two transmitted signals are aligned at the receiver's grid without aliasing.
In particular, the top two signal levels (or partitions) of $X_{1}$ and the top three partitions of $X_{2}$ can be ``heard" by the receiver (i.e., they are above the noise level), and the overall observable partitions is thus three. It can be easily inferred that the following GDoF region is achievable for this MAC:
\begin{align*}
0 \leq d_{1} &\leq 0.8\\
0 \leq d_{2} &\leq 1.2\\
d_{1}+d_{2} &\leq 1.2.
\end{align*}
This in fact is also exactly the GDoF region of the Gaussian MAC under consideration. 

\subsection{GDoF of the MAC-IC-MAC Explained}

Let us apply the above reasoning to the setting of the MAC-IC-MAC next. Consider the achievability of the three key inflection points
in the per-cell sum symmetric GDoF curve, namely, $(1-\frac{1}{K},1)$, 
$(1,\frac{K}{K+1})$ and $(1+\frac{1}{K},1)$. 

\begin{figure}[tbh]
\centering{}\includegraphics{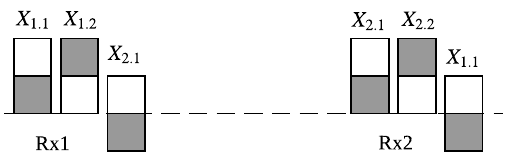}\caption{Achieving sum symmetric GDoF 1 per-cell when $\alpha= \frac{1}{2}$ for $(2,2)$ Gaussian MAC-IC-MAC.\label{fig:mim-dtn-1}}
\end{figure}

\begin{figure}[tbh]
\centering{}\includegraphics{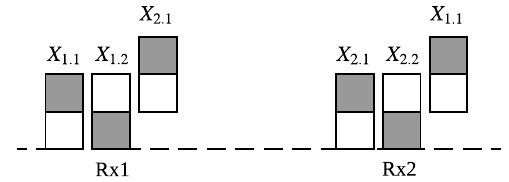}\caption{Achieving sum symmetric GDoF 1 per-cell when $\alpha=\frac{3}{2}$ for $(2,2)$ Gaussian MAC-IC-MAC.\label{fig:mim-dtn-2}}
\end{figure}

\begin{figure}[tbh]
\centering{}\includegraphics{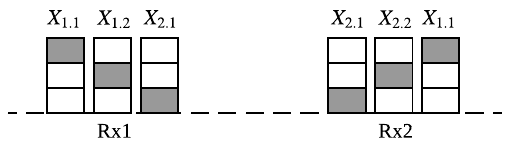}\caption{Achieving sum symmetric GDoF $\frac{2}{3}$ per-cell when $\alpha=1$ for $(2,2)$ Gaussian MAC-IC-MAC.\label{fig:mim-dtn-3}}
\end{figure}

In what follows, we pick $K=2$ as an example throughout. Assume that the transmitted power at each transmitter is divided into multiple partitions such that every signal partition carries the same DoF. Fig.\,\ref{fig:mim-dtn-1}-\ref{fig:mim-dtn-3}, on their left and right sides respectively, show the aligned received signals from Receiver 1 and Receiver 2's perspectives, one figure for each of the three key values of $\alpha$. The dashed  horizontal line represents the noise floor. The symbols $X_{i.j}$ on top of each column at each receiver denotes that the partitions below it correspond to the partitions of Transmitter $i.j$ as seen at that receiver. The shadowed signal partitions are the ones used to transmit information whereas the white partitions are unused.

When $\alpha=\frac{1}{2}$, a transmitted signal is partitioned into two levels (so that $p=q=4$ for the cross-links) so each can carry 1/2 DoF. As shown in Fig. \ref{fig:mim-dtn-1}, the lower of the two levels of the interference will be under the noise floor. Hence, it is best to let both interfering transmitters Tx$i.1$ use the lower power
level, to obtain $\frac{1}{2}$ DoF per transmitter. This leaves the empty upper power level to the non-interfering transmitters Tx$i.2$, each of which thus achieves $\frac{1}{2}$ DoF as well. Without the non-interfering transmitters Tx$i.2$, the upper partition of the received signal at the receivers remains unused in this scheme wherein interference is treated as noise. But it is known \cite{etkin2008gaussian} that the symmetric GDoF is 1/2 in the 2-user IC when $\alpha=\frac{1}{2}$, and that that symmetric GDoF is achieved by treating interference as noise at each receiver. Hence, the above scheme without the non-interfering transmitters Tx$i.2$ cannot be improved upon in the 2-user IC. This example explains how sum symmetric GDoF per cell is {\em improved} from 1/2 to 1 by the presence of the non-interfering transmitters  in the MAC-IC-MAC. 

Consider next the case of strong interference with $\alpha=\frac{3}{2}$ in which it is also possible to achieve sum symmetric GDoF of 1 per cell. This case is illustrated in Fig.\,\ref{fig:mim-dtn-2}. Here, the non-interfering transmitters use the lower (of two) signal levels and the interfering transmitters use the higher level for sending information. At each receiver, there are three signal levels with the highest level containing interference which is decoded first, followed by the signal in the second level and then the third so that both desired signals are decoded by their intended receiver. In this scheme, interference is fully decoded at each receiver.

When $\alpha=1$, a sum symmetric GDoF per cell of $\frac{2}{3}$ is achievable. Each transmitted signal is partitioned into three levels so that each can carry $\frac{1}{3}$ DoF. The signal partitions at each of the receivers is shown in Fig.\,\ref{fig:mim-dtn-3}. Rx$i$ can successively decode the information from Tx$i.1$,
Tx$i.2$ and Tx$i^{'}.1$ (strongest level first, weakest last). This implies that $\frac{1}{3}$ DoF for each of the four transmitters is hence achievable. Note that the interference is fully decoded at Receiver 2 and it can be treated as noise at Receiver 1. Also, in this case, the power allocation for common and private information is asymmetric between the two transmitters.

\begin{figure}[tbh]
\centering{}\includegraphics{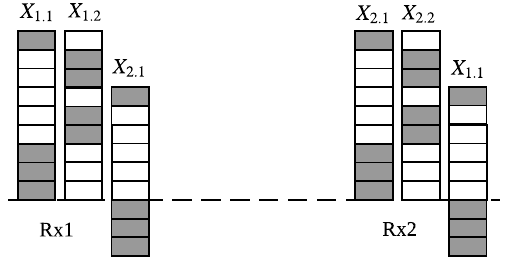}\caption{Achieving sum symmetric GDoF $\frac{8}{9}$ per-cell when $\alpha=\frac{2}{3}$ for $(2,2)$ Gaussian MAC-IC-MAC.\label{fig:mim-dtn-4}}
\end{figure}

Note that the achievable schemes for the three key inflection points did not require strictly partial interference decoding in that interference is either treated as noise or decoded completely at each of the two receivers. In the next example, we study the case when $\alpha = \frac{2}{3}$. In this case, according to Corollary \ref{cor:mim-sym-gdof}, each transmitter can transmit at DoF $\frac{4}{9}$. Fig. \ref{fig:mim-dtn-4} demonstrates the power allocation scheme of each transmitted signal from the two receivers' perspectives. In this case, the interfering transmitters use the highest signal level (out of 9) to send common information that will be decoded uniquely at both receivers (it arrives at level 4). Private information at each interfering transmitter is encoded at the lowest three levels so it can be treated as noise at the other (non-intended) receiver. This leaves 4 signal levels unoccupied at each receiver if you just consider the interfering transmitters. These four levels are opportunistically filled by the non-interfering transmitter in each cell, so that each transmitter can achieve $\frac{4}{9}$ DoF for a per-cell sum symmetric DOF of $\frac{8}{9}$.

Next, we consider achievability using time-sharing between the respective GDoF-optimal strategies in the embedded 2-user IC and 2 non-interfering point-to-point links (recall $K=2$) in the MAC-IC-MAC. Hence, Tx$1.1$ and Tx2.1 transmit (in the embedded 2-user IC) while Tx1.2 and Tx2.2 are turned off for some fraction of time and in the other strategy Tx1.2 and Tx2.2 transmit while Tx$1.1$ and Tx2.1 are turned off during the remaining fraction of time. 

\begin{figure}[tbh]
\begin{centering}
\includegraphics[width=3.4in]{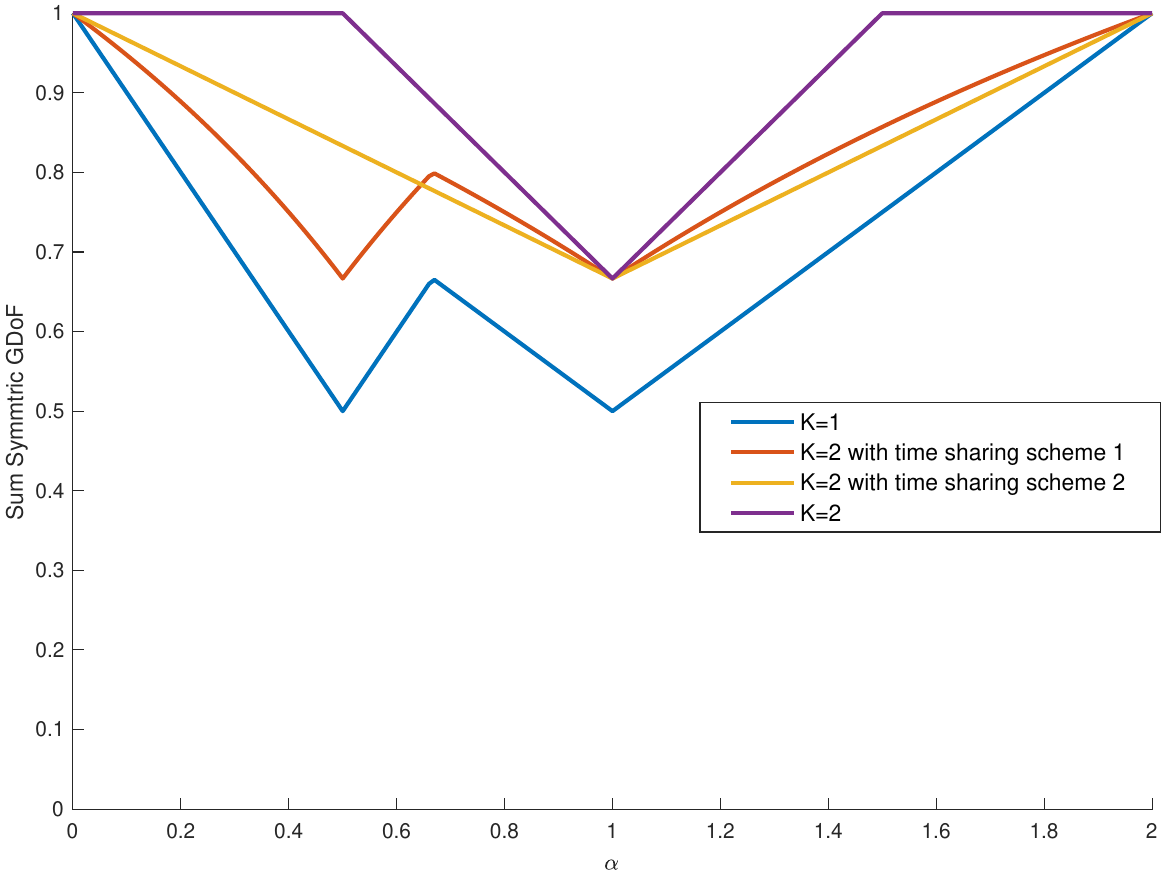} 
\par\end{centering}
\caption{An illustration of the sub-optimality of two time-sharing schemes for the (2,2) MAC-IC-MAC. In scheme 1, time-sharing occurs between the two interfering transmitters and the two non-interfering transmitters. In scheme 2, time-sharing occurs between the two embedded MAC-Z-P2Ps. Both these time-sharing schemes are sub-optimal. The fundamental sum-symmetric GDoF per cell of (2,2) MAC-IC-MAC as well as the symmetric GDoF of the two-user IC are also depicted for comparison.} \label{fig:mim-time-sharing-gdof}
\end{figure}

Let $d_{\rm sym}(1,\alpha)$ be the symmetric DoF of the two-user IC
at $\mathsf{SNR}=\rho$ and $\mathsf{INR}=\rho^{\alpha}$. To share
the DoF evenly, transmitters Tx$i.1$ have to use $\frac{1}{d_{{\rm sym}}(1,\alpha)+1}$ of time and leave the rest $\frac{d_{{\rm sym}}(1,\alpha)}{d_{{\rm sym}}(1,\alpha)+1}$ portion to transmitters Tx$i.2$. This way we get per-cell sum symmetric GDoF of $\frac{2d_{{\rm sym}}(1,\alpha)}{d_{{\rm sym}}(1,\alpha)+1}$ which is plotted in Fig.\,\ref{fig:mim-time-sharing-gdof} as the red curve (with the legend ``K=2 with time-sharing scheme 1") in comparison to the fundamental sum symmetric GDoFs in the 2-user IC and the (2,2) MAC-IC-MAC. Clearly, the time-sharing scheme is sub-optimal, except when $\alpha= 0, 1, 2$. The reason for this can be observed from Fig. \ref{fig:mim-dtn-1}. Say when $\alpha=\frac{1}{2}$, the role of Tx$i.2$ in the optimal scheme is to fill the unused signal levels left by transmitters Tx$i.1$ and Tx$i^{'}.1$ instead of time sharing DoF with them.

Time-sharing can be performed in a different way. Continuing with the (2,2) MAC-IC-MAC as an example, let Tx$1.1$, Tx$1.2$ and Tx$2.2$ transmit for the first half-time, and let Tx$1.2$, Tx$2.1$ and Tx$2.2$ transmit in the other half-time. Due to symmetry, the GDoF Tx1.2 achieves in the first half-time will be exactly the GDoF Tx2.2 achieves in the second half-time. Therefore, to get the same GDoF for all transmitters, we should let $d_{1.1} = d_{1.2}+d_{2.2}$ and $d_{2.1}=0$ in the first half-time, then $d_{2.1} = d_{1.2}+d_{2.2}$ and $d_{1.1}=0$ in the second half-time.
Following the fundamental GDoF region given in Theorem \ref{thm:mim-gdof} and performing linear programming to compute the maximum sum symmetric GDoF, the resulting GDoF curve is plotted in yellow in Fig.\,\ref{fig:mim-time-sharing-gdof} (with the legend ``K=2 with time-sharing scheme 2"). This time-sharing scheme yields a V-shape sum-symmetric GDoF defined by the points (0,1), (1,2/3) and (2,1). At these three points and these three points only, this time-sharing scheme coincides with the fundamental per cell sum-symmetric GDoF. Hence, time-sharing between the two embedded MAC-Z-P2Ps is not sufficient to achieve the fundamental sum-symmetric GDoF either.

\begin{rem}
To summarize, (a) superposition coding helps reserve certain signal-level partitions for each interfering transmitter at its intended receiver (b) it reserves certain signal-level partitions for the common sub-message of an interfering transmitter at its unintended receiver (c) there are in general still unused signal-level partitions at each receiver that are not accessible to interfering transmitters, intended or not, but which are open to be exploited by non-interfering transmitters and (d) when there are more than one non-interfering transmitters in each cell (say when $K=3$) the signal-level partitions that are inaccessible to interfering transmitters can be time-shared by the intended non-interfering transmitters toward the achievement of optimal GDoF.
\end{rem}

\section{Conclusions\label{sec:mim-conclusions}}

Classical results on the capacity of the DM and Gaussian multiple-access channels of \cite{liao1972multiple,ahlswede1973multi,wyner1974recent} and those on the two-user interference channel of \cite{carleial1978interference,han1981new} as well as recent results on capacity bounds and approximations for the interference channel are unified and generalized in this work by obtaining bounds and capacity approximations for the two-cell, $(K_1, K_2)$-user MAC-IC-MAC.
In particular, we generalize the inner bounds of \cite{kobayashi2007further} and \cite{chong2008han} from the 2-user DM IC to the $(K_1, K_2)$-user DM MAC-IC-MAC, bound the capacity region of the semi-deterministic DM MAC-IC-MAC to within a quantifiable gap, thus extending the work of \cite{telatar2007bounds} on the semi-deterministic two-user DM IC, and finally, we obtain capacity approximations to within a constant gap and hence up to GDoF accuracy for the Gaussian MAC-IC-MAC, thereby generalizing the results of \cite{etkin2008gaussian}. 
 
 A single coding scheme in the Gaussian setting is shown to have an achievable rate region that is within two bits of the capacity region. Improvements in cell spectrum efficiency with increasing number of users in the symmetric MAC-IC-MAC is quantified and illustrated via the per-cell sum-symmetric GDoF. Some examples of achieving GDoF via signal-level partitioning are discussed in detail to reveal the role of non-interfering transmitters in improving spectrum efficiency.

\appendices{}

\section{Proof of Theorem \ref{thm:mim-dm-inner-fix-dist}\label{sec:mim-dm-inner-fix-dist-proof}}

We prove achievability through a random coding argument. We fix
a coding distribution $P_{\mathrm{in}}\in\mathcal{P}_{\mathrm{in}}$, and obtain reliability
conditions in the form of partial sum-rate restrictions, which together give
the achievable region $\mathcal{R}_{\mathrm{in}}(P_{\mathrm{in}})$.

\subsection{The Achievable Scheme}
\begin{enumerate}
\item Generate time sharing sequence $q^{n}$ according to $p(q^{n})=\prod_{t=1}^{n}p(q_{t})$. 
\item Tx$i.j$, $j\neq1$, independently generates $2^{nR_{i.j}}$ sequences
$x_{i.j}^{n}$ according to $p(x_{i.j}^{n}|q^{n})=\prod_{t=1}^{n}p(x_{i.j,t}|q_{t})$
and indexes them by $k_{i.j}\in\{1,\cdots,2^{nR_{i.j}}\}$; 
\item Tx$i.1$ generates $2^{nR_{i.1c}}$ sequences $u_{i.1c}^{n}$ according
to $p(u_{i}^{n}|q^{n})=\prod_{t=1}^{n}p(u_{it}|q_{t})$ and indexes
them by $k_{i.1c}\in\{1,\cdots,2^{nR_{i.1c}}\}$. For each $u_{i}^{n}(k_{i.1c})$,
it generates $2^{n(R_{i.1}-R_{i.1c})}$ sequences $x_{i.1}^{n}$ according
to $p(x_{i.1}^{n}|u_{i}^{n}(k_{i.1c}),q^{n})=\prod_{t=1}^{n}p(x_{i.1t}|u_{it}(k_{i.1c}),q_{t})$
and indexes them by $(k_{i.1c},k_{i.1p})\in\{1,\cdots,2^{nR_{i.1c}}\}\times\{1,\cdots,2^{n(R_{i.1}-R_{i.1c})}\}$. 
\item Once the codebooks are generated, they are fixed for the duration
of communication and revealed to receivers Rx1 and Rx2. 
\item A $K_{i}$-tuple message $(m_{i.1},\cdots,m_{i.K_{i}})=((k_{i.1c},k_{i.1p}),k_{i.2},\cdots,k_{i.K_{i}})$
in cell-$i$ is encoded to ($(x_{i.1}^{n}(k_{i.1c},k_{i.1p})$, $x_{i.2}^{n}(k_{i.2})$,
$\cdots$, $x_{i.K_{i}}^{n}(k_{i.K_{i}})$) at Tx$i.1$, Tx$i.2$,
$\cdots$, Tx$i.K_{i}$, respectively, and sent over the channel. 
\item Upon receiving $y_{i}^{n}$, each Rx$i$ declares its decoded messages
$(\hat{m}_{i.1},\cdots,\hat{m}_{i.K_{i}})$ as the unique index-tuple
($(\hat{k}_{i.1c},\hat{k}_{i.1p})$, $\hat{k}_{i.2}$, $\cdots$,
$\hat{k}_{i.K_{i}}$) for which $q^{n}$, $x_{i.1}^{n}(\hat{k}_{i.1c},\hat{k}_{i.1p})$,
$x_{i.2}^{n}(\hat{k}_{i.2})$,$\cdots$, $x_{i.K_{i}}^{n}(\hat{k}_{i.K_{i}})$,
$u_{i^{'}}^{n}(\hat{k}_{i^{'}.1c})$ and $y_{i}^{n}$ are jointly
typical, for some $\hat{k}_{i^{'}.1c}\in\{1,\cdots,2^{nR_{i^{'}.1c}}\}$.
If such an index-tuple cannot be found, Rx$i$ declares an error. 
\end{enumerate}
Suppose $(k_{i.1c},k_{i.1p})=(1,1)$, $k_{i.2}=\cdots=k_{i.K_{i}}=1$
is sent. For any $\varUpsilon_{i}$ or $\varOmega_{i}$, we define
four classes of error events as 
\begin{itemize}
\item Class 1: $\hat{k}_{i.1c}=1$, $\hat{k}_{i.1p}\neq1$ , $\hat{k}_{i.j}\neq1$
$\forall i.j\in\varUpsilon_{i}\backslash\{i.1\}$, $\hat{k}_{i^{'}.1c}=1$; 
\item Class 2: $\hat{k}_{i.j}\neq1$ for all $i.j\in\varOmega_{i}\backslash\{i.1\}$
and if $i.1\in\varOmega_{i}$, then $\hat{k}_{i.1c}\neq1$; and $\hat{k}_{i^{'}.1c}=1$; 
\item Class 3: $\hat{k}_{i.1c}=1$, $\hat{k}_{i.1p}\neq1$ , $\hat{k}_{i.j}\neq1$
$\forall i.j\in\varUpsilon_{i}\backslash\{i.1\}$, $\hat{k}_{i^{'}.1c}\neq1$; 
\item Class 4: $\hat{k}_{i.j}\neq1$ for all $i.j\in\varOmega_{i}\backslash\{i.1\}$
and if $i.1\in\varOmega_{i}$, then $\hat{k}_{i.1c}\neq1$; and $\hat{k}_{i^{'}.1c}\neq1$. 
\end{itemize}
Any possible error event at Rx$i$ that affects messages sent by transmitters
Tx$i.j$, $\forall i.j\in\varOmega_{i}$ also belongs to one of the
four error event classes, depending on the correctness of each of
following components: (1) the common message $m_{i.1c}$ from Tx$i.1$,
if $i.1\in\varOmega_{i}$; (2) the private message $m_{i.1p}$ from
Tx$i.1$, if $i.1\in\varOmega_{i}$; (3) the common message $m_{i^{'}.1c}$
from Tx$i^{'}.1$. We illustrate the classification in Table \ref{tab:mim-error-events-omega}. On the other hand, for any $\varUpsilon_{i}$, there exists some $\varOmega_{i}$ such
that $\varOmega_{i}=\varUpsilon_{i}$. Since we always have
$i.1\in\varUpsilon_{i}$, any error event that involves the message sent by transmitters Tx$i.j$, $\forall i.j \in \varUpsilon_{i}$ should also fall into one
of the four cases given by the last two columns of Table \ref{tab:mim-error-events-omega}.

\begin{table*}[tbh]
\begin{centering}
\begin{tabular}{|c|c|c|c|}
\hline 
 & %
\begin{tabular}{c}
$i.1\notin\varOmega_{i}$\tabularnewline
\tabularnewline
$\hat{k}_{i.j}\neq1$, $\forall i.j\in\varOmega_{i}$\tabularnewline
\end{tabular} & %
\begin{tabular}{c}
$\hat{k}_{i.1c}=1$ \tabularnewline
$\hat{k}_{i.1p}\neq1$\tabularnewline
$\hat{k}_{i.j}\neq1$, $\forall i.j\in\varOmega_{i}\backslash\{i.1\}$\tabularnewline
\end{tabular} & %
\begin{tabular}{c}
$\hat{k}_{i.1c}\neq1$\tabularnewline
$\hat{k}_{i.1p}$ arbitrary\tabularnewline
$\hat{k}_{i.j}\neq1$, $\forall i.j\in\varOmega_{i}\backslash\{i.1\}$\tabularnewline
\end{tabular}\tabularnewline
\hline 
\hline 
$\hat{k}_{i^{'}.1}=1$  & Class 2  & Class 1  & Class 2\tabularnewline
\hline 
$\hat{k}_{i^{'}.1}\neq1$  & Class 4  & Class 3  & Class 4\tabularnewline
\hline 
\end{tabular}
\par\end{centering}
\caption{Error events affecting messages sent by transmitters Tx$i.j$, $\forall i.j\in\varOmega_{i}$.\label{tab:mim-error-events-omega}}
\end{table*}

The probabilities of each of the four classes can be bounded via a
careful application of the joint typicality lemma (cf. Chapter 2,
\cite{el2011network}), by the set functions $\mathsf{A}$, $\mathsf{B}$,
$\mathsf{E}$ and $\mathsf{G}$ as following, respectively, corresponding
to each of the two receivers. Consequently, including the non-negativity
of $R_{i.1c}$ and $\sum_{i.j\in\varUpsilon_{i}}R_{i,j}-R_{i.1c}$,
we have

\begin{align}
\sum_{1.j\in\varUpsilon_{1}}R_{1.j}-R_{1.1c} & \leq\mathsf{A}_{\varUpsilon_{1}}\label{eq:mim-inner-init-a-1}\\
\sum_{1.j\in\varOmega_{1}}R_{1.j} & \leq\mathsf{B}_{\varOmega_{1}}\label{eq:mim-inner-init-b-1}\\
\sum_{1.j\in\varUpsilon_{1}}R_{1.j}-R_{1.1c}+R_{2.1c} & \leq\mathsf{E}_{\varUpsilon_{1}}\label{eq:mim-inner-init-e-1}\\
\sum_{1.j\in\varOmega_{1}}R_{1.j}+R_{2.1c} & \leq\mathsf{G}_{\varOmega_{1}}\label{eq:mim-inner-init-g-a}\\
-R_{1.1c} & \leq0\label{eq:mim-inner-init-c-1}\\
R_{1.1c}-\sum_{1.j\in\varUpsilon_{1}}R_{1.j} & \leq0\label{eq:mim-inner-init-d-1}\\
\nonumber \\
\sum_{2.j\in\varUpsilon_{2}}R_{2.j}-R_{2.1c} & \leq\mathsf{A}_{\varUpsilon_{2}}\label{eq:mim-inner-init-a-2}\\
\sum_{2.j\in\varOmega_{2}}R_{2.j} & \leq\mathsf{B}_{\varOmega_{2}}\label{eq:mim-inner-init-b-2}\\
\sum_{2.j\in\varUpsilon_{2}}R_{2.j}-R_{2.1c}+R_{1.1c} & \leq\mathsf{E}_{\varUpsilon_{2}}\label{eq:mim-inner-init-e-2}\\
\sum_{2.j\in\varOmega_{2}}R_{2.j}+R_{1.1c} & \leq\mathsf{G}_{\varOmega_{2}}\label{eq:mim-inner-init-g-2}\\
-R_{2.1c} & \leq0\label{eq:mim-inner-init-c-2}\\
R_{2.1c}-\sum_{2.j\in\varUpsilon_{2}}R_{2.j} & \leq0\label{eq:mim-inner-init-d-2}
\end{align}

\subsection{Fourier Motzkin Elimination}

Inequalities \eqref{eq:mim-inner-init-a-1}-\eqref{eq:mim-inner-init-d-2}
group many inequalities that all have the same structure together
into one class of inequalities. This allows us to manipulate each
class of inequalities as if it were one inequality. This elegant structure
of the initial inequality system given in \eqref{eq:mim-inner-init-a-1}-\eqref{eq:mim-inner-init-d-2}
allows us to apply Fourier Motzkin elimination analytically. Without
loss of generality, we first eliminate $R_{1.1c}$. Note all the lower
bounds to $R_{1.1c}$ are contributed by the classes of inequalities
\eqref{eq:mim-inner-init-a-1}, \eqref{eq:mim-inner-init-e-1} and
\eqref{eq:mim-inner-init-c-1}, and upper bounds by the classes of
inequalities \eqref{eq:mim-inner-init-d-1}, \eqref{eq:mim-inner-init-e-2}
and \eqref{eq:mim-inner-init-g-2}. Now, $R_{1.1c}$ can be eliminated
by having the minimum of its upper bounds be greater than the maximum
of the lower bounds. The system of inequalities after eliminating $R_{1.1c}$
becomes 
\begin{align*}
0 & \leq\mathsf{A}_{\varUpsilon_{1}}\qquad*\\
\sum_{1.j\in\varUpsilon_{1}}R_{1.j}+\sum_{2.j\in\varUpsilon_{2}}R_{2.j}-R_{2.1c} & \leq\mathsf{A}_{\varUpsilon_{1}}+\mathsf{E}_{\varUpsilon_{2}}\\
\sum_{1.j\in\varUpsilon_{1}}R_{1.j}+\sum_{2.j\in\varOmega_{2}}R_{2.j} & \leq\mathsf{A}_{\varUpsilon_{1}}+\mathsf{G}_{\varOmega_{2}}\\
R_{2.1c} & \leq\mathsf{E}_{\varUpsilon_{1}}\\
\sum_{1.j\in\varUpsilon_{1}}R_{1.j}+\sum_{2.j\in\varUpsilon_{2}}R_{2.j} & \leq\mathsf{E}_{\varUpsilon_{1}}+\mathsf{E}_{\varUpsilon_{2}}\\
\sum_{1.j\in\varUpsilon_{1}}R_{1.j}+R_{2.1c}+\sum_{2.j\in\varOmega_{2}}R_{2.j} & \leq\mathsf{E}_{\varUpsilon_{1}}+\mathsf{G}_{\varOmega_{2}}\\
-\sum_{1.j\in\varUpsilon_{1}}R_{1.j} & \leq0\qquad*\\
\sum_{2.j\in\varUpsilon_{2}}R_{2.j}-R_{2.1c} & \leq\mathsf{E}_{\varUpsilon_{2}}\qquad*\\
\sum_{2.j\in\varOmega_{2}}R_{2.j} & \leq\mathsf{G}_{\varOmega_{2}}\qquad*\\
\sum_{1.j\in\varOmega_{1}}R_{1.j}+R_{2.1c} & \leq\mathsf{G}_{\varOmega_{1}}\\
\sum_{2.j\in\varUpsilon_{2}}R_{2.j}-R_{2.1c} & \leq\mathsf{A}_{\varUpsilon_{2}}\\
-R_{2.1c} & \leq0\\
R_{2.1c}-\sum_{2.j\in\varUpsilon_{2}}R_{2.j} & \leq0.
\end{align*}
The starred inequalities are clearly redundant by the non-negativity of rate and mutual information as well as the fact that, by the chain rule of mutual information, $\mathsf{A}_{\varUpsilon_{i}}\leq\mathsf{E}_{\varUpsilon_{i}}$
and $\mathsf{B}_{\varOmega_{i}}\leq\mathsf{G}_{\varOmega_{i}}$. Removing
these redundancies, we proceed to the elimination of $R_{2.1c}$.
Lower bounds on $R_{2.1c}$ are provided by the terms with right hand
side valued $\mathsf{A}_{\varUpsilon_{1}}+\mathsf{E}_{\varUpsilon_{2}}$,
$\mathsf{E}_{\varUpsilon_{2}}$, $\mathsf{A}_{\varUpsilon_{2}}$ and
0, upper bounds by $\mathsf{E}_{\varUpsilon_{1}}$, $\mathsf{E}_{\varUpsilon_{1}}+\mathsf{G}_{\varOmega_{2}}$,
$\mathsf{G}_{\varOmega_{1}}$, and 0. Eliminating $R_{2.1c}$, a new
system shows up as: 
\begin{align*}
0 & \leq\mathsf{E}_{\varUpsilon_{1}}\quad*\\
\sum_{2.j\in\varUpsilon_{2}}R_{2.j} & \leq\mathsf{A}_{\varUpsilon_{2}}+\mathsf{E}_{\varUpsilon_{1}}\\
\sum_{1.j\in\varUpsilon_{1}}R_{1.j}+\sum_{2.j\in\varUpsilon_{2}}R_{2.j} & \leq\mathsf{A}_{\varUpsilon_{1}}+\mathsf{E}_{\varUpsilon_{2}}+\mathsf{E}_{\varUpsilon_{1}}\quad*\\
-\sum_{2.j\in\varUpsilon_{2}}R_{2.j} & \leq0\quad*\\
0 & \leq\mathsf{A}_{\varUpsilon_{2}}\quad*\\
\sum_{1.j\in\varUpsilon_{1}}R_{1.j} & \leq\mathsf{A}_{\varUpsilon_{1}}+\mathsf{E}_{\varUpsilon_{2}}\\
\sum_{1.j\in\varOmega_{1}}R_{1.j} & \leq\mathsf{G}_{\varOmega_{1}}\quad*\\
\sum_{2.j\in\varUpsilon_{2}}R_{2.j}+\sum_{1.j\in\varOmega_{1}}R_{1.j} & \leq\mathsf{A}_{\varUpsilon_{2}}+\mathsf{G}_{\varOmega_{1}}\\
\sum_{1.j\in\varUpsilon_{1}}R_{1.j}+\sum_{2.j\in\varUpsilon_{2}}R_{2.j}\qquad\\
+\sum_{1.j\in\varOmega_{1}}R_{1.j} & \leq\mathsf{A}_{\varUpsilon_{1}}+\mathsf{E}_{\varUpsilon_{2}}+\mathsf{G}_{\varOmega_{1}}\\
\sum_{1.j\in\varUpsilon_{1}}R_{1.j}+\sum_{2.j\in\varOmega_{2}}R_{2.j} & \leq\mathsf{E}_{\varUpsilon_{1}}+\mathsf{G}_{\varOmega_{2}}\quad*\\
\sum_{2.j\in\varUpsilon_{2}}R_{2.j}+\sum_{1.j\in\varUpsilon_{1}}R_{1.j}\qquad\\
+\sum_{2.j\in\varOmega_{2}}R_{2.j} & \leq\mathsf{A}_{\varUpsilon_{2}}+\mathsf{E}_{\varUpsilon_{1}}+\mathsf{G}_{\varOmega_{2}}\\
\sum_{1.j\in\varUpsilon_{1}}R_{1.j}+\sum_{2.j\in\varUpsilon_{2}}R_{2.j}\qquad\\
+\sum_{1.j\in\varUpsilon_{1}}R_{1.j}+\sum_{2.j\in\varOmega_{2}}R_{2.j} & \leq\mathsf{A}_{\varUpsilon_{1}}+\mathsf{E}_{\varUpsilon_{2}}+\mathsf{E}_{\varUpsilon_{1}}+\mathsf{G}_{\varOmega_{2}}\quad*\\
\sum_{1.j\in\varUpsilon_{1}}R_{1.j}+\sum_{2.j\in\varOmega_{2}}R_{2.j} & \leq\mathsf{A}_{\varUpsilon_{1}}+\mathsf{G}_{\varOmega_{2}}\\
\sum_{1.j\in\varUpsilon_{1}}R_{1.j}+\sum_{2.j\in\varUpsilon_{2}}R_{2.j} & \leq\mathsf{E}_{\varUpsilon_{1}}+\mathsf{E}_{\varUpsilon_{2}}.
\end{align*}
The starred inequalities can again be easily identified as redundancies.
Eventually, we obtain a region given by 
\begin{align}
\sum_{1.j\in\varOmega_{1}}R_{1.j} & \leq\mathsf{B}_{\varOmega_{1}}\label{eq:mim-inner-fix-dist-1}\\
\sum_{1.j\in\varUpsilon_{1}}R_{1.j} & \leq\mathsf{A}_{\varUpsilon_{1}}+\mathsf{E}_{\varUpsilon_{2}}\label{eq:mim-inner-fix-dist-2}\\
\sum_{2.j\in\varOmega_{2}}R_{2.j} & \leq\mathsf{B}_{\varOmega_{2}}\label{eq:mim-inner-fix-dist-3}\\
\sum_{2.j\in\varUpsilon_{2}}R_{2.j} & \leq\mathsf{A}_{\varUpsilon_{2}}+\mathsf{E}_{\varUpsilon_{1}}\label{eq:mim-inner-fix-dist-4}\\
\sum_{1.j\in\varUpsilon_{1}}R_{1.j}+\sum_{2.j\in\varOmega_{2}}R_{2.j} & \leq\mathsf{A}_{\varUpsilon_{1}}+\mathsf{G}_{\varOmega_{2}}\label{eq:mim-inner-fix-dist-5}\\
\sum_{1.j\in\varOmega_{1}}R_{1.j}+\sum_{2.j\in\varUpsilon_{2}}R_{2.j} & \leq\mathsf{G}_{\varOmega_{1}}+\mathsf{A}_{\varUpsilon_{2}}\label{eq:mim-inner-fix-dist-6}\\
\sum_{1.j\in\varUpsilon_{1}}R_{1.j}+\sum_{2.j\in\varUpsilon_{2}}R_{2.j} & \leq\mathsf{E}_{\varUpsilon_{1}}+\mathsf{E}_{\varUpsilon_{2}}\label{eq:mim-inner-fix-dist-7}\\
\sum_{1.j\in\varUpsilon_{1}}R_{1.j}+\sum_{1.j\in\varOmega_{1}}R_{1.j}\qquad\nonumber \\
+\sum_{2.j\in\varUpsilon_{2}}R_{2.j} & \leq\mathsf{A}_{\varUpsilon_{1}}+\mathsf{G}_{\varOmega_{1}}+\mathsf{E}_{\varUpsilon_{2}}\label{eq:mim-inner-fix-dist-8}\\
\sum_{1.j\in\varUpsilon_{1}}R_{1.j}+\sum_{2.j\in\varUpsilon_{2}}R_{2.j}\qquad\nonumber \\
+\sum_{2.j\in\varOmega_{2}}R_{2.j} & \leq\mathsf{E}_{\varUpsilon_{1}}+\mathsf{A}_{\varUpsilon_{2}}+\mathsf{G}_{\varOmega_{2}}.\label{eq:mim-inner-fix-dist-9}
\end{align}
The non-negative achievable ($K_{1}+K_{2}$) rate tuples ($R_{\varTheta_{1}},R_{\varTheta_{2}}$)
that satisfy \eqref{eq:mim-inner-fix-dist-1}-\eqref{eq:mim-inner-fix-dist-9}
are precisely the ones that are defined as $\mathcal{R}_{{\rm in}}(P_{\mathrm{in}})$
in \eqref{eq:mim-inner-fix-dist}. Hence, the region $\mathcal{R}_{{\rm in}}(P_{\mathrm{in}})$
is an inner bound to the capacity region and Theorem \ref{thm:mim-dm-inner-fix-dist}
is proved.

\section{Proof of Theorem \ref{thm:mim-dm-inner-compact} \label{sec:mim-dm-inner-proof}}

In this proof, we need to show that inequalities \eqref{eq:mim-inner-fix-dist-2}
and \eqref{eq:mim-inner-fix-dist-4} in Appendix \ref{sec:mim-dm-inner-fix-dist-proof}
are redundant when we take the inner bound as the union of fixed pmf
inner bounds over all admissible input distributions. We demonstrate
the redundancies by contradiction that the region $\mathcal{R}_{1}=\mathcal{R}_{\mathrm{c}}\backslash\mathcal{R}_{\mathrm{in}}$
is empty. 

Suppose $\mathcal{R}_{1} \neq \emptyset $. We construct a
region $\mathcal{R}_{2}$ inside $\mathcal{R}_{\mathrm{in}}$ by setting
$U_{1}$ or $U_{2}$ to be trivial (a constant). Then it must be true
that $\mathcal{R}_{1}\cap\mathcal{R}_{2}=\emptyset$. However, through
an explicit computation of $\mathcal{R}_{1}$ and $\mathcal{R}_{2}$,
we show that $\mathcal{R}_{1}\subseteq\mathcal{R}_{2}$, which means $\mathcal{R}_1\cap\mathcal{R}_{2} \neq \emptyset$, a contradiction. Hence, $\mathcal{R}_{1} = \emptyset $.

We first prove the redundancy of \eqref{eq:mim-inner-fix-dist-2}.
Define $\mathcal{R}_{2}=\{(R_{\varTheta_{1}},R_{\varTheta_{2}})\in\mathcal{R}_{\mathrm{in}}:U_{1}=\emptyset\}$.
Since $\mathcal{R}_{2}\subseteq\mathcal{R}_{\mathrm{in}}$, it is
clear that 
\begin{equation}
\mathcal{R}_{1}\cap\mathcal{R}_{2}=\emptyset\label{eq:mim-inner-proof-r2-r3-exclusive}
\end{equation}
Region $\mathcal{R}_{2}$ can be obtained by setting $U_{1}=\emptyset$
in the initial system \eqref{eq:mim-inner-init-a-1}-\eqref{eq:mim-inner-init-d-2}.
It can then be shown by performing Fourier-Motzkin elimination procedure
as we have done in Appendix \ref{sec:mim-dm-inner-fix-dist-proof}
to eliminate $R_{2.1c}$, that the region $\mathcal{R}_{2}$ contains
all rate pairs that satisfy just the following three groups of inequalities,

\begin{align}
\sum_{1.j\in\varOmega_{1}}R_{1.j} & \leq I(X_{\varOmega_{1}};Y_{1}|X_{\bar{\varOmega}_{1}},U_{2},Q)\label{eq:mim-inner-proof-r3-1}\\
\sum_{2.j\in\varOmega_{2}}R_{2.j} & \leq I(X_{\varOmega_{2}};Y_{2}|X_{\bar{\varOmega}_{2}},Q)\label{eq:mim-inner-proof-r3-2}\\
\sum_{1.j\in\varOmega_{1}}R_{1.j}+\sum_{2.j\in\varUpsilon_{2}}R_{2.j} & \leq I(X_{\varOmega_{1}},U_{2};Y_{1}|X_{\bar{\varOmega}_{1}},Q)\nonumber \\
 & \qquad+I(X_{\varUpsilon_{2}};Y_{2}|X_{\bar{\varUpsilon}_{2}},U_{2},Q).\label{eq:mim-inner-proof-r3-3}
\end{align}
On the other hand, region $\mathcal{R}_{1}$ is constituted by all
the inequalities for region $\mathcal{R}_{\mathrm{c}}$ and $\sum_{1.j\in\varUpsilon_{1}}R_{1.j}\geq\mathsf{A}_{\varUpsilon_{1}}+\mathsf{E}_{\varUpsilon_{2}}$. We explicitly write them as 
\begin{align}
\sum_{1.j\in\varOmega_{1}}R_{1.j} & \leq\mathsf{B}_{\varOmega_{1}}\label{eq:mim-inner-proof-r2-1}\\
-\sum_{1.j\in\varUpsilon_{1}}R_{1.j} & \leq-\mathsf{A}_{\varUpsilon_{1}}-\mathsf{E}_{\varUpsilon_{2}}\label{eq:mim-inner-proof-r2-2}\\
\sum_{2.j\in\varOmega_{2}}R_{2.j} & \leq\mathsf{B}_{\varOmega_{2}}\label{eq:mim-inner-proof-r2-3}\\
\sum_{1.j\in\varUpsilon_{1}}R_{1.j}+\sum_{2.j\in\varOmega_{2}}R_{2.j} & \leq\mathsf{A}_{\varUpsilon_{1}}+\mathsf{G}_{\varOmega_{2}}\label{eq:mim-inner-proof-r2-4}\\
\sum_{1.j\in\varOmega_{1}}R_{1.j}+\sum_{2.j\in\varUpsilon_{2}}R_{2.j} & \leq\mathsf{G}_{\varOmega_{1}}+\mathsf{A}_{\varUpsilon_{2}}\label{eq:mim-inner-proof-r2-5}\\
\sum_{1.j\in\varUpsilon_{1}}R_{1.j}+\sum_{2.j\in\varUpsilon_{2}}R_{2.j} & \leq\mathsf{E}_{\varUpsilon_{1}}+\mathsf{E}_{\varUpsilon_{2}}\label{eq:mim-inner-proof-r2-6}\\
\sum_{1.j\in\varUpsilon_{1}}R_{1.j}+\sum_{1.j\in\varOmega_{1}}R_{1.j}\qquad\nonumber \\
+\sum_{2.j\in\varUpsilon_{2}}R_{2.j} & \leq\mathsf{A}_{\varUpsilon_{1}}+\mathsf{G}_{\varOmega_{1}}+\mathsf{E}_{\varUpsilon_{2}}\label{eq:mim-inner-proof-r2-7}\\
\sum_{1.j\in\varUpsilon_{1}}R_{1.j}+\sum_{2.j\in\varUpsilon_{2}}R_{2.j}\qquad\nonumber \\
+\sum_{2.j\in\varOmega_{2}}R_{2.j} & \leq\mathsf{E}_{\varUpsilon_{1}}+\mathsf{A}_{\varUpsilon_{2}}+\mathsf{G}_{\varOmega_{2}}.\label{eq:mim-inner-proof-r2-8}
\end{align}
From inequality \eqref{eq:mim-inner-proof-r2-1}, we know 
\begin{align}
\sum_{1.j\in\varOmega_{1}}R_{1.j} & \leq\mathsf{B}_{\varOmega_{1}}=I(X_{\varOmega_{1}};Y_{1}|X_{\bar{\varOmega}_{1}},U_{2},Q)\label{eq:mim-inner-proof-r2-9}
\end{align}
adding inequalities \eqref{eq:mim-inner-proof-r2-2} and \eqref{eq:mim-inner-proof-r2-4},
we have 
\begin{align}
\sum_{2.j\in\varOmega_{2}}R_{2.j} & \leq\mathsf{G}_{\varOmega_{2}}-\mathsf{E}_{\varUpsilon_{2}}\nonumber \\
 & =I(X_{\varOmega_{2}},U_{1};Y_{2}|X_{\bar{\varOmega}_{2}},Q)\nonumber \\
 & \qquad-I(X_{\varUpsilon_{2}},U_{1};Y_{2}|X_{\bar{\varUpsilon}_{2}},U_{2},Q)\nonumber \\
 & =H(Y_{2}|X_{\bar{\varOmega}_{2}},Q)-H(Y_{2}|X_{\bar{\varUpsilon}_{2}},U_{2},Q)\nonumber \\
 & \leq H(Y_{2}|X_{\bar{\varOmega}_{2}},Q)-H(Y_{2}|X_{\bar{\varUpsilon}_{2}},X_{2.1},Q)\nonumber \\
 & =H(Y_{2}|X_{\bar{\varOmega}_{2}},Q)-H(Y_{2}|X_{\bar{\varUpsilon}_{2}},X_{\varUpsilon_{2}},Q)\nonumber \\
 & =H(Y_{2}|X_{\bar{\varOmega}_{2}},Q)-H(Y_{2}|X_{\varTheta_{2}},Q)\nonumber \\
 & =H(Y_{2}|X_{\bar{\varOmega}_{2}},Q)-H(Y_{2}|X_{\varOmega_{2}},X_{\bar{\varOmega}_{2}},Q)\nonumber \\
 & =I(X_{\varOmega_{2}};Y_{2}|X_{\bar{\varOmega}_{2}},Q)\label{eq:mim-inner-proof-r2-10}
\end{align}
adding inequalities \eqref{eq:mim-inner-proof-r2-2} and \eqref{eq:mim-inner-proof-r2-7},
we have 
\begin{align}
 & \quad\thinspace\sum_{2.j\in\varUpsilon_{2}}R_{2.j}+\sum_{1.j\in\varOmega_{1}}R_{1.j}\leq\mathsf{G}_{\varOmega_{1}}\nonumber \\
 & =I(X_{\varOmega_{1}},U_{2};Y_{1}|X_{\bar{\varOmega}_{1}},Q)\nonumber \\
 & \leq I(X_{\varOmega_{1}},U_{2};Y_{1}|X_{\bar{\varOmega}_{1}},Q)+I(X_{\varUpsilon_{2}};Y_{2}|X_{\bar{\varUpsilon}_{2}},U_{2},Q).\label{eq:mim-inner-proof-r2-11}
\end{align}
Note inequalities \eqref{eq:mim-inner-proof-r3-1}-\eqref{eq:mim-inner-proof-r3-3}
are identical to \eqref{eq:mim-inner-proof-r2-9}-\eqref{eq:mim-inner-proof-r2-11},
which means a rate-tuple in $\mathcal{R}_{1}$ is in $\mathcal{R}_{2}$,
i.e., 
\begin{equation}
\mathcal{R}_{1}\subseteq\mathcal{R}_{2}\label{eq:mim-inner-proof-r2-in-r3}
\end{equation}
therefore contradicts \eqref{eq:mim-inner-proof-r2-r3-exclusive},
so \eqref{eq:mim-inner-fix-dist-2} is redundant.

Due to the symmetry of the channel, \eqref{eq:mim-inner-fix-dist-4}
is redundant too, and so are both of them together. Hence, \eqref{eq:mim-inner-fix-dist-2}
and \eqref{eq:mim-inner-fix-dist-4} are both redundancies, which
completes the proof.

\section{Clarifications in \cite{pang2013bounds} \label{sec:mim-errors}}

We clarify some points in the conference version of this paper \cite{pang2013bounds}
in this appendix which considers the $K_{1}=2$ and $K_{2}=1$ case.

In \cite{pang2013bounds}, Theorem 1 (which is Corollary \ref{cor:mim-PMAIC}
 in this paper) correctly states that $\cup_{p(\cdot)\in\mathcal{P}}\mathfrak{R}_{inner}(p)$,
which is equivalent to $\mathcal{R}_{\mathrm{c}}$ in this paper,
is achievable. Its proof in Appendix A of \cite{pang2013bounds} is
not complete. The missing part is to prove the redundancies of two
extra inequalities after Fourier-Motzkin elimination of the initial
system given in Appendix A of \cite{pang2013bounds}. The missing
proof is contained in Appendix \ref{sec:mim-dm-inner-proof} of this
paper.

The probability distribution given in Definition 8 of \cite{pang2013bounds}
should not contain $u_{1}$ and $u_{2}$. In Theorems 11 and 12 of
\cite{pang2013bounds}, the probability distribution should be the
one which is defined in Definition 10, instead of $p(\cdot)\in\mathcal{P}$.

The phrase ``For a given $p(\cdot)\in\mathcal{P}$'' should be removed
in Definitions 14 and 16, because both these bounds are single
region bounds resulting from one fixed input distribution.

In Theorem 18, the outer bound is again a single region bound. There
should be no union over $\mathcal{P}$.

\section{Proof of Theorem \ref{thm:mim-sd-outer}\label{sec:mim-outer-proof}}

The work by Telatar and Tse in \cite{telatar2007bounds} has given
an outer bound for two-user semi-deterministic IC, which is a special
case of semi-deterministic MAC-IC-MAC. In their proof, each multi-letter
mutual information term is expressed as the difference of corresponding
conditional entropy terms. The genie information $T_{i}^{n}$ is chosen
to be distributed as \eqref{eq:mim-sd-genie} to ensure that the multi-letter
negative entropy terms which do not appear in the form of $H(\text{output}|\text{input})$
(and are therefore hard to single-letterize) can be eventually canceled.
This technique is extended to the semi-deterministic MAC-IC-MAC in
this appendix.

We show in our proof to follow, that in the presence of the non-interfering
transmitters, the intra-cell sum rate $\sum_{i.j\in\varUpsilon_{i}}R_{i.j}$
and $\sum_{i,j\in\varOmega{}_{i}}R_{i,j}$ can be upper bounded into
four classes by grouping their transmitted signals as $X_{\varUpsilon_{i}}$
(or $X_{\varOmega_{i}}$), and feeding the remaining transmit signals
$X_{\bar{\varUpsilon}_{i}}$ together with the genie information $T_{i}^{n}$
(if $i.1\in\varUpsilon_{i}$ or $i.1\in\varOmega_{i}$, resp.) to
the receiver. Notably, the algebraic structure of these four classes
of bounds still allows the cancellation of the negative conditional
entropies of the form $H(\text{output}|\text{input})$. The details
are given shortly.

Moreover, giving a subset of intended signals $X_{\bar{\varUpsilon}_{i}}^{n}$
(or $X_{\bar{\varOmega}_{i}}^{n}$), genie information $T_{i}^{n}$
or the non-intended signal $X_{i^{'}.1}^{n}$ to help Rx$i$ decode
will not decrease the capacity region of the channel. For any considered
subset $\varUpsilon_{i}$ (or $\varOmega_{i}$), we always give $X_{\bar{\varUpsilon}_{i}}^{n}$
( or $X_{\bar{\varOmega}_{i}}^{n}$) to help Rx$i$ to decode. Besides,
if we give both the genie information $T_{i}^{n}$ and the interference
signal $X_{i^{'}.1\rightarrow i}^{n}$, the partial sum rate $\sum_{i.j\in\varUpsilon_{i}}R_{i.j}$
can be upper bounded by $\overline{\mathsf{A}}_{\varUpsilon_{i}}$;
if we give only genie information $T_{i}^{n}$, the partial sum rate
$\sum_{i.j\in\varUpsilon_{i}}R_{i.j}$ can be upper bounded by $\overline{\mathsf{E}}_{\varUpsilon_{i}}$;
if we only give the interference signal $X_{i^{'}.1}^{n}$, the partial
sum rate $\sum_{i.j\in\varOmega_{i}}R_{i.j}$ (with or without $i.1$
in $\varOmega_{i}$) can be upper bounded by $\overline{\mathsf{B}}_{\varOmega_{i}}$;
if we give neither $T_{i}^{n}$ nor $X_{i^{'}.1}^{n}$, the partial
sum rate $\sum_{i.j\in\varOmega_{i}}R_{i.j}$ can be upper bounded
by $\overline{\mathsf{G}}_{\varOmega_{i}}$.

Next, we provide the promised details. For some fixed $P_{\mathrm{o}}^{\mathrm{sd}}\in\mathcal{P}_{\mathrm{o}}^{\mathrm{sd}}$,
applying Fano's inequality, chain rule, independence and Markov chain
property, we obtain following four classes of upper bounds:

\begin{figure}[tbh]
\begin{centering}
\includegraphics{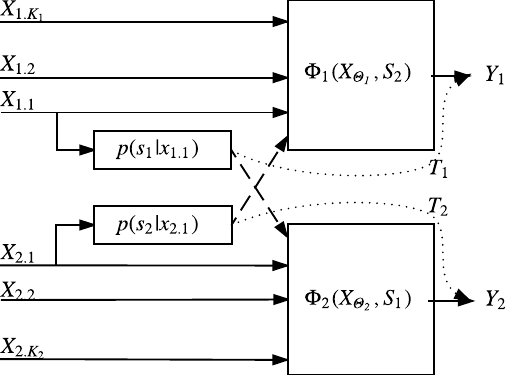} 
\par\end{centering}
\caption{Genie aided semi-deterministic MAC-IC-MAC, where a genie provides
information $T_{1}$ to Rx$1$ and $T_{2}$ to Rx$2$. \label{fig:mim-sd-outer-proof}}
\end{figure}

\begin{align}
 & \quad\thinspace n\sum_{1.j\in\varUpsilon_{1}}R_{1.j}\nonumber \\
 & \stackrel{(a)}{\leq}I(X_{\varUpsilon_{1}}^{n};Y_{1}^{n},X_{\bar{\varUpsilon}_{1}}^{n},T_{1}^{n},X_{2.1}^{n})+n\epsilon_{n}\nonumber \\
 & \stackrel{(b)}{=}I(X_{\varUpsilon_{1}}^{n};T_{1}^{n}|X_{\bar{\varUpsilon}_{1}}^{n},X_{2.1}^{n})\nonumber \\
 & \qquad+I(X_{\varUpsilon_{1}}^{n};Y_{1}^{n}|X_{\bar{\varUpsilon}_{1}}^{n},T_{1}^{n},X_{2.1}^{n})+n\epsilon_{n}\nonumber \\
 & \stackrel{(c)}{=}I(X_{\varUpsilon_{1}}^{n};T_{1}^{n})+I(X_{\varUpsilon_{1}}^{n};Y_{1}^{n}|X_{\bar{\varUpsilon}_{1}}^{n},T_{1}^{n},X_{2.1}^{n})+n\epsilon_{n}\nonumber \\
 & =H(T_{1}^{n})-H(T_{1}^{n}|X_{\varUpsilon_{1}}^{n})+H(Y_{1}^{n}|X_{\bar{\varUpsilon}_{1}}^{n},T_{1}^{n},X_{2.1}^{n})\nonumber \\
 & \qquad-H(Y_{1}^{n}|X_{\varTheta_{1}}^{n},T_{1}^{n},X_{2.1}^{n})+n\epsilon_{n}\nonumber \\
 & \stackrel{(d)}{=}H(T_{1}^{n})-H(T_{1}^{n}|X_{1.1}^{n})+H(Y_{1}^{n}|X_{\bar{\varUpsilon}_{1}}^{n},T_{1}^{n},X_{2.1}^{n})\nonumber \\
 & \qquad-H(S_{2}^{n}|X_{2.1}^{n})+n\epsilon_{n}\nonumber \\
 & \leq H(T_{1}^{n})+\sum_{t=1}^{n}\left[H(Y_{1t}|X_{\bar{\varUpsilon}_{1}t},T_{1t},X_{2.1t})\right.\nonumber \\
 & \qquad\left.-H(T_{1t}|X_{1.1t})-H(S_{2t}|X_{2.1t})\right]+n\epsilon_{n}\nonumber \\
 & =n\overline{\mathsf{A}}_{\varUpsilon_{1}}-nH(T_{1}|X_{1.1})+H(T_{1}^{n})+n\epsilon_{n}\label{eq:mim-sd-outer-proof-a-1}
\end{align}

\begin{align}
 & \quad\thinspace n\sum_{1.j\in\varOmega_{1}}R_{1.j}\nonumber \\
 & \stackrel{(a)}{\leq}I(X_{\varOmega_{1}}^{n};Y_{1}^{n},X_{\bar{\varOmega}_{1}}^{n},X_{2.1}^{n})+n\epsilon_{n}\nonumber \\
 & \stackrel{(b)}{=}I(X_{\varOmega_{1}}^{n};Y_{1}^{n}|X_{\bar{\varOmega}_{1}}^{n},X_{2.1}^{n})+n\epsilon_{n}\nonumber \\
 & =H(Y_{1}^{n}|X_{\bar{\varOmega}_{i}}^{n},X_{2.1}^{n})-H(Y_{i}^{n}|X_{\varOmega_{1}}^{n},X_{\bar{\varOmega}_{1}}^{n},X_{2.1}^{n})+n\epsilon_{n}\nonumber \\
 & \stackrel{(d)}{=}H(Y_{1}^{n}|X_{\bar{\varOmega}_{1}}^{n},X_{2.1}^{n})-H(S_{2}^{n}|X_{2.1}^{n})+n\epsilon_{n}\nonumber \\
 & \leq\sum_{t=1}^{n}\left[H(Y_{1t}|X_{\bar{\varOmega}_{1}t},X_{2.1,t})-H(S_{2t}|X_{2.1t})\right]+n\epsilon_{n}\nonumber \\
 & =n\overline{\mathsf{B}}_{\varOmega_{1}}+n\epsilon_{n}\label{eq:mim-sd-outer-proof-b-1}
\end{align}
\begin{align}
 & \quad\thinspace n\sum_{1.j\in\varUpsilon_{1}}R_{1.j}\nonumber \\
 & \stackrel{(a)}{\leq}I(X_{\varUpsilon_{1}}^{n};Y_{1}^{n},X_{\bar{\varUpsilon}_{1}}^{n},T_{1}^{n})+n\epsilon_{n}\nonumber \\
 & \stackrel{(b)}{=}I(X_{\varUpsilon_{1}}^{n};T_{1}^{n}|X_{\bar{\varUpsilon}_{1}}^{n})+I(X_{\varUpsilon_{1}}^{n};Y_{1}^{n}|X_{\bar{\varUpsilon}_{1}}^{n},T_{1}^{n})+n\epsilon_{n}\nonumber \\
 & \stackrel{(c)}{=}I(X_{\varUpsilon_{1}}^{n};T_{1}^{n})+I(X_{\varUpsilon_{1}}^{n};Y_{1}^{n}|X_{\bar{\varUpsilon}_{1}}^{n},T_{1}^{n})+n\epsilon_{n}\nonumber \\
 & =H(T_{1}^{n})-H(T_{1}^{n}|X_{\varUpsilon_{1}}^{n})+H(Y_{1}^{n}|X_{\bar{\varUpsilon}_{1}}^{n},T_{1}^{n})\nonumber \\
 & \qquad-H(Y_{1}^{n}|X_{\varTheta_{1}}^{n},T_{1}^{n})+n\epsilon_{n}\nonumber \\
 & \stackrel{(d)}{=}H(T_{1}^{n})-H(T_{1}^{n}|X_{1.1}^{n})+H(Y_{1}^{n}|X_{\bar{\varUpsilon}_{1}}^{n},T_{1}^{n})\nonumber \\
 & \qquad-H(S_{2}^{n})+n\epsilon_{n}\nonumber \\
 & \leq\sum_{t=1}^{n}\left[H(Y_{1t}|X_{\bar{\varUpsilon}_{1}t},T_{1t})-H(T_{1t}|X_{1.1t})\right]\nonumber \\
 & \qquad+H(T_{1}^{n})-H(S_{2}^{n})+n\epsilon_{n}\nonumber \\
 & =n\overline{\mathsf{E}}_{\varUpsilon_{1}}+nH(S_{2}|X_{2.1})-nH(T_{1}|X_{1.1})\nonumber \\
 & \qquad+H(T_{1}^{n})-H(S_{2}^{n})+n\epsilon_{n}\label{eq:mim-sd-outer-proof-e-1}
\end{align}
and 
\begin{align}
 & \quad\thinspace n\sum_{1.j\in\varOmega_{1}}R_{1.j}\nonumber \\
 & \stackrel{(a)}{\leq}I(X_{\varOmega_{1}}^{n};Y_{1}^{n},X_{\bar{\varOmega}_{1}}^{n})+n\epsilon_{n}\nonumber \\
 & \stackrel{(b)}{=}I(X_{\varOmega_{1}}^{n};Y_{1}^{n}|X_{\bar{\varOmega}_{1}}^{n})+n\epsilon_{n}\nonumber \\
 & =H(Y_{1}^{n}|X_{\bar{\varOmega}_{i}}^{n})-H(Y_{i}^{n}|X_{\varOmega_{1}}^{n},X_{\bar{\varOmega}_{1}}^{n})+n\epsilon_{n}\nonumber \\
 & \stackrel{(d)}{=}H(Y_{1}^{n}|X_{\bar{\varOmega}_{1}}^{n})-H(S_{2}^{n})+n\epsilon_{n}\nonumber \\
 & \leq\sum_{t=1}^{n}H(Y_{1t}|X_{\bar{\varOmega}_{1}t})-H(S_{2}^{n})+n\epsilon_{n}\nonumber \\
 & =n\overline{\mathsf{G}}_{\varOmega_{1}}+nH(S_{2}|X_{2.1})-H(S_{2}^{n})+n\epsilon_{n}.\label{eq:mim-sd-outer-proof-g-1}
\end{align}
The steps (a)-(d) hold because: $(a)$ giving $X_{\bar{\varUpsilon}_{i}}^{n}$,
$X_{\bar{\varOmega}_{i}}^{n}$, $X_{i^{'},1}^{n}$, or $T_{i}^{n}$
to help Rx$i$ decode will not decrease the capacity region; (b) chain
rule of conditional mutual information and the independence of input
symbols; (c) $T_{i}^{n}$ is independent of $X_{\bar{\varUpsilon}_{i}}^{n}$
and $X_{i^{'}.1}^{n}$; (d) semi-deterministic property.

Similarly, we could have 4 upper bounds on rate $R_{\varTheta_{2}}$
too, 
\begin{align}
n\sum_{2.j\in\varUpsilon_{2}}R_{2.j} & \leq n\overline{\mathsf{A}}_{\varUpsilon_{2}}-nH(T_{2}|X_{2.1})+H(T_{2}^{n})+n\epsilon_{n}\label{eq:mim-sd-outer-proof-a-2}\\
n\sum_{2.j\in\varOmega_{2}}R_{2.j} & \leq n\overline{\mathsf{B}}_{\varOmega_{2}}+n\epsilon_{n}\label{eq:mim-sd-outer-proof-b-2}\\
n\sum_{2.j\in\varUpsilon_{2}}R_{2.j} & \leq n\overline{\mathsf{E}}_{\varUpsilon_{2}}+nH(S_{1}|X_{1.1})-nH(T_{2}|X_{2.1})\nonumber \\
 & \qquad+H(T_{2}^{n})-H(S_{1}^{n})+n\epsilon_{n}\label{eq:mim-sd-outer-proof-e-2}\\
n\sum_{2.j\in\varOmega_{2}}R_{2.j} & \le n\overline{\mathsf{G}}_{\varOmega_{2}}+nH(S_{1}|X_{1.1})-H(S_{1}^{n})+n\epsilon_{n}.\label{eq:mim-sd-outer-proof-g-2}
\end{align}
By construction (c.f. Definition \ref{def:mim-sd-outer}), we have
\begin{equation}
H(T_{i}|X_{i.1})=H(S_{i}|X_{i.1})\label{eq:mim-sd-outer-key-1}
\end{equation}
and
\begin{equation}
H(T_{i})=H(S_{i}).\label{eq:mim-sd-outer-key-2}
\end{equation}
Taking \eqref{eq:mim-sd-outer-key-1} and \eqref{eq:mim-sd-outer-key-2}
into consideration, the seven inequalities in Theorem \ref{thm:mim-sd-outer}
can be obtained by following linear combinations of inequalities:
\eqref{eq:mim-sd-outer-proof-b-1}, \eqref{eq:mim-sd-outer-proof-b-2},
\eqref{eq:mim-sd-outer-proof-a-1}+\eqref{eq:mim-sd-outer-proof-g-2},
\eqref{eq:mim-sd-outer-proof-g-1}+\eqref{eq:mim-sd-outer-proof-a-2},
\eqref{eq:mim-sd-outer-proof-e-1}+\eqref{eq:mim-sd-outer-proof-e-2},
\eqref{eq:mim-sd-outer-proof-a-1}+\eqref{eq:mim-sd-outer-proof-g-1}+\eqref{eq:mim-sd-outer-proof-e-2}
and \eqref{eq:mim-sd-outer-proof-a-2}+\eqref{eq:mim-sd-outer-proof-g-2}+\eqref{eq:mim-sd-outer-proof-e-1}.

\section{Proof of Lemma \ref{lem:mim-sd-gap-1}\label{sec:mim-sd-gap-1-proof}}

To prove the lemma, we show that for any admissible $\varUpsilon_{1},\varUpsilon_{2}$,
there exists a pair $\varOmega_{1},\varOmega_{2}$ such that the gap
between $B_{\varOmega_{i}}$ and $A_{\varUpsilon_{i}}+E_{\varUpsilon_{i^{'}}}$
is within $I(X_{i.1};S_{i}|U_{i})$. Without loss of generality, we
prove this for $i=1$.

If the set $\varUpsilon_{1}$ contains only one user, i.e., $\varUpsilon_{1}=\{1.1\}$,
\begin{align*}
\mathsf{A}_{\{1.1\}}+\mathsf{E}_{\varUpsilon_{2}} & =I(X_{1.1};Y_{1}|U_{1},U_{2},Q)\\
 & \qquad+I(X_{\varUpsilon_{2}},U_{1};Y_{2}|X_{\bar{\varUpsilon}_{2}},U_{2},Q)\\
 & \stackrel{(a)}{=}H(Y_{1}|U_{1},U_{2},Q)-H(S_{2}|U_{2},Q)\\
 & \qquad+H(Y_{2}|X_{\bar{\varUpsilon}_{2}},U_{2},Q)-H(S_{1}|U_{1},Q)\\
 & \stackrel{(b)}{=}H(Y_{1}|U_{1},U_{2},Q)-H(S_{2}|U_{2},Q)\\
 & \qquad+H(Y_{2}|X_{\varTheta_{2}},Q)-H(S_{1}|U_{1},Q)\\
 & \stackrel{(a)}{=}H(Y_{1}|U_{1},U_{2},Q)-H(S_{2}|U_{2},Q)\\
 & \qquad+H(S_{1}|Q)-H(S_{1}|U_{1},Q)\\
 & \stackrel{(c)}{=}H(Y_{1}|U_{1},U_{2},Q)-H(S_{2}|U_{2},Q)\\
 & \qquad+H(U_{1}|Q)-H(S_{1}|U_{1},Q)\\
 & \stackrel{(d)}{=}H(Y_{1}|U_{1},U_{2},Q)-H(S_{2}|U_{2},Q)\\
 & \qquad+H(U_{1}|U_{2},Q)-H(S_{1}|U_{1},Q)\\
 & =H(Y_{1},U_{1}|U_{2},Q)-H(S_{2}|U_{2},Q)\\
 & \qquad-H(S_{1}|U_{1},Q)\\
 & =H(Y_{1}|U_{2},Q)+H(U_{1}|Y_{1},U_{2},Q)\\
 & \qquad-H(S_{2}|U_{2},Q)-H(S_{1}|U_{1},Q)\\
 & =H(Y_{1}|U_{2},Q)+H(U_{1}|Y_{1},Q)\\
 & \qquad-H(S_{2}|U_{2},Q)-H(S_{1}|U_{1},Q)\\
 & \overset{(e)}{\geq}H(Y_{1}|U_{2},Q)+H(U_{1}|X_{1.1},Q)\\
 & \qquad-H(S_{2}|U_{2},Q)-H(S_{1}|U_{1},Q)\\
 & \geq H(Y_{1}|U_{2},Q)+H(S_{1}|X_{1.1},Q)\\
 & \qquad-H(S_{2}|U_{2},Q)-H(S_{1}|U_{1},Q)\\
 & =H(Y_{1}|U_{2},Q)+H(S_{1}|X_{1.1},U_{1},Q)\\
 & \qquad-H(S_{2}|U_{2},Q)-H(S_{1}|U_{1},Q)\\
 & =I(X_{1.1};Y_{1}|U_{2},Q)-I(X_{1.1};S_{1}|U_{1},Q).
\end{align*}
The steps (a), (b), (c) and (d) hold true because: (a) given $X_{\varTheta_{i}}$,
the mapping from $S_{i^{'}}$ to $Y_{i}$ is one-to-one, since the
channel is semi-deterministic (cf. Definition \ref{def:mim-sd-model});
(b) conditioning reduces entropy; (c) the coding scheme forces $p(u_{i}|x_{i.1},q)=p(s_{i}|x_{i.1},q)$,
therefore $H(S_{i}|X_{i.1},Q)=H(U_{i}|X_{i.1},Q)$ and $H(S_{i.1}|Q)=H(U_{i.1}|Q)$;
(d) random variables $U_{1}$ and $U_{2}$ are independent conditioned
on $Q$ and (e) according to the definition of the channel, the output
$Y_{i}$ only relies on $X_{\varTheta_{i}}$ and $X_{\varTheta_{i^{'}}}$,
i.e. $p(y_{i}|x_{\varTheta_{i}},x_{\varTheta_{i^{'}}},u_{i})=p(y_{i}|x_{\varTheta_{i}},x_{\varTheta_{i^{'}}})$.
When only the interfering transmitter Tx1.1 is considered, the conditional
distribution $p(y_{1}|x_{1.1},u_{1})$ is computed as 
\begin{align*}
p(y_{1}|x_{1.1},u_{1}) & =\sum_{x_{2.1}}p(y_{1},x_{2.1}|x_{1.1},u_{1})\\
 & =\sum_{x_{2.1}}p(x_{2.1}|x_{1.1},u_{1})p(y_{1}|x_{1.1},x_{2.1},u_{1})\\
 & =\sum_{x_{2.1}}p(x_{2.1}|x_{1.1})p(y_{1}|x_{1.1},x_{2.1})\\
 & =\sum_{x_{2.1}}(y_{1},x_{2.1}|x_{1.1})\\
 & =p(y_{1}|x_{1.1}).
\end{align*}
Hence, we have $U_{1}-\circ-X_{1.1}-\circ-Y_{1}$.

If $\varUpsilon_{1}$ has more than one user, the gap can be shown
as follows
\begin{align*}
\mathsf{A}_{\varUpsilon_{1}}+\mathsf{E}_{\varUpsilon_{2}} & =I(X_{\varUpsilon_{1}};Y_{1}|X_{\bar{\varUpsilon}_{1}},U_{1},U_{2},Q)\\
 & \qquad+I(X_{\varUpsilon_{2}},U_{1};Y_{2}|X_{\bar{\varUpsilon}_{2}},U_{2},Q)\\
 & \stackrel{(a)}{=}H(Y_{1}|X_{\bar{\varUpsilon}_{1}},U_{1},U_{2},Q)-H(S_{2}|U_{2},Q)\\
 & \qquad+H(Y_{2}|X_{\bar{\varUpsilon}_{2}},U_{2},Q)-H(S_{1}|U_{1},Q)\\
 & \stackrel{(b),(c)}{\geq}H(Y_{1}|X_{\bar{\varUpsilon}_{1}},U_{1},X_{2.1},Q)-H(S_{2}|U_{2},Q)\\
 & \qquad+H(Y_{2}|X_{\bar{\varUpsilon}_{2}},U_{2},Q)-H(S_{1}|U_{1},Q)\\
 & \stackrel{(b)}{\geq}H(Y_{1}|X_{\bar{\varUpsilon}_{1}},U_{1},X_{2.1},Q)-H(S_{2}|U_{2},Q)\\
 & \qquad+H(Y_{2}|X_{\bar{\varUpsilon}_{2}},X_{\varUpsilon_{2}},U_{2},X_{1.1},Q)\\
 & \qquad-H(S_{1}|U_{1},Q)\\
 & =H(Y_{1}|X_{\bar{\varUpsilon}_{1}},U_{1},X_{2.1},Q)-H(S_{2}|U_{2},Q)\\
 & \qquad+H(Y_{2}|X_{\varTheta_{2}},X_{1.1},Q)-H(S_{1}|U_{1},Q)\\
 & \stackrel{(a)}{\geq}H(Y_{1}|X_{\bar{\varUpsilon}_{1}},U_{1},X_{2.1},Q)-H(S_{2}|U_{2},Q)\\
 & \qquad+H(S_{1}|X_{1.1},Q)-H(S_{1}|U_{1},Q)\\
 & \stackrel{(c)}{=}H(Y_{1}|X_{\bar{\varUpsilon}_{1}},U_{1},X_{2.1},Q)-H(S_{2}|U_{2},Q)\\
 & \qquad+H(S_{1}|X_{1.1},U_{1},Q)-H(S_{1}|U_{1},Q)\\
 & =H(Y_{1}|X_{\bar{\varUpsilon}_{1}},U_{1},X_{2.1},Q)-H(S_{2}|U_{2},Q)\\
 & \qquad-I(S_{1};X_{1.1}|U_{1},Q)\\
 & \stackrel{(b)}{\geq}H(Y_{1}|X_{\bar{\varUpsilon}_{1}},X_{1.1},X_{2.1},Q)-H(S_{2}|U_{2},Q)\\
 & \qquad-I(S_{1};X_{1.1}|U_{1},Q)\\
 & \stackrel{(d)}{=}H(Y_{1}|X_{\bar{\varOmega}_{1}},X_{2.1},Q)-H(S_{2}|U_{2},Q)\\
 & \qquad-I(S_{1};X_{1.1}|U_{1},Q)\\
 & \geq\mathsf{B}_{\varOmega_{1}}-I(S_{1};X_{1.1}|U_{1},Q).
\end{align*}
The steps (a), (b), (c) and (d) hold true because: (a) given $X_{\varTheta_{i}}$,
the mapping from $S_{i^{'}}$ to $Y_{i}$ is one-to-one, since the
channel is semi-deterministic (cf. Definition \ref{def:mim-sd-model});
(b) conditioning reduces entropy; (c) random variables $Q-\circ-U_{i}-\circ-X_{i.1}$
forms a Markov chain; (d) for a given $\varUpsilon_{1}$ with more
than one user, we can always find some $\bar{\varOmega}_{1}\in2^{\varTheta_{1}}$
such that $\bar{\varOmega}_{1}=\bar{\varUpsilon}_{1}\cup\{1.1\}$
and $\varOmega\neq\emptyset$, which completes the proof.

\section{Proof of Theorem \ref{thm:.mim-gs-outer}\label{sec:mim-gs-outer-proof}}

According to Theorem \ref{thm:mim-sd-outer}, we characterize the
outer bound on the capacity region of Gaussian MAC-IC-MAC by maximizing
the set functions in Definition \ref{thm:mim-sd-outer}. The genie
random variable $T_{i}$ is chosen as $T_{i}=h_{i.1\rightarrow i^{'}}X_{i.1}+Z_{i^{'}}^{'}$,
where $Z_{i^{'}}^{'}\sim\mathcal{CN}(0,1)$ and is independent of
$Z_{i^{'}}$ and $X_{\varTheta_{i}}$. For conciseness, we compute
$\overline{A}_{\varUpsilon_{i}}$ for instance, 
\begin{align*}
\overline{\mathsf{A}}_{\varUpsilon_{i}} & =h(Y_{i}|X_{\bar{\varUpsilon}_{i}},T_{i},X_{i^{'}.1},Q)-h(S_{i^{'}.1}|X_{i^{'}.1},Q)\\
 & =\frac{1}{n}\sum_{t=1}^{n}\left[h(Y_{it}|X_{\bar{\varUpsilon}_{i}t},T_{it},X_{i^{'}.1,t})-h(S_{i^{'}t}|X_{i^{'}.1,t})\right]+\epsilon_n\\
 & =\frac{1}{n}\sum_{t=1}^{n}\bigg[h\big(\sum_{i.j\in\varUpsilon_{i}}|h_{i.j\rightarrow i}|^{2}X_{i.j,t}+Z_{it},|h_{i.1\rightarrow i^{'}}|^{2}X_{i.1,t}\\
 & \qquad+Z_{i^{'}t}^{'}\big)-h(|h_{i.1\rightarrow i^{'}}|^{2}X_{i.1,t}+Z_{i^{'}t}^{'})-h(Z_{i^{'}t})\bigg]+\epsilon_n\\
 & \stackrel{(a)}{\leq}\frac{1}{n}\sum_{t=1}^{n}\log\left(1+\sum_{i.j\in\varUpsilon_{i}\backslash\{i.1\}}\left(|h_{i.j\rightarrow i}|^{2}P_{i.j,t}\right)\right.\\
 & \qquad\left.+\frac{|h_{i.1\rightarrow i}|^{2}P_{i.1,t}}{1+|h_{i.1\rightarrow i^{'}}|^{2}P_{i.1,t}}\right)+\epsilon_n\\
 & \stackrel{(b)}{\leq}\log\bigg(1+\sum_{i.j\in\varUpsilon_{i}\backslash\{i.1\}}|h_{i.j\rightarrow i}|^{2}\frac{1}{n}\sum_{t=1}^{n}P_{i.j,t}\\
 & \qquad+\frac{|h_{i.1\rightarrow i}|^{2}\frac{1}{n}\sum_{t=1}^{n}P_{i.1,t}}{1+|h_{i.1\rightarrow i^{'}}|^{2}\frac{1}{n}\sum_{t=1}^{n}P_{i.1,t}}\bigg)+\epsilon_n\\
 & \stackrel{(c)}{\leq}\log\left(1+\sum_{i.j\in\varUpsilon_{i}\backslash\{i.1\}}|h_{i.j\rightarrow i}|^{2}P_{i.j}\right.\\
 & \qquad\left.+\frac{|h_{i.1\rightarrow i}|^{2}P_{i.1}}{1+|h_{i.1\rightarrow i^{'}}|^{2}P_{i.1}}\right)+\epsilon_n\\
 & =\log\left(1+\sum_{i.j\in\varUpsilon_{i}\backslash\{i.1\}}\mathsf{SNR}_{i.j\rightarrow i}+\frac{\mathsf{SNR}_{i.1\rightarrow i}}{1+\mathsf{INR}_{i.1\rightarrow i^{'}}}\right)\\
 & \qquad+\epsilon_n.
\end{align*}
Step (a) holds because the Gaussian distribution maximize conditional entropy,
(hence, the optimality of Gaussian input distribution); Step (b)
holds due to the log-sum inequality; Step (c) states the optimality of
full power transmission. Other terms can be verified in a similar way, which completes the proof.

\bibliographystyle{IEEEtran}
\bibliography{Documents/CURef}

\begin{thebibliography}{10}
\providecommand{\url}[1]{#1}
\csname url@samestyle\endcsname
\providecommand{\newblock}{\relax}
\providecommand{\bibinfo}[2]{#2}
\providecommand{\BIBentrySTDinterwordspacing}{\spaceskip=0pt\relax}
\providecommand{\BIBentryALTinterwordstretchfactor}{4}
\providecommand{\BIBentryALTinterwordspacing}{\spaceskip=\fontdimen2\font plus
\BIBentryALTinterwordstretchfactor\fontdimen3\font minus
  \fontdimen4\font\relax}
\providecommand{\BIBforeignlanguage}[2]{{%
\expandafter\ifx\csname l@#1\endcsname\relax
\typeout{** WARNING: IEEEtran.bst: No hyphenation pattern has been}%
\typeout{** loaded for the language `#1'. Using the pattern for}%
\typeout{** the default language instead.}%
\else
\language=\csname l@#1\endcsname
\fi
#2}}
\providecommand{\BIBdecl}{\relax}
\BIBdecl

\bibitem{liao1972multiple}
H.~H.-J. Liao, ``Multiple access channels.'' Ph.D. dissertation, HAWAII UNIV
  HONOLULU, 1972.

\bibitem{ahlswede1973multi}
R.~Ahlswede, ``Multi-way communication channels,'' in \emph{Second
  International Symposium on Information Theory: Tsahkadsor, Armenia, USSR,
  Sept. 2-8, 1971}, 1973.

\bibitem{wyner1974recent}
A.~Wyner, ``Recent results in the shannon theory,'' \emph{IEEE Transactions on
  information Theory}, vol.~20, no.~1, pp. 2--10, 1974.

\bibitem{carleial1978interference}
A.~Carleial, ``Interference channels,'' \emph{IEEE Transactions on Information
  Theory}, vol.~24, no.~1, pp. 60--70, 1978.

\bibitem{han1981new}
T.~S. Han and K.~Kobayashi, ``A new achievable rate region for the interference
  channel,'' \emph{IEEE transactions on information theory}, vol.~27, no.~1,
  pp. 49--60, 1981.

\bibitem{chong2006comparison}
H.-F. Chong, M.~Motani, and H.~K. Garg, ``A comparison of two achievable rate
  regions for the interference channel,'' in \emph{Proc. Information Theory and
  Applications Workshop}, 2006, pp. 6--10.

\bibitem{kobayashi2007further}
K.~Kobayashi and T.~Han, ``A further consideration on the {HK} and the {CMG}
  regions for the interference channel,'' in \emph{Proc. Information Theory and
  Applications Workshop}, 2007.

\bibitem{telatar2007bounds}
E.~Telatar and D.~Tse, ``Bounds on the capacity region of a class of
  interference channels,'' in \emph{Information Theory, 2007. ISIT 2007. IEEE
  International Symposium on}.\hskip 1em plus 0.5em minus 0.4em\relax IEEE,
  2007, pp. 2871--2874.

\bibitem{chong2008han}
H.-F. Chong, M.~Motani, H.~K. Garg, and H.~E. Gamal, ``On the han-kobayashi
  region for the interference channel,'' \emph{IEEE Transactions on Information
  Theory}, vol.~54, no.~7, pp. 3188--3194, 2008.

\bibitem{etkin2008gaussian}
R.~H. Etkin, D.~N. Tse, and H.~Wang, ``Gaussian interference channel capacity
  to within one bit,'' \emph{Information Theory, IEEE Transactions on},
  vol.~54, no.~12, pp. 5534--5562, 2008.

\bibitem{karmakar2012generalized}
S.~Karmakar and M.~K. Varanasi, ``The generalized degrees of freedom region of
  the mimo interference channel and its achievability,'' \emph{Information
  Theory, IEEE Transactions on}, vol.~58, no.~12, pp. 7188--7203, 2012.

\bibitem{kramer2006review}
G.~Kramer, ``Review of rate regions for interference channels,'' in \emph{2006
  international Zurich seminar on communications}, 2006.

\bibitem{karmakar2013capacity}
S.~Karmakar and M.~K. Varanasi, ``The capacity region of the mimo interference
  channel and its reciprocity to within a constant gap,'' \emph{Information
  Theory, IEEE Transactions on}, vol.~59, no.~8, pp. 4781--4797, 2013.

\bibitem{suh2008interference}
C.~Suh and D.~Tse, ``Interference alignment for cellular networks,'' in
  \emph{Communication, Control, and Computing, 2008 46th Annual Allerton
  Conference on}.\hskip 1em plus 0.5em minus 0.4em\relax IEEE, 2008, pp.
  1037--1044.

\bibitem{perron2009interference}
E.~Perron, S.~Diggavi, and E.~Telatar, ``The interference-multiple-access
  channel,'' in \emph{Communications, 2009. ICC'09. IEEE International
  Conference on}.\hskip 1em plus 0.5em minus 0.4em\relax IEEE, 2009, pp. 1--5.

\bibitem{chaaban2011capacity}
A.~Chaaban and A.~Sezgin, ``On the capacity of the 2-user gaussian mac
  interfering with a p2p link,'' in \emph{Wireless Conference 2011-Sustainable
  Wireless Technologies (European Wireless), 11th European}.\hskip 1em plus
  0.5em minus 0.4em\relax VDE, 2011, pp. 1--6.

\bibitem{gherekhloo2014sub}
S.~Gherekhloo, A.~Chaaban, C.~Di, and A.~Sezgin, ``({S}ub-)optimality of
  treating interference as noise in the cellular uplink with weak
  interference,'' \emph{IEEE Transactions on Information Theory}, vol.~62,
  no.~1, pp. 322--356, Jan 2016.

\bibitem{chaaban2011capacitya}
A.~Chaaban and A.~Sezgin, ``Capacity results for a primary mac in the presence
  of a cognitive radio,'' in \emph{Global Telecommunications Conference
  (GLOBECOM 2011), 2011 IEEE}.\hskip 1em plus 0.5em minus 0.4em\relax IEEE,
  2011, pp. 1--5.

\bibitem{buhler2012multiple}
J.~B{\"u}hler and G.~Wunder, ``The multiple access channel interfering with a
  point to point link: Linear deterministic sum capacity,'' in
  \emph{Communications (ICC), 2012 IEEE International Conference on}.\hskip 1em
  plus 0.5em minus 0.4em\relax IEEE, 2012, pp. 2365--2369.

\bibitem{fritschek2014upper}
R.~Fritschek and G.~Wunder, ``Upper bounds and duality relations of the linear
  deterministic sum capacity for cellular systems,'' in \emph{Communications
  (ICC), 2014 IEEE International Conference on}.\hskip 1em plus 0.5em minus
  0.4em\relax IEEE, 2014, pp. 1884--1889.

\bibitem{fritschek2015constant}
------, ``Constant-gap sum-capacity approximation of the deterministic
  interfering multiple access channel,'' in \emph{Information Theory (ISIT),
  2015 IEEE International Symposium on}.\hskip 1em plus 0.5em minus 0.4em\relax
  IEEE, 2015, pp. 2643--2647.

\bibitem{niesen2013interference}
U.~Niesen and M.~A. Maddah-Ali, ``Interference alignment: From degrees of
  freedom to constant-gap capacity approximations,'' \emph{Information Theory,
  IEEE Transactions on}, vol.~59, no.~8, pp. 4855--4888, 2013.

\bibitem{fritschek2014enabling}
R.~Fritschek and G.~Wunder, ``Enabling the multi-user generalized degrees of
  freedom in the gaussian cellular channel,'' in \emph{Information Theory
  Workshop (ITW), 2014 IEEE}.\hskip 1em plus 0.5em minus 0.4em\relax IEEE,
  2014, pp. 107--111.

\bibitem{zhu2014capacity}
F.~Zhu, X.~Shang, B.~Chen, and H.~V. Poor, ``On the capacity of
  multiple-access-z-interference channels,'' \emph{Information Theory, IEEE
  Transactions on}, vol.~60, no.~12, pp. 7732--7750, 2014.

\bibitem{pang2013bounds}
Y.~Pang and M.~K. Varanasi, ``Bounds on the capacity region of a class of
  multiple access interference channels,'' in \emph{Communication, Control, and
  Computing (Allerton), 2013 51st Annual Allerton Conference on}.\hskip 1em
  plus 0.5em minus 0.4em\relax IEEE, 2013, pp. 599--606.

\bibitem{gamal1982capacity}
A.~Gamal and M.~Costa, ``The capacity region of a class of deterministic
  interference channels (corresp.),'' \emph{IEEE Transactions on information
  Theory}, vol.~28, no.~2, pp. 343--346, 1982.

\bibitem{el2011network}
A.~El~Gamal and Y.-H. Kim, \emph{Network information theory}.\hskip 1em plus
  0.5em minus 0.4em\relax Cambridge university press, 2011.

\end{thebibliography}

\begin{IEEEbiographynophoto}{Yimin Pang} (S'11) received his Bachelor's degree in Electrical Engineering from Xidian University, Xi'an, China, in 2008 and his Master's degree from Zhejiang University, Hangzhou, China, in 2011. He is currently working towards his doctoral degree in the Department of Electrical, Computer, and Energy Engineering at the University of Colorado Boulder, CO. His research interests are in information theory and wireless communications.
\end{IEEEbiographynophoto}

\begin{IEEEbiographynophoto}{Mahesh K. Varanasi}
(S'87-M'89-SM'95-F'10) received the B.E. degree in electronics and communications engineering from Osmania University, Hyderabad, India, in 1984 and the M.S. and Ph.D. degrees in electrical engineering from Rice University, Houston, TX, USA, in 1987 and 1989, respectively. 

Since 1989, he has been a member of the faculty of the University of Colorado, Boulder, CO, USA, where he is currently a Professor of Electrical, Computer, and Energy Engineering. He has been a Professor
(by courtesy) of the Department of Mathematics since 2010 and an affiliated faculty member of the Department of Applied Mathematics since 2013.
His research and teaching interests are in the areas of network information theory, wireless communications and coding, statistical inference and learning theory, and signal processing. 

Dr. Varanasi is a Highly Cited Researcher according to the ISI Web of Science. He is the recipient of the 2017 and 2018 Qualcomm Faculty Awards. Dr. Varanasi served as an Editor of the IEEE TRANSACTIONS ON WIRELESS
COMMUNICATIONS during 2007-2009. He was an Associate Editor of the IEEE TRANSACTIONS ON INFORMATION THEORY during 2014-2017. He was a General Chair of the 2018 IEEE International Symposium on Information Theory (ISIT), Vail, Colorado.
\end{IEEEbiographynophoto}

\end{document}